\newif\ifpreprint

\def\journal{\preprintfalse}
\journal      % Uncomment if you want the ELSEVIER version.
\ifpreprint
   \documentstyle[aps,epsfig,floats]{revtex} 
\else
   \documentclass[11pt]{elsart}
   \usepackage{graphicx,amssymb}
   \usepackage{multicol}
   \usepackage[titles]{tocloft}
   \journal{Physics Reports}
\fi

\renewcommand{\cftsecfont}{\small}
\renewcommand{\cftdot}{}
\newcommand {\be} {\begin{equation}} 
\newcommand {\ee} {\end{equation}} 
\newcommand {\Be}{\begin{eqnarray*}}
\newcommand {\Ee} {\end{eqnarray*}}
\newcommand {\bey} {\begin{eqnarray}} 
\newcommand {\eey} {\end{eqnarray}} 
\newcommand{\bearr}{\begin{array}}
\newcommand{\enarr}{\end{array}}

\def\<{\langle}
\def\>{\rangle}
\def\~{\tilde}

\def\vq{{\bf q}}

\def\vp{{\bf p}}
\def\vr{{\bf r}}
\def\vx{{\bf x}}
\def\vj{{\bf j}}
\def\vJ{{\bf J}}

\def\dekinchin {{d\sigma\over \parallel\nabla {\mathcal H}\parallel}}
\def\pquadro {\sum_{i=1}^{N} p_i^2\over {N m}}

\def\#{\nonumber}

\begin{document}

\ifpreprint
  \title{\bf Thermal conduction in classical low-dimensional lattices}
  \author{Stefano Lepri$^{1,2}$, Roberto Livi$^{3,2}$, Antonio Politi$^{4,2}$}
 
  \address{$^1$ Dipartimento di Energetica ``S. Stecco", via S. Marta 3 I-50139
  Florence, Italy \\
  $^2$ Istituto Nazionale di Fisica della Materia-Unit\`a di
  Firenze, Via G. Sansone 1 I-50019 Sesto Fiorentino Italy \\
  $^3$ Dipartimento di Fisica, Via G. Sansone 1 I-50019 Sesto Fiorentino Italy \\
  $^4$ Istituto Nazionale di Ottica Applicata, largo E. Fermi 6 I-50125
  Florence, Italy \\}
  \date{\today}
  \maketitle                                                                      
\else
  \begin{frontmatter}
  \title{\Large Thermal conduction in classical low-dimensional lattices}
  \author[DE,INFM]{Stefano Lepri}
  \author[DF,INFM]{Roberto Livi}
  \author[INO,INFM]{Antonio Politi}
 
  \address[DE]{Dipartimento di Energetica ``S. Stecco", via S. Marta 3 I-50139
  Florence, Italy}
  \address[INFM]{Istituto Nazionale di Fisica della Materia-Unit\`a di
  Firenze, Via G. Sansone 1 I-50019 Sesto Fiorentino, Italy}
  \address[DF]{Dipartimento di Fisica,  Via G. Sansone 1 I-50019 
  Sesto Fiorentino, Italy}
  \address[INO]{Istituto Nazionale di Ottica Applicata, largo E. Fermi 6 I-50125
  Florence, Italy}
  \date{\today}

\vspace{0.75cm}
\begin{minipage}{\textwidth}
\begin{multicols}{2}
\tableofcontents
\end{multicols}

\fi

\begin{abstract}
Deriving macroscopic phenomenological laws of irreversible thermodynamics
from simple microscopic models is one of the tasks of non-equilibrium
statistical mechanics. We consider stationary energy transport in crystals
with reference to simple mathematical models consisting of coupled oscillators
on a lattice. 
The role of lattice dimensionality on the breakdown of the
Fourier's law is discussed and some universal quantitative aspects are
emphasized: the divergence of the finite-size thermal conductivity is
characterized by universal laws in one and two dimensions. 
Equilibrium and non-equilibrium molecular dynamics
methods are presented  along with a critical survey of previous numerical results.
Analytical results for the non-equilibrium dynamics can be 
obtained in the harmonic chain where the role of disorder and localization 
can be also understood.
The traditional kinetic approach, based on
the Boltzmann-Peierls equation is also briefly sketched with reference to
one-dimensional chains. 
Simple toy models can be defined in which the conductivity is finite.
Anomalous transport in integrable nonlinear systems is briefly discussed.                 
Finally, possible future research themes are outlined.

\end{abstract}

\ifpreprint
  \tableofcontents
\else
  \begin{keyword}
  Thermal conductivity, classical lattices
  \PACS 63.10.+a \sep 05.60.-k \sep 44.10.+i
  \end{keyword}
  \end{minipage}
  \end{frontmatter}
  \newpage
\fi

\section{Introduction}
The customary macroscopic approach to non-equilibrium phenomena relies
crucially on the definition of transport coefficients through 
phenomenological constitutive equations. Under the hypothesis of being close
enough to global equilibrium, this is usually accomplished by postulating
the proportionality among fluxes and thermodynamic forces \cite{DGM}. For
instance when dealing with energy transport in a solid one defines the
thermal conductivity $\kappa$ through the {\it Fourier's law}\footnote{The 
thermal conductivity should be represented in general as a tensor.
Here we assume to consider a simple cubic crystal, in the absence of
any external force field. Accordingly, $\kappa$  has a
diagonal representation with equal diagonal elements.}
\begin{equation}
\vJ_Q  \;=\; -\kappa \nabla T \quad,
\label{fourier} 
\end{equation}
where the heat flux $\vJ_Q$ is the amount of heat transported through the
unit surface per unit time and $T(\vx,t)$ is the local temperature. Such a
phenomenological relation was first proposed in 1808 by J.B.J. Fourier  as
an attempt to explain the thermal gradient present inside the Earth  - a
problem that had raised a long and controversial debate inside the
scientific community at that time. Eq.~(\ref{fourier}) is assumed to be
valid close to equilibrium. Actually, the very definition of local 
energy flux $\vJ_Q(\vx,t)$ and of temperature field $T(\vx,t)$ relies
on the {\it local equilibrium hypothesis} i.e. on the possibility of
defining a local temperature for a macroscopically small but
microscopically large volume at each location $\vx$ for each time $t$. 

The ultimate goal of a complete theory would be to derive an equation like 
(\ref{fourier}) from some statistical-mechanics calculation, a task which may
be formidably difficult. For insulating crystals where heat is transported
by lattice vibrations, the first and most elementary attempt to give a
microscopic foundation to Fourier's law dates back to Debye. By rephrasing the
results of the kinetic theory for the (diluted) phonon gas, he found that
the thermal conductivity should be proportional to $Cv\ell$ where $C$ is the 
heat capacity and $v$, $\ell$ are the phonon mean velocity and free path,
respectively. In 1929, R. Peierls further extended this idea and formulated a
Boltzmann equation \cite{peierls} that shows how anharmonicity is necessary to
obtain genuine diffusion of the energy through the so-called {\it Umklapp}
processes. Since then, the Boltzmann-Peierls approach became one the
cornerstones in the theory of lattice thermal conductivity. Standard
methods, like the relaxation--time approximation, allow to
compute, say, the temperature dependence of $\kappa$.  

From a more fundamental point of view, there are however basic questions that
go beyond the actual calculation of the transport coefficient. For example,
under what condition is local equilibrium realized? Can we ensure that a unique
non-equilibrium stationary state is attained on physically accessible time 
scales? In this respect, simple mathematical models are an invaluable 
theoretical playground to provide a more firm foundation to heat conductivity 
and to understand more deeply the hypotheses underlying Eq.~(\ref{fourier}).
Admittedly, this program is still nowadays far from being accomplished, at
least from a mathematically rigorous point of view \cite{BLR00}. On the other
hand, even in the absence of solvable examples, one can rely on
numerical simulations as a tool to investigate many of those items.  

As usual in theoretical physics, the guiding criterion of mathematical 
simplicity leads naturally to consider 1d or 2d periodic lattices (i.e. chains 
or planes) of point-like atoms interacting with their neighbors through 
nonlinear forces. The hope is to reproduce realistic thermodynamic properties 
of their three-dimensional counterparts without having to refer
to specific structures. This brings to the fore the following question: 
what are the minimal requirements for a dynamical model to satisfy or not 
Eq.~(\ref{fourier})?  Although it may appear surprising, this
issue has been addressed in the literature already in the late 60s without
yet receiving a definite answer. To a large extent, the present review aims at 
settling this question by reconciling the very many (and sometimes
contrasting) numerical results that appeared since then. In fact, several 
times in the past wrong interpretations have been given to
the outcome of numerical simulations. In the absence of a general theoretical 
framework, it has been overemphasized the role of deterministic chaos in
ensuring a normal heat transport. Indeed, while ergodicity (implied
to some extent by a chaotic dynamics) is certainly a necessary condition
to establish energy diffusion, it has become clear that it is not at all 
sufficient, as too-rapidly claimed more than a decade ago \cite{M84}.

As it is known in
the context of fluids \cite{PR75}, much of the difficulties arise from the
fact that transport coefficients in low spatial dimensions may {\it not
exist at all}, thus implying a breakdown of usual hydrodynamics. In the
present context, this usually manifests itself as: {\it (i)} a slow decay of
equilibrium correlations of the heat current; {\it (ii)} a divergence of the
{\it finite-size conductivity} $\kappa(N,T)$ in the thermodynamic limit
$N\to\infty$ (where $N$ is the  number of atoms in the sample). One of our
concerns will thus be to clarify, through the analysis of several examples,
under what conditions this should occur. Particular emphasis has been put
on the universality of quantitative data like the decay law of 
equilibrium correlation of the flux.

A by-no-means side issue that is also considered in the present review is the
coupling with thermal baths. It is only after having properly set the
interaction between the system of interest and thermal reservoirs that one
can be sure that a physically meaningful non-equilibrium regime is established. 
Ideally, a thermal bath is a set of (infinitely) many degrees of freedom, so 
that it can either absorb or release energy, without appreciably changing its 
own state.  Unfortunately, such a type of reservoir is very difficult to 
treat analytically and too much time demanding in numerical simulations. 
Accordingly, various shortcuts have been proposed and are here recalled, 
ranging from stochastic to nonlinear deterministic rules. 

Up to now we have mainly emphasized the theoretical issues that motivate the
study of transport in low-dimensional lattices. Of course, a further
relevant motivation is the existence of a variety of real systems that could
be, at least in principle, effectively described by 1d or 2d models. For
instance, reduced dimensionality has been indeed invoked to explain
experiments on heat transport in anisotropic crystals \cite{MHSU86,SLM96} or
magnetic systems \cite{SGOVR01}. Remarkably, a dependence of thermal
conductivity on the chain length of solid polymers has also been
experimentally observed \cite{FA84}. More generally, modern experimental
techniques \cite{TWR97} allow to directly probe the transport properties of
semiconductor films \cite{LP98,HGRK98}  and single-walled nanotubes
\cite{HWPZ99,KSMM01} that markedly display two-dimensional features at low
temperatures. Some theoretical investigations  of thermal conductance for a
quantum wire in ballistic \cite{RK98} and  anharmonic \cite{LW00} regimes
have been also recently undertaken. Another important example is the
problem of heat conductivity in quantum spin chains \cite{STM00}.

The present work is not simply a review paper in the customary sense:
many results were not previously published and older results are critically 
reconsidered. This is of course particularly crucial when 
dealing with numerical data.
The plan of the article is the following. In Chapter 2 we present the 
simple lattice models that will be considered throughout the review. To be as 
self-contained as possible, we derive the microscopic expression of the 
heat flux with reference to the specific case of one-dimensional systems
with nearest-neighbor interactions. The advantage of this presentation is 
twofold: it provides the expression to be referred to in the following and
allows to understand the hypotheses behind its derivation without having 
to dwell into more involved notations. As already mentioned, an important 
point for non-equilibrium molecular dynamics is the way thermal reservoirs are 
modeled. Chapter 3 contains a brief survey of some simple schemes that have 
been used in the literature on the topic. The relevant differences among the
most widely used methods are also discussed. 
Most of our understanding of energy transport in lattices relies on the 
harmonic approximation for the microscopic dynamics. One major advantage of 
treating the simple harmonic chain is that non-equilibrium properties
can be derived in a non-perturbative way. This is reviewed in Chapter 4.
Harmonic chains are also presumably the only class of systems in which the 
consequences of the presence of quenched disorder can be effectively studied.
A discussion of chains with isotopic disorder and of the role of localization 
on the heat conduction is given.

A very sketchy discussion of the two ``traditional'' approaches, the 
Boltzmann-Peierls kinetic theory for phonons and the Green-Kubo 
method, is presented in Chapter 5. Since detailed accounts 
exist in many textbooks, we limited ourselves to recall those issues
that are relevant in the present context.
In Chapter 6 we present a detailed account of the many numerical studies
performed with models where total momentum is conserved. Both equilibrium and
non-equilibrium simulations are discussed and compared. Chapter 7 is, instead,
devoted to the ``complementary'' class of systems where the interaction with
an external substrate breaks total momentum conservation. This turns out to be
a crucial difference in ensuring normal heat transport. 

The peculiar behavior of integrable systems is briefly summarized in
Chapter 8, while the role of dimensionality of the physical space can be
appreciated in Chapter 9, where we illustrate the behavior of some $2d$
models. The last chapter is finally devoted to summarizing the key points that
have been so far understood and, more importantly, to the still open
questions. The more technical discussions have been confined to the
Appendices in order not to downgrade the readability of the main body of the
text.

\section{Definitions}
\subsection{Models}
The present Report deals mainly with classical arrays of coupled oscillators. 
To be more specific, we introduce the models for the one-dimensional case.
The generalization to two dimension is rather straightforward and will be
recalled later when needed. 

\begin{figure}
\begin{center}
\includegraphics*[width=8cm]{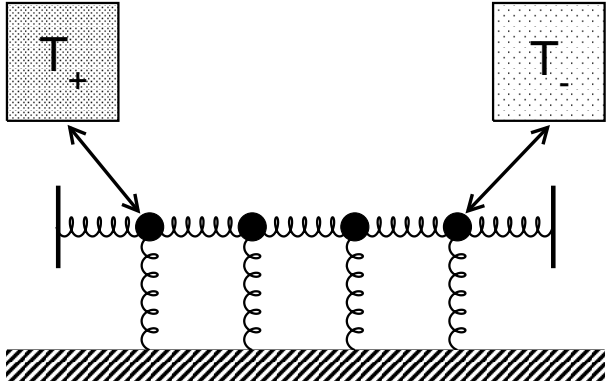}
\caption{A pictorial representation of a chain of $N=4$ mutually coupled
oscillators in interaction with an external substrate and coupled with two
thermal reservoirs working at different temperatures.}
\label{setup}
\end{center}
\end{figure}

A schematic setup of the systems that will be mostly studied in the following
is drawn in Fig.~\ref{setup}, where we have depicted a chain of $N$
coupled atoms, the first and the last of which interact also with a 
thermal bath. 
Let $m_l$ and $x_l$ be respectively the mass and the position of the $l$-th
particle. Only nearest-neighbor interactions will be considered for simplicity.
The first class of models we wish to consider are defined by an Hamiltonian  of
the form ($p_l=m_l\dot x_l$)
\begin{equation}
{\mathcal H}\;=\; \sum_{l=1}^N \left[{p_l^2\over 2m_l}+
V(x_{l+1}-x_{l})\right] \quad .
\label{acoustic}
\end{equation}
Boundary conditions need also to be specified by defining $x_{0}$ and 
$x_{N+1}$. Typical choices are periodic, fixed or free boundaries.
As only internal forces, that depend on relative positions, are present, the 
total momentum is conserved and thus a zero (Goldstone) mode exist. In the 
harmonic limit, model (\ref{acoustic}) admits at least a phonon branch whose 
frequency vanishes for vanishing wavenumber. Long-wavelength waves move at 
the sound velocity and  for this reason one sometimes refer to (\ref{acoustic}) 
as {\it acoustic} models.                              

An important example is the well-known Lennard-Jones potential
\begin{equation}
V(z)\;=\; \epsilon\left[\left({a\over z}\right)^{12} -
2\left({a\over z}\right)^6\right] \quad.
\label{lenjo}
\end{equation} 
In this formulation $a$ is the equilibrium distance and $\epsilon$ the 
well depth. The other example we will often consider is the celebrated 
Fermi-Pasta-Ulam (FPU) potential \cite{FPU65}
\begin{equation}
V(z)\;=\; {g_2\over 2}\,(z-a)^2+ {g_3\over 3}\,(z-a)^3 + 
{g_4\over 4}\, (z-a)^4 \, ,
\label{fpu}
\end{equation} 
that can be regarded as resulting from an expansion of $V$  close to its
equilibrium position $z=a$. Due to its simple algebraic form, the model is 
computationally very convenient. Two important particular cases are worth
mentioning: the quadratic plus cubic ($g_4=0$) and quadratic plus quartic
($g_3=0$) potentials that, for historical reasons, are referred to as the
FPU-$\alpha$ and FPU-$\beta$ models, respectively. In the former one,
sufficiently small coupling constant $g_3$ and/or energies must be considered
to avoid runaway instability of trajectories.  

Models like (\ref{acoustic}) are a very drastic idealization of a real
crystal. Natural low-dimensional lattice structures are usually embedded in 
three-dimensional matrices that couple them to the environment. Furthermore, 
artificial arrays of atoms can be constructed by growing them on a substrate 
exerting a pinning force on the atoms in such a way to stabilize the lattice 
(in general, this is not necessary in the three-dimensional case). At the 
simplest level of modelization, this can be described by adding an external, 
on-site, potential. For instance, neglecting
the transverse motion leads to one-dimensional models of the form
\begin{equation}
{\mathcal H} = \sum_{l=1}^N \left[{p_l^2\over 2m_l} + U(x_l) +
V(x_{l+1}-x_{l})\right] \quad .
\label{optical}
\end{equation}
The substrate potential $U$ breaks the invariance $x_l \to x_l+const.$ of 
(\ref{acoustic}) and
the total momentum is no longer a constant of the motion. Accordingly, all
branches of the dispersion relation have a gap at zero wavenumber. We 
therefore refer to (\ref{optical}) as {\it optical} models.                    

Dimensionless variables will be used throughout whenever possible, especially 
when reporting simulation data. The choice of the most natural units is 
usually dictated by the particular model at hand. For example, for the FPU 
model with $m_l=m$, it is convenient to set
$a$, $m$ and the angular frequency $\omega_0=\sqrt{g_2/m}$ to unity. This 
implies, for instance, that the sound velocity $a\omega_0$ is also unity and 
that the energy is measured in units of $m\omega_0^2 a^2$.    

\subsection{Temperature}
The first problem that has to be solved in order to interpret 
molecular-dynamics simulations in a thermodynamic perspective is the 
definition of 
temperature in terms of dynamical variables. In appendix A, we show that
the problem can be tackled rigorously, although at the expense of introducing 
some technicalities (see also \cite{R97,GL99}). Here below, we follow
the traditional approach based on the virial theorem,
\begin{equation} 
T = \left\langle {\bf u} \cdot \nabla {\mathcal H} \right\rangle _\mu  
\label{tempvir}
\end{equation}
where $\bf u$ is any vector fulfilling the condition $\nabla\cdot{\bf u}=1$,
the symbol $\langle \cdot \rangle _\mu$ indicates the $\mu$-canonical ensemble 
average, and units are chosen in such a way that the Boltzmann constant 
is $k_B=1$. 

$\mu$-canonical averages are the most appropriate ones whenever an isolated 
system is numerically investigated, but as soon as the system is put
in contact with one or more heat baths, canonical averages should be 
considered instead. Fortunately, it is well known, though only partially 
proved, that averages are independent of the chosen ensemble in the 
thermodynamic limit. Nevertheless, when working with finite systems one has 
to be aware of the existence of finite-size corrections as discussed in the 
celebrated paper by Lebowitz, Percus and Verlet \cite{LPV67}.

Moreover, in molecular-dynamics simulations, averages are more conveniently 
computed 
by following single trajectories over time. This is not a problem whenever 
ergodicity can be invoked, so as to ensure that ensemble and time averages 
are equivalent to one another. However, time averages of quantities 
corresponding to thermodynamic observables have been found to converge to 
the expected ensemble averages even in systems that are known not to be ergodic in a 
strictly mathematical sense (as e.g. the FPU-$\beta$ model at sufficiently 
small energy values) and even when fluctuations around the mean value are not
consistent with equilibrium predictions. This suggests that a weaker condition 
than ergodicity might suffice to ensure the equivalence of time and
ensemble averages of the physically relevant observables\footnote{
A possible candidate might be the weak ergodicity criterion, proposed by 
Khinchin \cite{Khin}, that is equivalent to
assume a sufficiently fast decay of the time--correlation functions
at least for thermodynamically relevant observables, like temperature
internal energy, specific heat etc.}. 

According to Eq.~(\ref{tempvir}), many formally different, but physically 
equivalent, definitions of temperature are possible. For instance, the choice 
${\bf u} = (0,\ldots ,0,p_1/N,\ldots ,p_N/N)$ yields 
the usual definition adopted in the canonical ensemble. 
\begin{equation} 
T =  \left\langle {\pquadro} \right\rangle _{\mu} \quad , 
\end{equation}
while the choice ${\bf u} = (0,\ldots ,0,p_i,0,\ldots ,0)$ 
leads to a local definition of temperature,
\begin{equation}
T  = \left\langle \frac{p_i^2}{m} \right \rangle_{\mu} .
\end{equation}
The identification of an optimal definition of temperature to be adopted in 
numerical studies is strictly related to the convergence properties of  
time-averages. In this sense, it has been observed that definitions like
the above ones involving only momenta converge always quite rapidly, 
also when the
dynamics is weakly chaotic, while definitions involving an explicit
dependence on space coordinates may converge over much longer time
scales \cite{GL99}.

\subsection{Flux}
The goal of this section is to give a meaningful definition of heat flux
\cite{C63,H63}. This requires some care since it involves the transformation
of an implicit ``mesoscopic'' definition into a workable microscopic 
definition. For the sake of simplicity, we discuss the problem with reference 
to one-dimensional systems with nearest-neighbor interactions, the extension 
to the more general case
being more technically involved but conceptually equivalent. The heat flux $j(x,t)$
at time $t$ in the spatial position $x$ is nothing but the energy current, 
implicitly defined by the continuity equation
\begin{equation} 
\frac{d h(x,t)}{d t} + \frac{\partial j(x,t)}{\partial x} \; = \; 0
\label{hydro}
\end{equation}
where $h(x,t)$ is the energy density.  It is important to realize that 
the energy flux defined as above {\it does not}, in general, coincide with
heat flux, as the former arises also from macroscopic motion \cite{DGM}. 
Nonetheless, in solids and one-dimensional fluids no steady motion can 
occour, so that the two fluxes do coincide and we can 
interchangeably use both names.

With reference to an ensemble of interacting particles, we can write the 
microscopic energy density as the sum of the isolated contributions located 
in the instantaneous position of each particle
\begin{equation}
h(x,t)  = \sum_n h_n \delta(x-x_n) \quad ,
\label{e:sumdel}
\end{equation}
where $\delta(x)$ is the Dirac distribution and
\begin{equation}
h_n =  \frac{p_n^2}{2m_n} + U(x_n) + {1\over 2} \bigg[
V(x_{n+1} - x_n)  + V(x_n - x_{n-1}) \bigg] 
\label{hamilton}
\end{equation}
is the energy contribution of the $n$th particle.
The first two terms on the r.h.s. correspond to the kinetic energy and,
respectively, to the potential  energy $U(x_n)$ associated with the (possible)
interaction with an external  field. The last term amounts to half of the
potential energy of the pairwise interactions with the neighboring particles.
In a similar way, we can write the heat flux as the sum of 
localized contributions,
\begin{equation}
j(x,t)  = \sum_n j_n \delta(x-x_n) \quad .
\label{e:sumdel2}
\end{equation}
The problem amounts therefore to give a definition of the local heat flux 
$j_n$. One should keep in mind that the latter quantity has not the same 
physical dimensions of $j(x,t)$. 

In the limit of small oscillations around the equilibrium position, density 
fluctuations can be neglected and $h_n$ is equal to the energy density times
the lattice spacing $a$. The time derivative of $h_n$ 
\begin{equation} 
\frac{d h_n}{d t} = m_n {\dot x}_n {\ddot x}_n + {\dot x}_n U'(x_n) -
 {1\over 2} \bigg[ ({\dot x}_{n+1} - {\dot x}_n) F(x_{n+1} - x_n) +
 ({\dot x}_n - {\dot x}_{n -1}) F(x_n - x_{n-1}) \bigg] ,
\label{dhdt}
\end{equation} 
where the prime denotes derivative with respect to the argument and
$F(x) = -V'(x)$  can be rewritten, with the help of the equations of motion
derived from (\ref{optical})
\begin{equation} 
m_n {\ddot x}_n = - U'(x_n) - F(x_{n+1} - x_n) + F(x_n - x_{n-1})  \, ,
\label{eqmot}
\end{equation} 
as
\begin{equation} 
\frac{d h_n}{d t} = 
 -{1\over 2} \bigg[ ({\dot x}_{n+1} + {\dot x}_n) F(x_{n+1} - x_n) -
 ({\dot x}_n + {\dot x}_{n -1}) F(x_n - x_{n-1}) \bigg] .
\label{hf1}
\end{equation} 
This equation can, in turn, be rewritten as
\begin{equation} 
\frac{d h_n}{d t} + {j _n - j_{n-1} \over a} = 0
\label{hydrolat}
\end{equation}
where   
\begin{equation} 
j_n = a\phi_n := {1\over 2} a ({\dot x}_{n+1} + {\dot x}_n) \, F(x_{n+1} - x_n) 
\label{hfdef}
\end{equation} 
which can thus be interpreted as the local heat flux.

More in general, if density fluctuations cannot be neglected, a different
approach has to be followed in order to determine a workable expression
for $j_n$. The key idea consists in Fourier transforming Eq.~(\ref{hydro}),
\begin{equation} 
\frac{d {\tilde h}(k,t)}{d t} \;=\; -i k  \, {\tilde j}(k,t) ,
\label{hydroFo}
\end{equation}
where
\begin{equation}
{\tilde h}(k,t) = \int dx \, h(x,t) e^{-i kx}  \quad \quad 
{\tilde j}(k,t) = \int dx \, j(x,t) e^{-i kx}  
\label{thd}
\end{equation} 
In fact, according to the idea that the heat flux is a macroscopic observable,
one can define it as the component of ${\tilde h}(k,t)$ that is proportional 
to the wave-vector $k$, i.e. the leading contribution over sufficiently large
scales. From Eq.~(\ref{e:sumdel}),
\begin{equation} 
\frac{d {\tilde h}(k,t)}{d t} = \sum_{n}{\bigg( \frac{d h_n}{d t} 
-i k {\dot x}_n h_n \bigg) e^{- i kx_n}} \quad .
\end{equation}
From Eq.~(\ref{hf1}), the first sum in the r.h.s. can be written by a
suitable shift of indexes  as
\begin{equation} 
\sum_n  \frac{d h_n}{d t} e^{- i kx_n}  = \frac{1}{2}\sum_n (\phi_n -
 \phi_{n -1})
e^{- i k x_n} = \frac{1}{2}\sum_n \phi_n e^{- i kx_n}
\bigg(1 - e^{- i k (x_{n+1} - x_n)} \bigg)  \quad .
\label{e:lwq}
\end{equation}
In the low-$k$ limit, the above expression reduces to 
\begin{equation} 
\sum_n \frac{d h_n}{d t} e^{- i kx_n} \approx
 -\frac{ik}{2}\sum_n (x_{n+1} - x_n) \phi_n e^{- i k x_n} \quad .
\end{equation}
By replacing this expression back into Eq.~(\ref{e:lwq}) and with the help
of Eq.~(\ref{hydroFo}), we find that
\begin{equation} 
j_n = \frac{1}{2} (x_{n+1} - x_n) ({\dot x}_{n+1} + {\dot x}_n) \,
     F(x_{n+1} - x_n) + {\dot x}_n h_n \, .
\label{hf2}
\end{equation}
In the limit of small oscillations (compared to the lattice spacing), the
second term in the above formula can be neglected and $x_n-x_{n-1}\simeq a$,
so that Eq.~(\ref{hf2}) reduces to the definition (\ref{hfdef}). Another class 
of systems for which energy fluctuations can be neglected is that for which 
the canonical variable ``$x_n$'' does not describe the longitudinal motion 
along the chain, but corresponds to different degrees of freedom (e.g.,
transversal oscillations or the rotation of a classical spin). In all such 
cases, the position where the energy $h_n$ is localized along the chain is 
fixed in time and, accordingly, no term proportional to $h_n$ can arise. 

In order to give a flavor of the relative weight of the two contributions
to the heat flux, we have studied a chain of equal-mass particles interacting 
through the Lennard-Jones potential (\ref{lenjo}).
In the low-temperature limit, the chain behaves indeed as a solid, and 
we expect that the ratio
\begin{equation} 
r_j = \frac{\langle{2\dot x}_nh_n \rangle}
{\langle(x_{n+1} - x_n) ({\dot x}_{n+1} + {\dot x}_n) F(x_{n+1} - x_n)\rangle} 
\label{rhf}
\end{equation}
goes to 0. In the opposite limit $T \to \infty$, the only relevant interaction
is the repulsive part of the potential that is responsible for elastic 
collisions, i.e. the system becomes equivalent to a hard-point gas. In 
this limit, the only relevant contribution to the heat flux arises from the 
kinetic term of $h_n$, i.e.
\begin{equation} 
  j_n  \;\approx\; \frac{1}{2} m_n{\dot x}_n^3 .
\label{lj612}
\end{equation}
This can be understood by looking at the integral of the flux over the average
time between consecutive collisions of the same two particles. The first 
term in the r.h.s. of Eq.~(\ref{hf2}) is not negligible only during the
infinitesimal collision time when, however, the distance $x_{n+1}-x_n$ 
vanishes. Therefore, in spite of the Dirac $\delta$ behaviour of the force, 
its integral contribution remains negligible. Analogously, the term 
${\dot x}_n V(x_{n+1}-x_n)$, arising from the potential energy, does not 
contribute, since $V$ remains finite during the collision. Such a conclusion 
is confirmed by numerical simulations which show that the average value of 
$j_n$ as defined in Eq.~(\ref{lj612}) coincides with the energy flux 
through the boundaries.
 
\begin{figure}
\begin{center}
\includegraphics*[width=7cm]{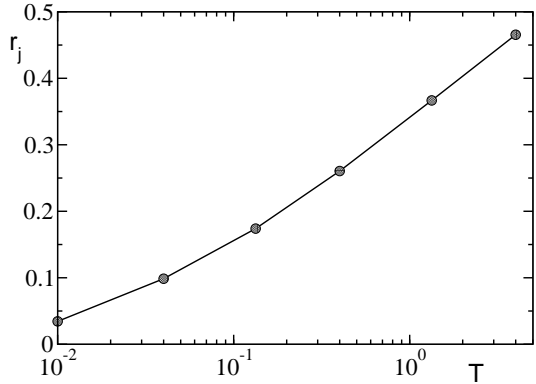}
\caption{The ratio $r_j$ between the two contributions to the heat flux versus
temperature in a chain of 32 particles with nearest-neighbor Lennard-Jones 
type interactions. $T$ is the average temperature of the two reservoirs while
the temperature difference $\Delta T$ is set equal to $T/2$.} 
\label{f:flura}
\end{center}
\end{figure}

Altogether, $r_j$ is expected to diverge for increasing temperature. This 
is illustrated in Fig.~\ref{f:flura},
where we report the data resulting from the simulation of a chain put in
contact with two heat reservoirs with a temperature difference 
$\Delta T = T/2$  and average temperature $T$.\footnote{The numerical results 
have been obtained by implementing a Nos\'e-Hoover thermostat with 
$\Theta = 1$ - see the next chapter for further details about the
thermostat scheme.}

Most nonlinear models, aiming at a characterization of solid-like structures
have been written with reference to the deviation $q_n = x_n - na$ from the 
equilibrium position $na$. This is, for instance, the case of the  FPU models
(\ref{fpu}) that  can be seen as the result of truncating the expansion of the
potential energy  in powers of $(q_n-q_{n+1})$. The result of introducing the
displacement  $q_n$ and getting rid of  the actual position $x_n$, is that
physical distances disappear from the model and the lattice spacing becomes an
arbitrary value. As a consequence, if $a$ is chosen too small, one can have
unphysical pictures of particles crossing each other and yet keeping the same
nearest neighbors.  A meaningful interpretation of such models is obtained
only by associating them with a sufficiently large spacing $a$ and thus by
computing the heat flux from the definition (\ref{hfdef}).

Finally, we want to introduce a less symmetric but more compact expression  for
$j_n$ that can be obtained by exploiting the equality  $\langle \dot
V(q_{n+1}-q_n)\rangle=0$ that holds in the stationary regime. By determining
the derivatives from the equations of motion, we find that
\begin{equation}
\langle \dot q_{n+1} F(q_{n+1}-q_n) \rangle =
\langle \dot q_n F(q_{n+1}-q_n)\rangle
\label{eq:id1}
\end{equation}
which allows expressing the average local heat flow as
\begin{equation}
\langle j_n \rangle = a \langle \dot q_{n+1} F(q_{n+1}-q_n)\rangle  .
\label{eq:newflux}
\end{equation}
In the stationary regime, it is easily seen that the equality 
$\langle \frac{\rm d}{{\rm d} t}(\dot q_{n})^2\rangle =0$ implies that
\begin{equation}
\langle \dot q_n F(q_{n+1}-q_n)\rangle =
\langle \dot q_n F(q_n-q_{n-1})\rangle
\label{eq:id2}
\end{equation}
The combined use of Eqs.~(\ref{eq:id1},\ref{eq:id2}) for suitable values
of $n$ shows that in the bulk 
\begin{equation}
\langle j_n \rangle \; =\; \langle j_{n-1} \rangle \;:=\; j .
\label{defjloc}
\end{equation} 
Accordingly, the average
local heat flux is constant along the chain as it should. 
At the boundaries,
one finds that, for whatever choice of the heat bath, the heat flux equals the
energy flow towards the corresponding reservoir. This can be seen by
just writing the balance equation for the kinetic energy of the first
(last) particle of the chain. 

The quantity that will be mostly used in the following is the total heat flux
\begin{equation}
J \;=\; \sum_n \, j_n \quad
\label{jtotal}
\end{equation}
Notice that, from definition (\ref{defjloc}), $\langle J \rangle = Nj$ 
in the stationary regime.
From definition (\ref{hf2}), $J$ is readily recognized to be the 
one-dimensional version of the general expression (see e.g. \cite{KT})
\begin{equation}
\vJ \;=\; V\vj \;=\; \sum_i \left[\dot \vx_i h_i +
\frac{1}{2}\sum_{j\ne i}(\vx_i-\vx_j)(\dot \vx_i+\dot \vx_j){\bf F}_{ij} 
\right]
\label{fluxgeneral}
\end{equation}  
which is valid for every state of the matter. 

To understand better the physical meaning of the definition given above
and for later reference, it is useful to consider the case of the simple 
harmonic chain, i.e. (\ref{fpu}) with $g_3=g_4=0$ and $m_n=m$. If 
periodic boundary conditions are assumed, 
the Hamiltonian is diagonalized by passing to the normal coordinates 
\begin{equation}
Q_k={1\over\sqrt{N}}\sum_{l=1}^N q_l \, e^{i{2\pi k\over N}l} \quad, \qquad
Q_{-k}=Q_k^*, \qquad k=-{N\over2}+1,\ldots, {N\over2} \quad .
\label{modi}
\end{equation}
Considering, for simplicity, definition (\ref{hfdef}) we get the expression
for the total heat flux (\ref{jtotal})
\begin{equation}
J_H \;=\;  im \, \sum_{k} \, v_k \omega_k Q_k\dot Q_k^* \quad ,
\label{eq:flmod}
\end{equation} 
where $v_k=\omega_k'$ is the group velocity of 
phonons and the mode frequencies are given by
\begin{equation}
\omega_k \;=\; 2 \sqrt{g_2 \over m} \left|\sin\left({\pi k \over N}\right)
\right|
\quad .
\label{omegak}
\end{equation} 
It is furthermore convenient to introduce the complex amplitudes $a_k$ 
through the standard transformations
\begin{equation}
Q_k \;=\; \frac{1}{2} \left( a_k e^{i\omega_k t} + a_{-k}^* e^{-i\omega_k t}
\right) ,
\qquad
\dot Q_k \;=\; \frac{i\omega_k}{2} 
\left( a_k e^{i\omega_k t} - a_{-k}^* e^{-i\omega_k t}.
\right) \quad. 
\label{ak}
\end{equation} 
A straightforward calculation shows that the heat flux can be expressed as
\begin{equation}
J_H \;=\; \sum_{k} \, v_k E_k \quad, \quad
\label{vpere}
\end{equation}
where $E_k=\frac{1}{2} m\omega_k^2 |a_k|^2$ is interpreted as the energy in the
$k-$th normal mode. This expression, originally  proposed by
Peierls \cite{peierls}, has a  simple intuitive interpretation and shows that
the heat flux is a constant of motion for the harmonic chain at equilibrium.  Notice
also that its equilibrium  average $\langle J_H \rangle=0$ as it should: this
stems from equipartition of energy $\langle E_k\rangle=k_BT$ and from the fact 
that $v_k=-v_{-k}$.

\section{Heat baths}
Theoretical and numerical investigations of statistical mechanical systems 
invariably rely upon a suitable modeling of the interaction with thermal 
reservoirs. At equilibrium, this is usually accomplished by well known
methods like micro-canonical molecular dynamics and Monte Carlo simulations. 
Out of equilibrium, the lack of a general theoretical framework forces to
define meaningful interactions with external thermal baths. The importance 
of such approach to simulations of transport processes
in crystalline solids has been already recognized since decades \cite{MT78}  
 
Conceptually, the correct way to proceed requires considering non-equilibrium
states of infinite systems. In the context of this review, one could imagine,
for example, an infinite chain with an initial condition such that all atoms
to the left and to the right of some prescribed subset (defining the system
of interest) are in equilibrium at
different temperatures. However, the only specific case in which such an 
approach can be effectively implemented is that of harmonic systems 
\cite{CL71,RG71,OL74,SL77}. In fact, the degrees of freedom corresponding
to the dynamics of the reservoirs can be traced out and, as a result, one 
can prove the existence of stationary non-equilibrium states and thereby 
determine the relevant thermodynamic properties. As soon as nonlinear
effects are included, it is no longer possible to reduce the evolution 
of the heat baths to a tractable model. One can, nevertheless, study
nonlinear chains by assuming that nonlinearity is restricted to the system
of interest, while still considering linear interactions for the semi-infinite
chains that correspond to the two reservoirs. By following this approach,
several results have been obtained (see Ref.~\cite{BLR00} and references
therein). In particular, it has been recently proved the existence of a unique 
invariant non-equilibrium measure in chains of highly nonlinear 
coupled oscillators 
subject to arbitrary large temperature gradients \cite{EPR99}. 

If the baths are modeled with linear wave equations (the continuum limit of 
the harmonic lattice), the Hamiltonian dynamics of the full system reduces
to the stochastic Markovian evolution of a few variables. In the simplest
coupling scheme, one can write \cite{BLR00}
\begin{eqnarray}
  m_n\ddot q_n &=& - F(q_{n+1} - q_n) + F(q_n - q_{n-1}) + r_+\delta_{n1}
                   + r_-\delta_{nN} \nonumber \nonumber \\
  \dot r_+ &=& -\gamma_+(r_+-\lambda_+q_1) + \xi_+ \\
  \dot r_- &=& -\gamma_-(r_--\lambda_-q_N) + \xi_- \nonumber 
\end{eqnarray}
where the $\xi_\pm$'s are independent Wiener processes 
with zero mean and variance $2\gamma_\pm\lambda_\pm T_\pm$. Moreover, 
$\lambda_\pm$ is the coupling strength between the chain and the 
corresponding bath, while $1/\gamma _\pm$ is the relaxation time. 
This approach is rather recent and we are not aware of any numerical study 
where it is implemented and compared with other methods. 

\subsection{Stochastic baths}

A traditional way to implement the interaction with reservoirs amounts to 
introducing simultaneously random forces and dissipation according to the
general prescription of fluctuation-dissipation theorem. This could be 
regarded as the limit case of the previous model when $\gamma_\pm$ becomes 
very large. Consequently, the reservoirs are not affected by the system 
dynamics. In the simple case of an equal-mass chain, this results in the 
following set of Langevin equations 
\begin{equation}
  m \ddot q_n =  F(q_n\negthinspace - \negthinspace q_{n-1}) \negthinspace -  
   \negthinspace  F(q_{n+1} \negthinspace - \negthinspace q_n) \negthinspace + 
   \negthinspace
  (\xi_+ \negthinspace -\negthinspace \lambda_+\dot q_n )\delta_{n1}
  \negthinspace + \negthinspace 
  (\xi_- \negthinspace - \negthinspace \lambda_-\dot q_n )\delta_{nN}  
\label{bathlang}
\end{equation}
where $\xi_\pm$'s are again independent Wiener processes with zero mean 
and variance $2\lambda_\pm k_BT_\pm$. In practice, this model too 
is amenable only to numerical investigation for nonlinear forces.

Once the nonequilibrium stationary state is attained, one evaluates the 
average flux $j$ defined in(\ref{defjloc}) and estimates the thermal 
conductivity as $\kappa=|j|L/\Delta T$ where $\Delta T=T_+-T_-$~.
The average flux can be obtained also directly from the
temperature profile. In fact, direct stochastic calculus shows that the 
average amount of energy exchanged between the first particle and the hot 
reservoir is
\begin{equation}
  j(\lambda,N) = \frac{\lambda_+}{m_{1}}( T_+ - T_{1} )
\label{eq:flusto}
\end{equation}
and an equivalent expression holds on the opposite boundary.
This formula states that the heat flux is proportional to the difference 
between the kinetic temperature of the particle in contact with the heat bath
and the temperature of the reservoir itself. 

A ``microscopic'' implementation that has been widely used amounts 
to imagining each reservoir as a one-dimensional ideal gas of particles 
of mass $M_\pm$ interacting with the chain through elastic 
collisions \cite{PRV67}. 
A simple strategy consists in selecting a random sequence of times $t_i$ 
when each thermostated atom collides with a particle of the corresponding
reservoir. A natural choice for the distribution $W(\tau)$ of the intervals 
$\tau$ between consecutive collisions would be the Poissonian 
\begin{equation}
W(\tau) = \frac{1}{\bar{\tau}}  \exp\left(-\frac{\tau}{\bar{\tau}}\right) ,
\end{equation}
where $\bar\tau$ is the average collision time.
The kinematics of the collision implies that
\begin{equation}
 \dot q_1 \longrightarrow \dot q_1 +  \frac{2M_+}{m+M_+}(v - \dot q_1)
\end{equation}
for the left reservoir (an analogous expression holds for the right one). 
The velocity $v$ of the gas particle is a random variable distributed
according to the Maxwellian distribution\footnote{In one-dimension, it
coincides with the Gaussian distribution.} 
\begin{equation}
P_+ (v) = \sqrt{\frac{M_+}{2\pi k_BT_+}} \exp\left
   (-\frac{M_+ v^2}{2k_BT_+} \right) .
\end{equation}
In the case $M_\pm = m$, the procedure reduces to assigning the velocity 
after the collision equal to random variable $v$ (the particles exchange 
their velocities). On the other hand, in the limit $M_\pm \ll m$,   
the interaction with the heat baths becomes Langevin-type as in 
Eq.~(\ref{bathlang}) with $\lambda_\pm = 2M_\pm/\bar\tau$.

This method is computationally very simple, as it avoids the problem of
dealing with stochastic differential equations: the integration can, in
fact, be performed by means of conventional algorithms. In particular, the
explicit absence of dissipation allows using symplectic integration schemes
\cite{MA92,C95} between collisions. Furthermore, a physically appealing feature of this
approach is that damping is not included a priori in the model, but is
self-consistently generated by the dynamics. 

A related but different scheme consists in determining the collision times
from the interaction with ``thermal walls'' suitably placed at the boundaries 
of the chain. This method has the advantage of allowing for the inclusion of
pressure effects in the molecular dynamics simulation.
In this case, the velocity of the thermostated particle is randomized
whenever it reaches the wall. While the sign of the velocity component normal 
to the wall has to be chosen in order to ensure that it is reflected, its
modulus has to be distributed according to a Maxwellian distribution at the
wall temperature (see Ref.\cite{TTKB98} for a discussion of possible pitfalls
of different choices).

\subsection{Deterministic baths}

In the attempt of providing a self-consistent description of
out-of-equilibrium processes, various types of deterministic heat baths
have been introduced \cite{EM}. This was also motivated by the need to overcome 
the difficulties of dealing with stochastic processes. The scheme that
has received the largest support within molecular-dynamics community is
perhaps the so-called Nos\'e-Hoover thermostat \cite{N84,H85}. More precisely,
the evolution of the particles in thermal contact with the bath $\alpha$ is
ruled by the equation
\begin{equation} 
m \ddot q_n = F(q_{n} - q_{n-1}) - F(q_{n+1} - q_{n}) -
\cases{\zeta_+\dot q_n & \hbox{if}  $n \in S_+ $ \cr
       \zeta_-\dot q_n & \hbox{if}  $n \in S_-$}
\label{nose1}
\end{equation} 
where $\zeta_\pm$ are two auxiliary variables modeling the microscopic
action of the thermostat, and $S_\pm$ denote two sets of $N_\pm$ particles
(at the beginning and the end of the chain, respectively) in contact with
the reservoirs. 

The dynamics of $\zeta_\pm$ is governed by the equation 
\begin{equation} 
\dot \zeta_\pm \;=\; {1\over \Theta^2_\pm}\left( 
      \frac{1}{k_B T_\pm N_\pm}
      \sum_{n\in S_\pm} m \dot q_n^2 - 1\right) 
\label{nose2}
\end{equation} 
where $\Theta_\pm$ are the thermostat response times. The action of the 
thermostat can be understood in the following terms. Whenever the (kinetic) 
temperature of the particles in $S_\pm$ is, say, larger than $T_\pm$, 
$\zeta_\pm$ increases becoming eventually positive. Accordingly, the 
auxiliary variable
acts as a dissipation in Eq.~(\ref{nose1}). Since the opposite occurs whenever
the temperature falls below $T_\pm$, this represents altogether a
stabilizing feedback around the prescribed temperature. Actually, the
justification of this scheme rests on a more solid basis. In fact, in Ref.
\cite{N84,H85}, it has been shown to reproduce the canonical equilibrium
distribution.

In the limit case $\Theta \to 0$, the model reduces to the so-called 
isokinetic (or Gaussian) thermostat: the kinetic energy is exactly 
conserved and the action of the thermal bath is properly described without the 
need to introduce a further dynamical variable, since $\zeta_\pm$ becomes an 
explicit function of the $\dot q$'s \cite{EM}:
\begin{equation}
\zeta_\pm = \frac{\sum_{n \in S_\pm} \dot q_n \left[ F(q_{n} - q_{n-1}) - 
   F(q_{n+1} - q_{n})\right]}{\sum_{n \in S_\pm} \dot q_n^2} .
\end{equation}
This latter thermostat scheme can be derived by variational methods 
after including the non-holonomic constraint due to the imposed kinetic 
energy conservations \cite{EM}.

More generally, it has been shown in Ref.\cite{N84,H85} that the dynamical 
equations of this entire class of deterministic thermostats possess a
Hamiltonian structure in a suitably enlarged phase-space. An interesting
property that is preserved by the projection onto the usual phase space
is time-reversibility. In fact, a simple inspection of the equations reveals
that they are invariant under time reversal composed with the involution $I$ 
\begin{equation}
\dot q_n\to -\dot q_n \quad n=1,\ldots,N \quad, \quad 
 \zeta_\pm \to -\zeta_\pm .
\end{equation}
This property represents the main reason for the success of this class of
thermostats, since dissipation is not included a priori, but it rather
follows self-consistently from the dynamical evolution. In particular,
at equilibrium $\langle \zeta_\pm \rangle = 0$, indicating that the action of
the bath does not break microscopic reversibility, while out of equilibrium
$\langle \zeta_+\rangle+langle \zeta_-\rangle > 0$ and this value has been 
connected with entropy production \cite{CHAOS}.

Another procedure can be defined by combining the idea of a thermostat
acting through collisions at the boundaries with that of a deterministic and 
reversible rule for the collisions themselves. The first context where a
scheme of this type has been successfully introduced is that of sheared
fluids \cite{CL97}; more recently, van Beijeren \cite{Vp} has proposed an
implementation that is suitable for one dimensional systems. The idea is
very similar to that of the above discussed thermal wall with the crucial
difference that the velocity $v'$ after the collision is a deterministic
and reversible function $ v' = \Phi(v)$ of the initial velocity $v$. A class
of reversible transformations is that defined as $\Phi = G R G^{-1}$, with
$R = R ^{-1}$. A further constraint is that the equilibrium distribution
be left invariant under the transformation $\Phi$. The choice adopted
in Ref.~\cite{GHN01} consists in fixing $G(v) = \exp(-mv^2/k_BT)$ and
$R(x) = 1-x$. According to this choice, $G$ transforms the Maxwellian
distribution into a uniform one on the unit interval, which, in turn, is
left invariant by $R$. Although this and the previous choices of reversible
thermostats do not correspond to any physical mechanism, they offer a
convenient framework for the application of dynamical-system
concepts \cite{CHAOS}.

In the above methods, the system is driven out of equilibrium by a suitable
forcing at the boundaries: an approach that is aimed at closely reproducing
experimental conditions. A different philosophy \cite{EM} consists in 
introducing an external field acting in the bulk of the system to keep it 
steadily out of equilibrium. The main advantage of this approach is the 
possibility to work with homogeneous systems with, e.g. a uniform 
temperature along the whole sample. In particular, periodic boundary
conditions can be enforced thus further reducing finite-size effects.
This is sometimes referred to as the Evans heat-flow algorithm and
has been applied to heat conduction in one-dimensional lattices 
\cite{GH85,MM95,ZIE00}.
 
More precisely, a fictitious heat field $F_e D_n$ is added to the 
equation of motion for the $n$th particle. The coupling $D_n$ must satisfy
two conditions: ({\it i}) the energy dissipation is proportional to $j F_e$, 
i.e., $\dot H = N j F_e$; ({\it ii}) phase-space flux remains divergence-free
(this is referred to as the adiabatic incompressibility of phase space 
\cite{EM}). Finally, in order to stabilize the dynamics at a prescribed
temperature, a thermostat rule has to be applied. The resulting equation
of motion reads as
\begin{equation} 
m \ddot q_n = F(q_{n} - q_{n-1}) - F(q_{n+1} - q_{n}) + F_eD_n -\zeta \dot q_n   
\end{equation} 
A definition of $D_n$ satisfying the above requirements is \cite{ZIE00}
\begin{equation}
D_n = {1 \over 2} [F(q_{n+1}-q_n)+F(q_{n}-q_{n-1})]  
-{1 \over N} \sum_{j=1}^N F(q_{j+1}-q_j),
\end{equation}
Moreover, we can consider a Gaussian thermostat:
\begin{eqnarray}
\zeta &=& {1 \over 2K_0} \sum_{j=1}^N \dot q_j [F(q_{n} - q_{n-1}) - 
  F(q_{n+1} - q_{n}) +F_e D_j]  \\
  K_0 &=& {m \over 2} \sum_{j=1}^N \dot q_j^2. \nonumber
\end{eqnarray}
Notice that at variance with the previous schemes, the thermostat acts 
on all particles (this is possible since we recall that the temperature
is uniform).

The thermal conductivity can then be determined from the ratio of the heat
flux to the applied heat field
\begin{equation}
\kappa = \lim_{F_e \rightarrow 0} {\langle j\rangle \over T F_e},
\end{equation}
where $\langle j\rangle$ can be equivalently interpreted as a time average
or a suitable ensemble average. Furthermore, the limit $F_e \rightarrow 0$
is dictated by the need to ensure the validity of the linear response theory
that is implicitly contained in the initial assumptions. This requirement
is all the way more substantial in view of the difficulties encountered
while working with too large heat fields \cite{ZIE00} (to some extent this
is also true for the Nos\`e-Hoover method descrived above \cite{FHLZ98}).

\subsection{Comparison of different methods}
In all schemes of heat baths there is at least one parameter controlling the
coupling strength: let us generically call it $g$. It can either be the inverse
of the average time between subsequent collisions, or the dissipation rate 
$\lambda$ in the Langevin equation, or the inverse of the time-constant 
$\Theta$ in the Nos\'e-Hoover scheme. Fixing $g$ is a 
practical question which is usually solved empirically by insuring that 
different choices do not appreciably affect the 
outcomes of a simulation. On general grounds, one should start 
by choosing $g$ to be of the order of some typical frequency of the 
system \cite{DR93}. In the present context, an interesting question immediately 
arises about the dependence of heat transport on $g$ in the various 
schemes.

In the case of stochastic reservoirs, the heat flux vanishes both in the weak- 
($\lambda \to 0$) and strong-coupling ($\lambda \to \infty$) limit. The 
first implication follows quite easily from the relation (\ref{eq:flusto}) 
and from the observation that $(T_+-T_1)$, cannot increase above $(T_+ - T_-)$ 
(more precisely, one expects that $T_1 \to (T_+ + T_-)/2$ for $\lambda \to 0$, 
since the profile should become increasingly flat). The opposite regime is less
trivial: here below, we explain why the same qualitative behavior can be found 
in two different heat-bath schemes. Let us first consider a reservoir acting 
through collisions separated by random intervals $\tau$. In this case the 
coupling constant $g$ is given by the inverse of the average collision time
$\overline \tau$. For small $\overline \tau$, one can rely on a perturbative
approach and write the kinetic energy $K_1$ a time $\overline \tau$ after
the collision as (it is sufficient to consider the first particle)
\begin{equation}
  K_1  = (v + F\overline \tau)^2 
\end{equation} 
where $v$ is the random, ``initial'' velocity. As a result, $K_1$, on the 
average, changes by an amount that is proportional to  $\overline \tau^2$,
since $v$ has zero average. This means that the deviation of the (kinetic)
temperature from the equilibrium value is also proportional to 
$\overline \tau^2$. In order to have the energy flux one should multiply this
contribution by the number $1/\overline \tau$ of collisions per unit time.
Accordingly, we find that the flux goes to zero as $1/g$ in the strong
coupling limit. It is interesting to notice that the quadratic dependence 
of the energy could also be inferred from the invariance under time
reversal of the microscopic equations: after a collision there should be
no difference between choosing the forward or backward direction for the
time axis: as a result we have to expect a quadratic behavior!

Rather similar is the analysis for thermal baths {\it \`a la} Langevin.
In that case, for large $\lambda$, $\dot q_1$ can be adiabatically eliminated,
giving rise to
\begin{equation}
 \dot q_1  = \frac{\xi}{\lambda} +\frac{F}{\lambda}
\end{equation} 
Again, one can see that the average value of the kinetic temperature deviates
${\mathcal O} (1/\lambda^2)$ from the equilibrium one.

\begin{figure} 
\begin{center} 
\includegraphics*[width=7cm]{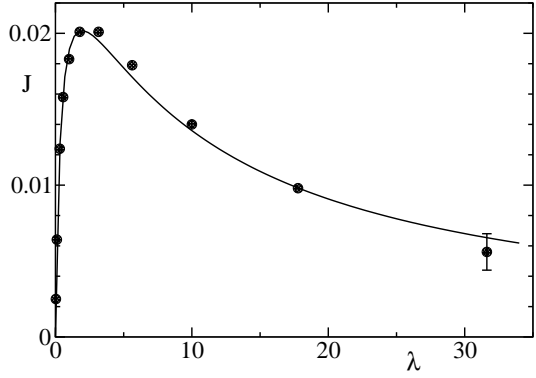}
\caption{Heat flux dependence on the coupling constant $\lambda$ for an
FPU chain in contact with two Langevin stochastic reservoirs at temperature
$T_+ = 1.1$ and $T_- = 0.9$. The chain length is $N=128$ and fixed b.c. have
been imposed.} 
\label{f:flucou} 
\end{center} 
\end{figure}

The simplest way to combine the behavior of the average flux in the 
strong and weak coupling limit is through the following heuristic formula
\begin{equation}
 j(\lambda)  = \frac{a \lambda}{1+ b_1\lambda + b_2\lambda^2} 
\label{eq:fit}
\end{equation}
that exhibits the expected $\lambda$ dependence in both limits $\lambda \to 0$
and $\lambda \to \infty$. 
In Fig.~\ref{f:flucou} we have plotted the flux for an FPU chain in 
contact with stochastic Langevin heat baths. Circles correspond to direct
numerical results, while the solid curve is a fit with formula (\ref{eq:fit}).
The parameter values turn out to be $a = 0.07$, $b_1 = 2.45$, and $b_2=0.26$.
The good agreement is to be considered as rather incidental, given the 
heuristic nature of the fitting formula.

\begin{figure} 
\begin{center} 
\includegraphics*[width=7cm]{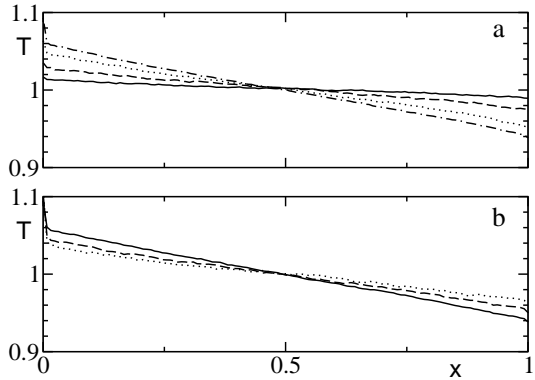}
\caption{Temperature profile in the same set up of the previous figure.
In panel (a), the profiles correspond to (from bottom to top in the left
part) $\lambda= 10^{-3/2}, 10^{-1}, 10^{-1/2}, 10^{0}$; panel (b) refers
to the strong coupling case. Again from bottom to top in the left part,
the curves correspond to $\lambda= 10^{1/4}, 10^{3/4}, 10$.}
\label{f:profcou} 
\end{center} 
\end{figure}

For what concerns the temperature profile, one can see from panel (a) of 
Fig.~\ref{f:profcou} that it is increasingly flat for $\lambda \to 0$.
Less obvious is the tendency of the profile to become flat also in the strong
coupling-limit. In fact, there is a small but crucial difference for
the temperature of the first and last particle. For $\lambda \to 0$
the temperature of the extremal particles obviously tends to be equal
to that in the bulk; on the contrary for $\lambda \to \infty$, such
temperatures converge to the temperature of the reservoir. In other words
the temperature-drop is observed between the boundary particle and its
neighbor. 

\begin{figure} 
\begin{center} 
\includegraphics*[width=7cm]{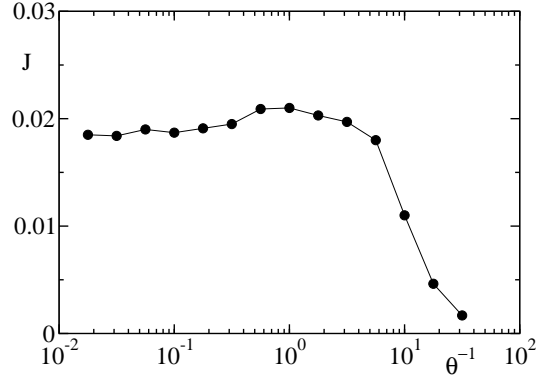}
\caption{Heat flux dependence on the coupling constant $1/\Theta$ for an
FPU chain in contact with two Nos\'e-Hoover reservoirs at temperature
$T_+ = 1.1$ and $T_- = 0.9$. The chain length is $N=128$ and fixed b.c. have
been imposed. The line is just a guide for the eyes.} 
\label{f:flucon} 
\end{center} 
\end{figure}

If we repeat the same analysis by using Nos\'e-Hoover thermostats, we find 
a similar behavior in the strong coupling limit. Indeed, in 
Fig.~\ref{f:flucon} we see that the heat flux vanishes when the response time 
$\Theta$ goes to 0. This is the limit of a strictly isokinetic bath and, one
is, therefore, bound to conclude that such a type of heat baths are unable to 
sustain heat transport (as long as one single particle is thermalized at each
extremum). Perhaps more surprising is the opposite limit $\Theta \to \infty$,
since we can see that the heat flux does not go to zero even though the action
of the heat baths becomes increasingly slow.

A partial understanding of this unexpected behavior comes from the observation
that the variable $\zeta$, though slowly, reaches the same asymptotic 
values for arbitrarily large time constants $\Theta$. This is illustrated 
in Fig.~{\ref{f:histoz} where we have plotted the histogram of $\zeta$ values
for $\Theta = 10$ (solid line) and 100 (dashed line). Both curves are clean
Gaussians centered around the same value $-0.018..$ but with standard
deviations differing by a factor 10 (the same factor existing between the
time constants). In practice, we are led to conclude that, in the limit of
$\Theta \to \infty$  the distribution of $\zeta$ values becomes a
$\delta$-function centered at a fixed dissipation value that depends only on
the chain length, the energy and the temperature drop. This result seems to be
at odds with the fluctuation-dissipation theorem as it seems to suggest that
for $\Theta \to 0$ fluctuations disappear while the dissipation remains
finite. However, one should notice that a correct measure of the amount of 
fluctuations is obtained by integrating over a sufficiently long time
to let correlations decay: the divergence of the time-constant still
ensures the validity of the fluctuation-dissipation theorem. 

\begin{figure} 
\begin{center} 
\includegraphics*[width=7cm]{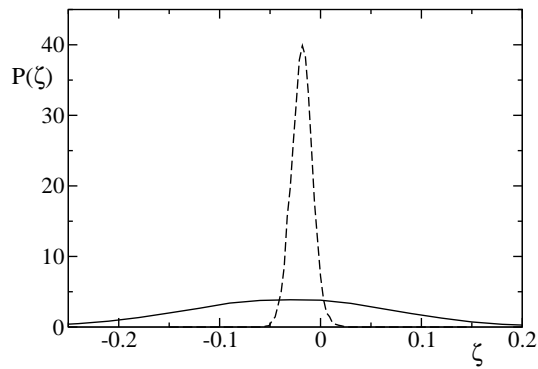}
\caption{Histogram $P(\zeta)$ of the left-heat variable in an FPU chain
in the same condition as in the previous figure and two different values
of $\Theta$: 10 (solid line), 100 (dashed line)}
\label{f:histoz} 
\end{center} 
\end{figure} 

As a result of this analysis we can conclude that values of $\Theta$ of order
1 (in the chosen dimensionless units) are the optimal choices for numerical 
simulations, since smaller values 
imply smaller heat fluxes, while larger ones would require longer simulation 
times (in order to ensure the decay of correlations).

\subsection{Boundary resistances}

Temperature discontinuities usually appear when heat flux is maintained across
an interface among two substances. This discontinuity is the result of a
boundary resistance, generally denoted as Kapitza resistance. Its origin is
traced back to the ``phonon mismatch" between two adjacent substances 
\cite{LSW78}. Such a phenomenon is invariably present in simulations (see e.g. 
Fig.~\ref{f:profcou} and Fig.~\ref{f:profs} below) and may actually 
reduce the accuracy of the measurements. 
The conductivity evaluated as $\kappa_{eff} = j L/\Delta T$ 
represents an effective transport coefficient that includes both boundary 
and bulk resistances. In practice, one may circumvent 
the problem by referring to a subchain far enough from the ends and compute 
a bulk conductivity as the ratio between $J$ and the actual 
temperature gradient in the subchain. Notice, however, that the advantage 
of reducing boundary effects may be partly reduced by the increased difficulty 
of dealing with very small temperature differences.

By denoting with $\delta T_\pm$ the temperature jumps at the edges $x=0$ and
$x=L$, the imposed temperature difference, $\Delta T=T_+-T_-$, can be written 
as  
$$
\Delta T \;=\;
\delta T_+ +\delta T_-
 + \int_0^L dx\,\nabla T  
$$  
Let us now express the jumps $\delta T_\pm$ as \cite{landau} 
\begin{equation} 
\delta T_\pm \;=\;  \varepsilon \ell_\pm \nabla T \big|_\pm
\label{defa}
\end{equation}
where $\nabla T|_\pm$ is the value of the temperature gradient extrapolated at 
the boundary ($x=0,L$), $\ell_\pm$ is the mean free path at the corresponding 
temperature; the phenomenological and dimensionless constant $\varepsilon$ 
measures the coupling strength between the thermostats and the system. 

This equation has been numerically tested by plotting $\delta T/\ell$ 
versus $\nabla T$ for different thermostats \cite{AK00,AK01}. From the
slope of the various curves it has been found that in the FPU-$\beta$ model
the coupling parameter may vary from $\varepsilon = 0.8$ for the thermostats
adopted in Ref.~\cite{AK01}, to $\varepsilon = 2.$ for Nos\'e-Hoover 
thermostats, to $\varepsilon = 400$ for stochastic reservoirs as evidenced
in the large boundary jumps seen, e.g., in Fig.~\ref{f:profcou}. This reflects
the empirical fact that the shape of the temperature profile for given
$T_\pm$ may depend on the choice of thermostats. Nevertheless, the scaling
is independent of this choice.

In the near-equilibrium regime, we can assume 
$\delta T_+\simeq\delta T_-$, $\ell_+\simeq \ell_-=\ell$
and the gradient to be constant, $\nabla T|_\pm=\nabla T$ obtaining
\begin{equation}
\label{eq-dt}
\left( 2\varepsilon\ell + L\right) \nabla T = \Delta T.
\end{equation}
If we further denote by $\kappa=j/\nabla T$ the bulk 
conductivity we obtain 
\begin{equation}
\kappa_{eff} \;=\;  {j \over \Delta T/L} 
\;=\;{\kappa \over 1 +{2\varepsilon\ell \over  L}} \quad .
\label{keff}
\end{equation}
When the mean free path is much larger than the system size $\ell\gg L$ (the
so-called Casimir limit \cite{landau}) energy flows almost freely through the
system, boundary scattering takes over, the system is almost harmonic 
with $\kappa_{eff} \propto L$ and a flat temperature profile occurs. In the 
opposite case $\ell\ll L$, $\kappa_{eff}\simeq \kappa$ and one is indeed 
probing the bulk interaction.

\section{Harmonic systems}
The simplest and almost unique class of systems for which one can perform
analytic calculations is represented by harmonic chains. Even though they  are
characterized by a peculiar dynamics, basically because of the  integrability
of the motion, their behavior can help shedding some light  on various aspects
of heat conductivity.  One of the properties of harmonic chains is the
possibility to decompose the heat flux into the sum of independent
contributions associated to the various eigenmodes. This analysis is
particularly useful to obtain a deeper insight about the role of boundary 
conditions.

We first discuss the dynamics of homogeneous chains, since it is possible  to
obtain an analytic expression for the invariant measure in the general case of
arbitrary coupling strength. The effect of disorder is discussed in the
subsequent section, where perturbative calculations of the relevant quantities
are illustrated. For completeness, we also recall the localization properties
of the eigenfunctions and self-averaging properties of several observables such
as the temperature profile and the heat flux. The approach that we have 
followed is 
mainly based on the Fokker-Planck equation and simple stochastic calculus. An
important alternative approach based on transmission coefficients can be found
in Ref. \cite{RG71}.

\subsection{Homogeneous chains}

We consider a homogeneous harmonic chain with fixed boundary conditions
in contact with stochastic Langevin heat baths. Eq.~(\ref{bathlang}) reduces
to
\begin{equation}
  \ddot q_n = \omega^2 (q_{n+1} - 2q_n + q_{n-1}) + 
 \delta_{n1}(\xi_+ - \lambda \dot q_1) + \delta_{nN}(\xi_- -\lambda \dot q_N) ,
\label{eq:homlin1}
\end{equation}   
where we have set $\lambda_+ = \lambda_-=\lambda$ to lighten the notations 
and assumed unitary masses. 
This set of stochastic equations can be solved \cite{RLL67} by passing to a 
phase-space description, i.e. by writing down the Liouville equation that, 
in this case, corresponds to the following Fokker-Planck equation
\begin{equation}
  \frac{\partial P}{\partial t} = A_{ij}\frac{\partial}{\partial x_i}
  (x_j P) + \frac{D_{ij}}{2} \frac{\partial^2 P}{\partial x_i
\partial x_j} 
\label{eq:liouv}
\end{equation}   
where here and in the following we adopt the summation convention 
(i.e., sums over repeated
indices are understood without explicitly writing down the summation sign); 
$x_i = q_i$ for 
$1\le i\le N$, $x_i = \dot q_i$ for $N < i\le 2N$. $A_{ij} $ and $D_{ij}$
are elements of the $2N\times2N$ matrices $\bf A$ and $\bf D$ that we write 
in terms of $N\times N$ blocks
\begin{equation}
 {\bf A} = \pmatrix { {\bf 0}     &&  -{\bf I}\cr \cr
           \omega^2 {\bf G}  &&  \lambda {\bf R}} \quad , \quad
 {\bf D} = \pmatrix {{\bf 0}     &&  {\bf 0}\cr \cr
           {\bf 0}  &&  2 \lambda k_B T ({\bf R}+\eta{\bf S})}
\label{eq:mats}
\end{equation}   
where we have introduced the average temperature $T = (T_++T_-)/2$ and
the rescaled temperature difference $\eta = (T_+-T_-)/T$. Moreover,
$\bf 0$ and $\bf I$ are the null and identity matrices, $\bf G$ is a 
tridiagonal matrix defined as 
$$
G_{ij} = 2\delta_{ij} -\delta_{i+1,j} - \delta_{i,j+1} \quad ,
$$ 
while $\bf R$ and $\bf S$ are defined as
\begin{eqnarray}
R_{ij} &=& \delta_{ij}(\delta_{i1} + \delta_{iN}) ,\nonumber \\
S_{ij} &=& \delta_{ij}(\delta_{i1} - \delta_{iN})  
\end{eqnarray}
The general solution of this equation can be sought of the form
\begin{equation}
  P(x) = \frac{\hbox{Det} \{{\bf C}^{-1/2}\}}{(2 \pi)^N}
         \exp \left[ - \frac{1}{2} C^{-1}_{ij}x_ix_j \right]
\label{eq:fpstat}
\end{equation}   
where $\bf C$ is the symmetric covariance matrix
\begin{equation}
   C_{ij} \equiv \langle x_ix_j \rangle \equiv \int dx P(x) x_ix_j
\label{eq:covmat}
\end{equation}   
By replacing the definition of $\bf C$ into Eq.~(\ref{eq:liouv}), one finds
that
\begin{equation}
   \dot {\bf C} = {\bf D} - {\bf A }{\bf C} - {\bf C}{\bf A}^\dag
\label{eq:eqcov}
\end{equation} 
where ${\bf A}^\dag$ denotes the transpose of $\bf A$. Accordingly, the
asymptotic stationary solutions can be determined from the following equation 
\cite{WU45}
\begin{equation}
   {\bf D} = {\bf A }{\bf C} + {\bf C}{\bf A}^\dag
\label{eq:FPasy}
\end{equation} 
In order to solve the problem, let us write $\bf C$ in terms of $N\times N$
blocks,
\begin{equation}
 {\bf C} = \pmatrix { \overline{\bf U}       &  \overline {\bf Z}\cr 
                      \overline{\bf Z}^\dag  &  \overline {\bf V}}
\label{eq:bmat}
\end{equation}   
where the matrices $\overline{\bf U}$, $\overline{\bf V}$, and 
$\overline{\bf Z}$ express the correlations among positions and velocities, 
\begin{equation}
 {\overline  U}_{ij} = \langle q_iq_j \rangle \quad , \quad 
 {\overline V}_{ij} = \langle \dot q_i \dot q_j \rangle \quad , \quad
 {\overline Z}_{ij} = \langle q_i \dot q_j \rangle \quad , \label{eq:corpv}
\end{equation}   
If the temperatures of the two heat baths coincide (i.e. $\eta = 0$), it can
be easily seen that 
\begin{equation}
  {\bf U}_e = \frac{k_BT}{\omega^2}{{\bf G}^{-1}} \quad , \quad
  {\bf V}_e = k_BT \ {\bf I} \quad , \quad {\bf Z}_e = 0 
\label{eq:coreq}
\end{equation}   
represent a meaningful solution, since it coincides with the equilibrium 
Boltzmann distribution $P(x) \approx \exp({-H/k_BT})$. 

The derivation of the stationary solution in the out-of-equilibrium case is
reported in App.~B. All relevant correlations can be expressed in terms of the
function $\phi(j)$ (see Eq.~(\ref{eq:newparal})) that decays exponentially with
the rate $\alpha$ defined in Eq.~(\ref{eq:musol}). As it can be seen by direct
inspection of correlations, $\alpha$ measures the length over which the
boundary reservoirs significantly affect the chain dynamics. As expected,
$\alpha$ diverges in the weak coupling limit 
($\nu =\omega^2/\lambda^2 \to \infty$). 

From Eqs.~(\ref{eq:usol},\ref{eq:Vexp}), it follows that position-position 
and velocity-velocity correlations are equal for all pairs of particles 
$(i,j)$ such that $i+j$ is constant. The qualitative explanation of this
property relies on the exponential decay of the boundary effects. In fact,
the amplitude of, e.g., $\langle q_i q_j \rangle$ decreasing exponentially with
both $i$ and $j$ has to depend on $i+j$. Less obvious is the left-right
antisymmetry ($V_{ij} = V_{N-i+1,N-j+1}$), which implies that the boundary
effects are exactly the same for the two thermostats, whatever is their
temperature. This is nicely reproduced by the temperature profile 
\begin{equation}
 T(i) =  T(1 + \eta V_{ii}) = \cases{ T_+ - \nu \eta T \phi(1) & $i=1$ \cr
         T[1 - \eta \nu \phi(2i-1)]  & $1 <i \le N/2$\cr
         T[1 + \eta \nu \phi(2(N-i)-1)]  & $N/2 < i < N$\cr
         T_- + \nu \eta T \phi(1) & $i=N$ }
\label{tlebo}
\end{equation}
which exhibits a further unexpected property (see Fig.~\ref{f:profana}): 
the temperature is higher
in the vicinity of the coldest reservoir (the only exception being
represented by the first and last particles)!. Because of the exponential
decay of $\phi(i)$, in the bulk, the temperature profile is constant
as if the system were at equilibrium at temperature $T$.
However, this is only superficially true, as position-velocity correlations 
significantly differ from the equilibrium ones. 

\begin{figure} 
\begin{center} 
\includegraphics*[width=7cm,angle=0]{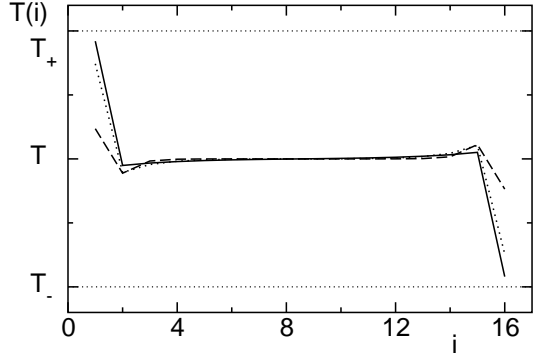}
\caption{
The temperature profile for the harmonic chain, formula (\ref{tlebo}), for coupling parameter
$\nu=0.05$, 0.2, 1.0  (solid, dotted and dashed lines respectively).
} 
\label{f:profana} 
\end{center} 
\end{figure}

Also the average stationary local flux can be computed explicitly:
\begin{equation}
j_i = \omega^2 {\overline Z}_{i-1,i} = 
   \frac{\omega^2 k_BT\eta}{\lambda} \phi(1)
\end{equation} 
Eq.~(\ref{eq:zsol}) implies that the value of $Z_{i,k}$ depends only $i-k$
rather than $i+k$, as before. Physically, this symmetry reflects  the fact that 
$j_i=j$ independent of the lattice position $i$ in the stationary state. 
In the limit of large $N$, Eqs.~(\ref{eq:musol},\ref{eq:newparal}) imply
\begin{equation}
j =  \frac{\omega^2 k_BT}{2 \lambda} \left[ 1 + \frac{\omega^2}{2\lambda^2}
-\frac{\omega}{\lambda}\sqrt{\frac{\omega^2}{4\lambda^2}+ 1} \right] (T_+-T_-) .
\end{equation}
Accordingly, the heat flux is proportional to the temperature difference
rather than to the gradient as it should be, were the Fourier law to be
satisfied. This proves that, as expected, homogeneous harmonic chains
do not exhibit normal transport properties since the effective conductivity 
$\kappa = j N/(T_+ -T_-) \propto N$, while the bulk conductivity diverges
exponentially, since the temperature gradient away from the baths is 
exponentially small.

For what concerns the dependence of the flux on $\lambda$, we see that 
$J$ vanishes both in the limit of large and small couplings. The asymptotic
expressions
\begin{equation}
 j = \cases{\frac{\omega^2}{2\lambda} k_B(T_+-T_-) & $\lambda \gg \omega$ \cr
           \frac{\lambda}{2} k_B(T_+-T_-) & $\lambda \ll \omega$},
\end{equation}
are thus consistent with the heuristic formula (\ref{eq:fit})
derived in the general case. The maximum flux is attained for 
$\lambda/\omega = \sqrt{3}/2$, a value that is close to
the one observed numerically in the nonlinear case (see again Fig. \ref{f:flucon}).

Let us conclude this section by recalling that a similar procedure can
be adopted to solve the problem for heat baths characterized by stochastic
elastic collisions. In Ref.~{\cite{RLL67}}, it is shown that very similar
expressions are found also in this case, with only minor quantitative
differences in the numerical factors.  Furthermore, it is worth mentioning 
the model of  self-consistent reservoirs introduced in Ref. \cite{BRV70},
that can be solved exactly. 

\subsection{Disordered chains}
While remaining in the realm of harmonic systems, we now consider the role
of disorder on transport properties. More specifically, we shall consider 
random-mass (or isotopically disordered) chains
\begin{equation}
  m_n \ddot q_n = q_{n+1} - 2q_n + q_{n-1} \quad . 
\label{eq:lin0}
\end{equation} 
As we shall see, boundary conditions play a crucial role, but, for the moment,
we leave them unspecified. 
Before entering in a more detailed discussion it is
worth mentioning the general results by  Lebowitz and collaborators
\cite{CL71,OL74}: they showed rigorously that the system approaches a unique
stationary non-equilibrium state for a large class of heat baths.

As it is known, the presence of disorder generally induces
localization of the normal modes of the chain and one may thus expect the
latter to behave as a perfect thermal insulator. Nonetheless, the actual 
situation turns out to be much more complicated, depending on boundary
conditions and on the properties of the thermostats.

{\it Localization of the eigenmodes -}
To understand the transport properties it is first useful to recall some basic
facts about localization. For illustration, let us consider the example of a
disordered chain with two evenly-distributed types of particles. Some of the
numerically computed eigenvectors are shown in Fig.~\ref{f:eigen}. Upon
ordering them  with increasing eigenfrequencies, a distinct difference in their
localization properties can be recognized. Indeed, for small frequencies (upper
panel of Fig.~\ref{f:eigen}), randomness induces only a relatively weak
modulation of the amplitude; a partial localization can be recognized in the
intermediate panel, while a clear evidence of localization is visible only for
the high-frequency eigenvector  reported in the bottom panel.

\begin{figure}[tcb]  
\begin{center} 
\includegraphics*[width=7cm]{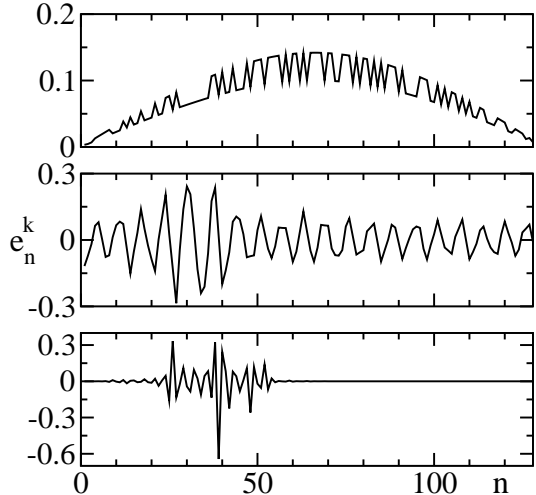} 
\caption{The first, 41st and
100th eigenvector (from top to bottom) in a chain of 130 particles with random
masses with an even distribution of 1 and 1/2 values. The increasing
localization with increasing eigenvalue is transparent.} 
\label{f:eigen} 
\end{center}  
\end{figure} 

A rigorous investigation can be performed by
applying the transfer matrix approach to the eigenvalue equation. 
After inserting the expression $q_n = v_n e^{i\omega t}$ in Eq.~(\ref{eq:lin0}),
we obtain
\begin{equation}
\label{eq:stat}
    -m_n \omega^2 v_n = v_{n-1} - 2v_n + v_{n+1} .
\end{equation}
It is well known that the spectral properties of linear operators
involving the discrete Laplacian can be determined from a 
recursive equation for the new variable $R_n = v_n/v_{n-1}$.
The most known example where this approach has been 
successfully employed is that of Anderson quantum localization in the 
tight-binding approximation (see, e.g. \cite{DG84})~. In the present context, 
Eq.~(\ref{eq:stat}) yields
\begin{equation}
    R_{n+1} = 2 - m_n \omega^2 - \frac{1}{R_n} ,
\label{eq:hyper}
\end{equation}
an equation that can be interpreted as a ``discrete time'' stochastic
equation. The mass $m_n$ plays the role of a noise source (with bias), 
whose strength is gauged by the frequency $\omega$. 
In particular, the inverse localization length $\gamma$ is given by
\begin{equation}
    \gamma = \langle \ln R_n \rangle ,
\label{eq:lyap}
\end{equation}
while the integrated density of states $I(\omega)$ follows from node counting
arguments, i.e. $I(\omega) = f$, where $f$ is the fraction of  negative $R_n$
values. In Fig.~\ref{f:local} it is shown that $I$ increases linearly for 
small $\omega$ and exhibits some irregular fluctuations at larger frequencies. 
The upper band edge (at $\omega \simeq 2.8$) is easily identifiable as the point above which
$I(\omega)$ remains constant and equal to 1. At 
variance with the standard Anderson problem, where all eigenmodes
are exponentially localized, here $\gamma$ tends to zero for $\omega \to 0$.
This can be easily understood from equation 
(\ref{eq:hyper}): Since $\omega^2$ multiplies the stochastic term, disorder
becomes less and less relevant in the  small frequency limit. In this limit,
one can thus resort to a perturbative approach. Let us start noticing that for 
$\omega= 0$, $R=1$ is a marginally stable  fixed point of the recursive equation
(\ref{eq:hyper}). For small $\omega$,  an intermittent process sets in: after a
slow drift driving $R_n$ below  one, a re-injection to values larger than one 
occurs
and nonlinearity become suddenly relevant. The process repeats again
and again. By writing  $R_n = 1 + r_n$ and expanding in powers of $r_n$,
we find that the  dynamics in the vicinity of $R_n=1$ is described by 
\begin{equation}
    r_{n+1} = r_n - r_n^2 - \omega^2 \langle m \rangle + \omega^2 \delta m  
\label{eq:hypap}
\end{equation}
where we have included only the first nonlinear correction and written
separately the average value of the noise term. In the limit of small 
$\omega$, this equation can be approximated by the Langevin equation 
\begin{equation}
    \dot r = - r^2 - \omega^2 \langle m \rangle + \omega^2 \delta m  
\qquad ,
\label{eq:hypapc}
\end{equation}
where, for the sake of simplicity, we have kept the same notations.
The corresponding Fokker-Planck equation writes
\begin{equation}
\frac{\partial P}{\partial t} = 
  \frac{\partial(r^2+ \omega^2 \langle m \rangle)P}{\partial r} + 
  \frac{\sigma_m^2}{2}
  \frac{\partial^2 P}{\partial r^2}
\label{eq:fokpla}
\end{equation}
where $\sigma_m^2=\langle m^2\rangle-\langle m\rangle^2$ stands for the 
variance of the mass distribution. Given the steady incoming and outcoming 
flow, the stationary solution can be obtained by imposing 
\begin{equation}
  (r^2+ \omega^2 \langle m \rangle)P + \frac{\sigma_m^2}{2}
  \frac{dP}{dr} = C
\end{equation}
where $C$ represents the probability flux to be determined by imposing 
the normalization of the probability density $P$. Notice also that $C$ can
be identified with the integrated density of states $I(\omega)$, since it
corresponds to the probability that, at each iterate, $R_n$ is re-injected
to the right, i.e. the probability of having a node in the eigenvector.  
In the absence of disorder ($\sigma_m=0$), 
\begin{equation}
    I(\omega) = C = \frac{\sqrt{\langle m \rangle}}{\pi} \omega
\label{eq:intden}
\end{equation}
and, correspondingly,
\begin{equation}
  P_0(r) = \frac{1}{\pi} \frac{\omega \sqrt{\langle m \rangle}}
         {r^2+ \omega^2 \langle m \rangle}
\end{equation}
This approximation is already sufficient to reproduce the behavior of
$I(\omega)$ at small frequencies, as in the limit $\omega \to 0$ the
variance of the disorder goes to zero faster than the average value.
This is confirmed by comparing the dotted line in Fig.~\ref{f:local}
(corresponding to the analytic expression (\ref{eq:intden})) with the 
numerically determined integrated density.

\begin{figure}
\begin{center}
\includegraphics*[width=7cm]{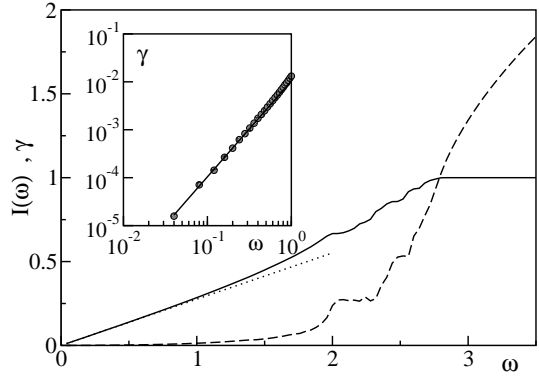}
\caption{Integrated density of states, $I(\omega)$ (solid line), and inverse 
localization length, $\gamma$ (dashed line), as a function of the frequency 
$\omega$ for a chain with mass disorder: the particles have either mass 1 or 
1/2 with equal probability. The dotted line corresponds to the analytic
expression (\ref{eq:intden}).  In the inset the inverse localization length 
is plotted in doubly logarithmic scales (circles) and compared
with the theoretical formula (\ref{eq:analoc}) - solid line.}
\label{f:local}
\end{center}
\end{figure}

On the other hand, the above approximation is not accurate enough to 
determine the localization length,
as disorder is totally disregarded. Indeed, the symmetry of $P_0$ implies
that $\gamma \approx \langle r \rangle = 0$. By going one step further, we
can write $P(r)$ as $P_0$ plus a small perturbation. A simple calculation 
shows that 
\begin{equation}
  P(r) =  P_0(r) + \frac{\omega^5 \sigma_m^2 r \sqrt{\langle m \rangle} }
         {\pi(r^2+ \omega^2 \langle m \rangle)^3}
\end{equation}
From expression (\ref{eq:lyap}) for the inverse localization
length $\gamma$, we find that 
\begin{equation}
  \gamma = \langle r \rangle = \frac{\omega^2 \sigma_m^2}{8 \langle m \rangle}
  ,\qquad {\rm for} \qquad \omega \longrightarrow 0
\label{eq:analoc}
\end{equation}
an equation derived in Ref.~\cite{MI70} (see Fig.~\ref{f:local}).

\begin{figure}
\begin{center}
\includegraphics*[width=7cm]{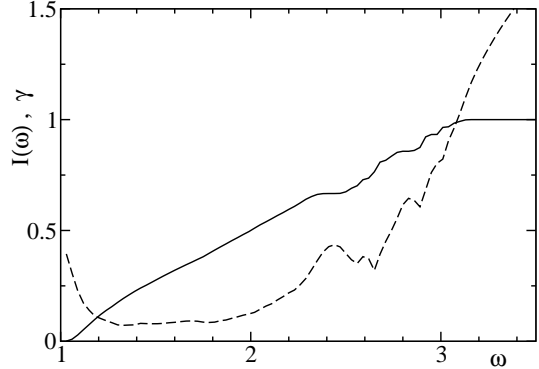}
\caption{Integrated density of states, $I(\omega)$ (solid line), and inverse 
localization length, $\gamma$ (dashed line) for a random-mass chain as in
Fig.~\ref{f:local} with the addition of a unit frequency on-site potential.} 
\label{f:local2}
\end{center}
\end{figure}

If we add a harmonic on-site potential to Eqs. (\ref{eq:lin0}), the 
corresponding scenario becomes analogous  to that of the 1d Anderson problem,
with all eigen-functions being exponentially localized. This is illustrated in
Fig.~\ref{f:local2}, where we have added harmonic springs with unit constant
(i.e. a force term $-q_n$ acting on the $n$-th particle) to the chain with
random masses considered above. The lower band-edge is now strictly bounded
away from zero and the inverse localization length does not vanish.
 
{\it The temperature profile - }
In order to study the non-equilibrium properties, we need to include the 
coupling with the thermal reservoirs. Here below, we consider Langevin-type heat 
baths, as they allow an analytic treatment, though limited to the weak-coupling 
regime.  The starting equation writes
\begin{equation}
  m_n \ddot q_n = q_{n+1} - 2q_n + q_{n-1} + \delta_{n1}(\xi_+ - \lambda 
  \dot q_1) + \delta_{nN}(\xi_- -\lambda \dot q_N)
\label{eq:lin1}
\end{equation} 
where for simplicity we have assumed $\lambda_+ = \lambda_-=\lambda$.
Although the equations are still linear, there is no general method
to derive an analytic solution for generic values of the coupling constant
$\lambda$. Accordingly, we restrict ourselves to considering the perturbative
regime  $\lambda \ll 1$. It is convenient to introduce the new variable
\begin{equation}
u_n = \sqrt{m_n} q_n \quad ,
\label{eq:change}
\end{equation}
which allows rewriting Eq.~(\ref{eq:lin1}) as
\begin{eqnarray}
  \ddot u_n = \frac{u_{n+1}}{\sqrt{m_n m_{n+1}}} - &2&\frac{u_n}{m_n} + 
              \frac{u_{n-1}}{\sqrt{m_n m_{n-1}}} + \nonumber \\ 
          & &  \frac{\delta_{n1}}{m_1}(\sqrt{m_1}\xi_+ -\lambda \dot u_1) + 
        \frac{\delta_{nN}}{m_N}(\sqrt{m_N}\xi_--\lambda \dot u_N) \, .
\label{eq:lin2}
\end{eqnarray}
The advantage of this representation is that the operator describing
the bulk evolution is symmetric and, accordingly, is diagonalized by an 
orthogonal transformation. In other words, upon denoting with $e_n^k$ 
the $n$-th component of the $k$-th eigenvector, it turns out that 
$\sum_n e_n^k e_n^h = \delta_{kh}$ and $\sum_k e_n^k e_j^k = \delta_{nj}$~. 

With reference to the new variables $U_k = \sum_n u_n e_n^k$~, the equations 
of motion write as
\begin{equation}
 \ddot U_k = - \omega^2_k U_k - \lambda \sum_j C_{kj} \dot U_j + 
   \frac{e_1^k}{\sqrt{m_1}} \xi_+ + \frac{e_N^k}{\sqrt{m_N}} \xi_- 
\label{eq:newbase}
\end{equation}
where $-\omega^2_k$ is the real, negative $k$-th eigenvalue of the unperturbed 
evolution operator and
\begin{equation}
  C_{kj} = \frac{e^k_1 e^j_1}{m_1} + \frac{e^k_N e^j_N}{m_N} \quad .
  \label{eq:matrix}
\end{equation}
Eq. (\ref{eq:newbase}) shows that the normal modes are coupled among 
themselves through the interaction with the reservoirs.
Standard stochastic calculus applied to the modal energy 
$E_k = (\dot U_k^2 + \omega_k^2U_k^2)/2$ shows that the stationarity
condition for the time average $\langle \dot E_k \rangle = 0$ implies 
\begin{equation}
 C_{kk} \langle \dot U_k^2 \rangle  + \sum_{j\ne k} C_{kj}
 \langle \dot U_k \dot U_j \rangle = T_+ \frac{(e_1^k)^2}{m_1}  +
    T_- \frac{(e_N^k)^2}{m_N} \qquad  .
\label{eq:tloc2}
\end{equation}
Let us now show that, in the small-coupling limit, this sum is negligible.
In fact, from the equality
\begin{equation}
\frac{d\langle \dot U_k U_h \rangle}{dt} = 0 \qquad ,
\end{equation}
we find that
\begin{equation}
\langle \dot U_k \dot U_h \rangle - \omega_k^2 \langle U_k U_h
 \rangle - \lambda \sum_j C_{jh} \langle \dot U_j U_h \rangle  = 0 
\end{equation}
By solving this equation together with the symmetric expression obtained by
exchanging $k$ and $h$, it is transparent that 
$\langle \dot U_k \dot U_h \rangle$ is proportional to $\lambda$  
for $k \ne h$. Accordingly, up to first order in $\lambda$, one has
\begin{equation}
 \langle \dot U_k^2 \rangle = \frac{1}{C_{kk}} 
   \left(T_+ \frac{(e_1^k)^2}{m_1} + T_- \frac{(e_N^k)^2}{m_N} \right)
\qquad ,
\label{eq:tloc3}
\end{equation}
As a consequence, by neglecting first order corrections, the local 
temperature $T_n$ reads 
\begin{equation}
T_n  = \left\langle \left(\sum_k \dot U_k e_n^k \right)^2 \right\rangle 
  \approx 
\sum_{k=1}^{N} \frac{(e_n^k)^2}{C_{kk}} \left( T_+ \frac{(e_1^k)^2}{m_1} + 
  T_- \frac{(e_N^k)^2}{m_N} \right)
\label{eq:simpl2}
\end{equation}
This is basically the expression derived by Matsuda and Ishii \cite{MI70}. 
As a consistency check, one can easily verify that if
$T_+ =T_- =T$, the local temperature $T_n$ is equal to $T$ for all values of 
$n$ (this follows from the normalization condition on the eigenvectors).
Furthermore, the profile is flat also when $T_+ > T_-$ and the amplitude of all eigenvectors
is the same at the two chain-ends: in this case, $T_n=(T_+ + T_-)/2$. An 
obvious limiting case is the homogeneous chain. Generally speaking, even though
the dynamics of a generic disordered chain is statistically invariant under
left-right symmetry, the same does not hold true for each individual eigenvector.
This induces some spatial dependence that we  shall investigate in the
following. 

Visscher \cite{V71} challenged equation (\ref{eq:simpl2}) by arguing that
quasi-resonances could be generic enough to affect typical realizations of the
disorder. In fact, a crucial assumption in the  derivation of the expression
for the temperature profile is that cross-correlations in Eq.~(\ref{eq:simpl2})
are negligible. This is basically  correct unless pairs of frequencies are
sufficiently close to each other, in which case the resonance phenomena should
be properly taken into account. Visscher indeed discussed particular examples
of mass distributions, where a more refined theory is needed. However, as
long as one is interested in generic realizations, the problem is whether
quasi-degeneracies in the spectrum are sufficiently frequent to significantly
affect the overall validity of formula (\ref{eq:simpl2}). In all the cases 
we have considered this issue turned out to be practically irrelevant.

Formula (\ref{eq:simpl2}) does not allow to obtain an analytic form of
the profile since it requires the knowledge of the eigenvectors and, on the
other hand, the localization length alone  does not suffice to predict their
amplitude at the boundaries. Therefore numerical diagonalization of the 
Hamiltonian for each different realization of the disorder is required. 
In Fig.~\ref{f:profran}, we have plotted the stationary temperature profile 
for a single realization of the disorder versus the rescaled lattice position
$x=n/N$. Strong fluctuations accompany an average decrease of the temperature 
from $T_+$ to $T_-$. 
 
\begin{figure} 
\begin{center} 
\includegraphics*[width=8cm]{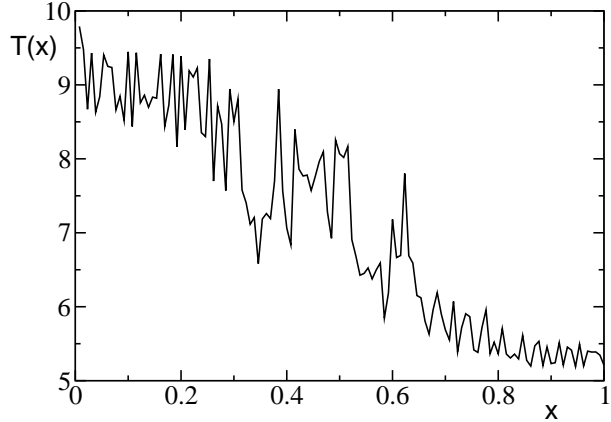}
\caption{Temperature profile as predicted from Eq.~(\ref{eq:simpl2}) for 
a given realization of disorder in a chain of length $N=128$, with $T_+=10$
and $T_-=5$.}
\label{f:profran} 
\end{center} 
\end{figure} 

This suggests averaging over independent realizations of the disorder in order
to better investigate the convergence properties with the system size.
In Fig.~\ref{f:profluc}a we have plotted the profile averaged over 1000 
realizations. Upon increasing the chain length, the profile seems to slowly 
attain a linear shape, but sizeable deviations are still present for chains 
as long as $N=512$. Such a slow convergence is
confirmed in Fig.~\ref{f:profluc}b, where we plotted the sample-to-sample
variance $\sigma_T^2$ of the temperature field. Although its asymptotic
behavior is even less clear, it is at least evident that fluctuations do not
vanish in the thermodynamic limit. This is tantamount to saying that the
temperature profile is not a self-averaging quantity. 

\begin{figure} 
\begin{center} 
\includegraphics*[width=8cm]{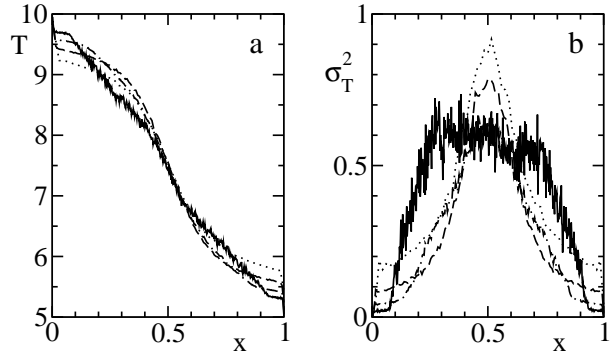}
\caption{(a) The disorder-averaged temperature profile as predicted from 
the formula (\ref{eq:simpl2}) for different chain lengths 
(dotted, dashed, dot-dashed and solid curves refer to $N=64$, 128, 256, 
and 512 respectively). (b) The variance of the temperature in the same 
notations as in panel (a).}
\label{f:profluc} 
\end{center} 
\end{figure}

\begin{figure} 
\begin{center}
\includegraphics*[width=7cm]{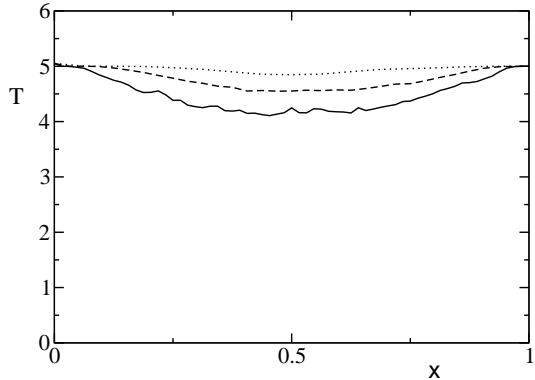}
\caption{Average temperature profile of random-mass chains for $N=16$, 32 
and 64 (dotted, dashed and solid curve, respectively). The coupling constant is 
$\lambda = 1$. The average is performed over 1000 realizations of the disorder 
each of $5\cdot 10^5$ time units.}
\label{f:prof-flat} 
\end{center}
\end{figure}

\begin{figure} 
\begin{center}
\includegraphics*[width=7cm]{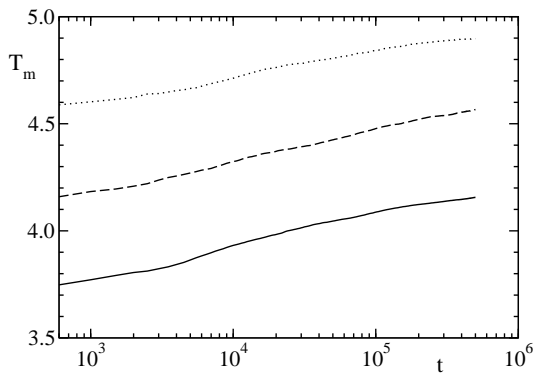}
\caption{Behavior of the time average of the temperature in the 5 central
sites for $N=16$, 32 and 64 (dotted, dashed and solid curve, respectively).}
\end{center}
\label{f:temp-flat} 
\end{figure} 

Such difficulties dramatically emerge when performing direct simulations of a
disordered chain. This issue is of great practical importance also in view of
more complex models where analytical results  are not available.  The major
problem is represented by the extremely slow convergence towards the asymptotic
regime that can be explained as follows. Eq.~(\ref{eq:newbase})  shows that
the effective coupling of each eigen-mode with the reservoirs is proportional to
its square amplitude at the extrema. Therefore, all eigenmodes that are
localized away from the boundaries can thermalize only in astronomically long
times. To be more specific, the coupling strength of an  eigen-mode
characterized by an inverse localization length $\gamma$ is of order
$\exp(-\gamma N)$, since it is localized at a distance equal, on the average, to
the half of the chain length. This implies that the asymptotic profile is
attained over times that grow exponentially with $N$. In other words, the
stationary state is never reached in the thermodynamic limit. To practically
illustrate the issue, we have simulated a chain in contact with two stochastic
heat baths operating at the same temperature $T=5$ (see
Fig.~\ref{f:prof-flat}). To further emphasize the slow convergence, all atoms
have been  initially set at rest in their equilibrium positions. Even for
relatively short chains ($N \sim {\mathcal O}(10^2)$~)
almost $10^6$ time units do not suffice to fully
thermalize the bulk.  A more direct way of looking at the convergence to the
flat temperature profile is by monitoring the cumulative time average $T_m$,
performed on the 5 central sites (and averaged also over different realizations
of the disorder). The data reported in Fig.~\ref{f:temp-flat}, with the choice
of a logarithmic scale for the time axis, give an idea of the time
needed to reach the equilibrium value $T_m = 5$.

{\it Heat flux - }
In the case of stochastic heat baths, like those considered in the previous
section, one can determine the total heat flux by using Eq.~(\ref{eq:flusto}).
By making use of Eq.~(\ref{eq:simpl2}) we obtain \cite{MI70,V71}
\begin{equation}
  j(\lambda,N) \; = \;\lambda (T_+-T_-)\sum_{k}
\frac{(e_1^k)^2(e_N^k)^2}{m_N(e_1^k)^2+m_1(e_N^k)^2} \; \equiv \;
 \sum_k J_k
\label{eq:fluxo}
\end{equation}
where the $k$-th addendum $J_k$ is naturally interpreted as the contribution 
of the $k$-th mode. As intuitively expected, the latter is 
larger for modes that have larger amplitudes at the boundaries and 
couple thus more strongly with the reservoirs. 
This interpretation can be justified from Eqs.~(\ref{eq:newbase}).
Indeed, in so far as cross-coupling can be neglected, the dynamics of the 
$k$-th eigenmode is approximately described by the equation
\begin{equation}
 \ddot U_k = - \omega^2_k U_k - \lambda C_{kk} \dot U_k + 
   \frac{e_1^k}{\sqrt{m_1}} \xi_+ + \frac{e_N^k}{\sqrt{m_N}} \xi_-
\qquad  .
\label{eq:newbase2}
\end{equation}
Standard stochastic calculus shows that, in the stationary regime, the 
energy exchanged per unit time with the two thermal baths is equal to $J_k$, 
where $J_k$ coincides with the expression implicitly defined by 
Eq.~(\ref{eq:fluxo}). 

However, heat transport is characterized by more subtle mechanisms than one 
could infer from this simple picture of independent modes. This is immediately 
understood if we look at the general expression for the local heat flux, 
Eq.~(\ref{eq:newflux}), that, in the case of harmonic chains, reduces to
\begin{equation}
j \;=\; \langle j_n \rangle = -\langle q_n \dot q_{n+1} \rangle .
\label{eq:nfluxhar}
\end{equation}
By expanding $q_n$ and $\dot q_{n+1}$ in eigenmodes, this equation 
can be rewritten as
\begin{equation}
\langle j_n \rangle = -\sum_{k,h=1}^N \frac{e^k_{n+1}e^{h}_{n}}
  {\sqrt{m_n m_{n+1}}} \langle U_k \dot U_h \rangle ,
\label{eq:nfluxex}
\end{equation}
an expression that, in spite of the explicit presence of the subscript $n$,
is independent of $n$. The interesting point that is made transparent by this
formula is that a non-vanishing heat flux is necessarily associated with the
existence of {\it correlations among different modes}. This is all the way 
more relevant, once we realize that diagonal terms with $k=h$
vanish, being $\langle U_k \dot U_{k} \rangle$ the average of the 
derivative of a bounded function. This observation seems to be in contrast 
with the derivation of Matsuda-Ishii formula itself, that is basically obtained 
by treating all modes as evolving independently of each other. Anyway, we
should notice that the heat flux is proportional to $\lambda$ and this is
compatible with the existence of weak modal correlations. In fact, 
``velocity-velocity'' or ``position-velocity'' correlations may arise from 
the fact that all eigenmodes are subject to the same noise source (except 
for a multiplicative factor) and this may well induce a sort of 
synchronization among them. 

The existence of this type of coherence has been numerically investigated 
and confirmed in Ref.~\cite{FMC98}, where the behavior of a homogeneous 
harmonic chain has been thoroughly studied. Here below, we proceed with our 
perturbative analysis by deriving an analytic expression. Let us start by
noticing that the equality $d\langle U_k U_h \rangle/dt = 0$ implies that 
\begin{equation}
\langle \dot U_k U_h\rangle=- \langle \dot U_h U_k\rangle \, .
\end{equation}
This antisymmetry property together with the further equality 
$d\langle \dot U_k \dot U_h\rangle/{dt}=0$ imply that 
\begin{equation}
\lambda C_{kh} \left( \langle \dot U_k^2  \rangle +
     \langle \dot U_h^2 \rangle \right) - \left(\omega_k^2-\omega_h^2\right) 
  \langle \dot U_k U_h \rangle = 2 \lambda 
  \left( T_+\frac{e^k_1 e^h_1}{m_1} + T_-\frac{e^k_N e^h_N}{m_N} \right) 
\end{equation}
After replacing the expression of $\langle \dot U_k^2 \rangle$ (see 
Eq.~(\ref{eq:tloc3})~) in the above equation, the latter can be solved for 
$\langle \dot U_k U_h\rangle$. 

Instead of discussing the general case, we prefer to illustrate the  presence
of these correlations coming back to the simpler case of a homogeneous harmonic chain (
see also Ref. \cite{FMC98}~)~. In
his context, one can, in principle, obtain a a general expression for the
correlations by transforming  Eq.~(\ref{eq:zsol}) (derived for an arbitrary
$\lambda$ value) in $k$ space.  However, the calculations, though
straightforward, are rather tedious. Therefore, we limit ourselves to
considering the weak-coupling limit.  The symmetry of the eigenmodes imply
that  if $\delta = h-k$ is an even number correlations vanish, while for  an
odd $\delta$ we have 
\begin{equation}
  \langle \dot U_k U_{k+\delta} \rangle = 2 \frac{\lambda (T_+-T_-)}{m}
  \frac{e^k_1 e^{k+\delta}_1}{\omega_{k+\delta}^2-\omega_k^2} 
\label{eq:corana}
\end{equation}
In Fig.~\ref{f:cormod}a we report the numerical results for a chain of length 
$N=128$ with fixed boundary conditions and interacting with two thermal baths 
at temperatures $T_+ = 75$ and $T_-=25$.  Apart from the residual statistical
fluctuations, a reasonable agreement with expression (\ref{eq:corana}) is found
upon letting   $\lambda = 0.056$ (approximately equal to the inverse of the
average separation between consecutive collisions).   Fig.~\ref{f:cormod}b
shows  the results for the same length but a stronger  coupling strength. The
different shape of the curves is a clear indication that higher order terms
must be taken into account, since the perturbative approach implies that the
coupling constant acts just as a multiplicative factor. It is anyhow
interesting that shape itself is invariant under change of $\delta$ as it can
be seen in the inset where the three curves are rescaled to their maximum
value.

\begin{figure} 
\begin{center}
\includegraphics*[width=7cm]{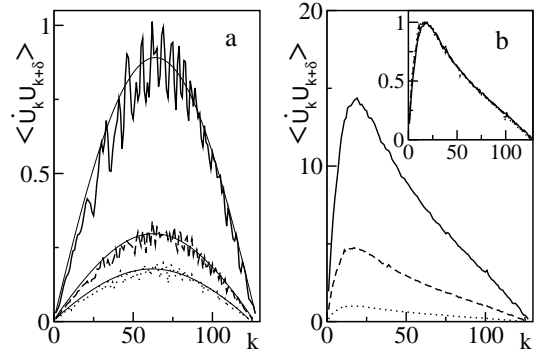}
\caption{
Modal correlations for a harmonic chain of length $N=128$ with fixed boundary
conditions.  Averaging has been performed on a time $t = 10^8$ units.  Solid,
dashed and dotted curves correspond to $\delta =1$, 3, and 5, respectively,
while the thin lines correspond to the analytic expressions (\ref{eq:corana}). 
(a) refers to a weak-coupling case: the times  between consecutive collisions
are  uniformly distributed in the interval $[19 \div 21]$. (b) corresponds to a
strong coupling: collision times distributed in $[0.9 \div 1.1]$.  The inset
contains  the same curves, after rescaling to the maximum values. } 
\label{f:cormod} 
\end{center}
\end{figure} 

{\it The thermal conductivity - }
If obtaining an accurate analytic estimate of the heat flux is as difficult as
determining the temperature profile, we can at least make use of
Eq.~(\ref{eq:fluxo}) to determine its scaling properties. In fact, since
high-frequency eigenmodes are strongly localized, it is clear that only the
first part of the spectrum contributes significantly to the heat flux. Let thus
$N_e$ be the number of modes whose localization length is larger
than the sample size $N$. From  Eqs.~(\ref{eq:intden},\ref{eq:analoc}), it
follows that  $\gamma \simeq \sigma^2_m I(\omega)^2/\langle m \rangle$. Upon
writing $I = N_e/N$ and imposing $\gamma = 1/N$, we find that
\begin{equation} 
   N_e \; = \; \frac{\langle m \rangle}{\sigma_m}\sqrt{N} \qquad .
\end{equation} 
At this point, it becomes crucial to specify the boundary conditions.
Let us first consider the case of free ones: the square amplitude of
an extended eigenmode in a lattice of size $N$ is of the order $1/N$.
This implies that the contribution to the heat flux of one of such modes is
$\lambda (T_+-T_-)/N$ and the heat flux in Eq. (\ref{eq:fluxo}) 
can be estimated as 
\begin{equation} 
   j_{free}(\lambda,N) \; \propto \;
   \lambda (T_+-T_-) \frac{\langle m \rangle}{\sigma_m}
   \frac{1}{\sqrt{N}} \qquad .
\end{equation} 
As a result, the conductivity diverges as 
\begin{equation} 
  \kappa_{free} \; \propto \;  
  \lambda \frac{\langle m \rangle}{\sigma_m}\sqrt{N} \qquad .
\end{equation} 
This scaling was first derived in Ref.~\cite{MI70} and later confirmed in
Ref.~\cite{KPW78} by means of a different approach. On the other hand, 
for fixed boundary conditions the result is completely different. In this 
case all eigenmodes must vanish for $n=0$ and $n=N+1$. By approximating the 
site-to-site variation of $e^k_n$ with the wavenumber $k/N$, we find that the 
square amplitude of $e^k_1$ and $e^k_N$ is of order $k^2/N^3$. As a
consequence, summing all such addenda up to $k=N_e$ in Eq. (\ref{eq:fluxo}) yields
\begin{equation} 
   j_{fix}(\lambda,N) \; \propto \;
   \lambda (T_+-T_-) \left(\frac{\langle m \rangle}{\sigma_m}\right)^3
   \frac{1}{{N}^{3/2}} \qquad .
\end{equation} 
and, accordingly, the thermal conductivity vanishes as  
\begin{equation} 
  \kappa_{fix} \; \propto \;  
  \lambda \left(\frac{\langle m \rangle}{\sigma_m}\right)^3
 \frac{1}{\sqrt{N}} \qquad .
\end{equation} 

The above estimates give only the leading orders in $N$. In view of
the previously encountered strong finite-size effects, it is   
crucial to check directly the convergence to the asymptotic results.
To this aim it is convenient to compute the effective exponent
\begin{equation}
\alpha_{eff}(N) \; = \;  \frac{d\ln \kappa}{d\ln N} \qquad .
\label{logder}
\end{equation}
The results are shown in Fig.~\ref{f:der-flux} for the case of 
fixed boundaries. 
For weak coupling, the conductivity has been evaluated by numerically 
computing the eigenvectors and averaging the Matsuda-Ishii formula
(\ref{eq:fluxo}) over 1000 realizations of the disorder. The asymptotic regime
$\alpha = -1/2$ is approached very slowly (see the circles): one should
consider $N$ values much  greater than $10^3$. Similar results are found at
stronger coupling by directly simulating chains that interact with stochastic
baths. The  data (diamonds in Fig.~\ref{f:der-flux}) suggest that a relatively
strong  coupling reduces the amplitude of finite-size corrections. Finally, it
is important to  realize that the small coupling of the exponentially localized
modes with the thermal baths does not cause any problem to the temporal
convergence of $j(\lambda,N)$, since independently of whether such modes have
reached  their stationary state, their contribution to the total heat flux is anyhow
negligible.

\begin{figure} 
\begin{center} 
\includegraphics*[width=7cm]{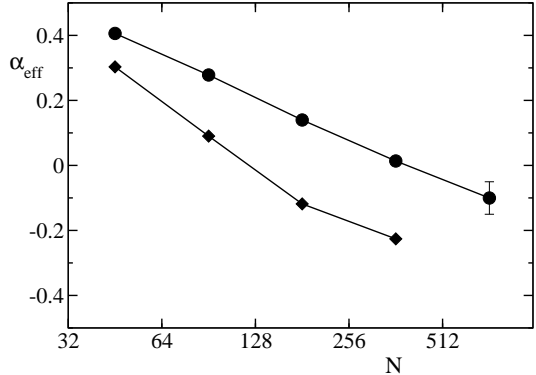}
\caption{The effective exponent defined in (\ref{logder}) 
versus the chain length. The logarithmic derivative has been evaluated
with finite differences, subsequent points correspond to chain of 
double length. Circles are obtained from the Matsuda-Ishii formula, 
while diamonds correspond to a direct simulation of a chain interacting 
with stochastic baths operating at $T_+ = 10$ and $T_-=5$, respectively. 
The collision times were uniformly distributed 
in the range $[1 \div 2]$.}
\label{f:der-flux} 
\end{center} 
\end{figure} 

In summary, not only boundary conditions affect the scaling behavior of
$\kappa$, but they give rise to qualitatively different scenarios: for free
boundaries, disordered harmonic chains exhibit an anomalous conductivity as it
diverges in the thermodynamic limit. On the contrary, a disordered chain with
fixed boundaries behaves as good insulator! This latter scenario is  brought to
an extremum if we add an on-site potential. In fact, we have already  mentioned
that all eigen-functions become exponentially localized and this implies that
conductivity is exponentially small in $N$. This is again very much reminiscent
of the electrical conductivity of the Anderson problem.

Dhar \cite{D01} went even further and showed how the scaling behavior of the
conductivity with the system size depends also {\it on the spectral properties of the
heat baths}. More precisely, if $\kappa \propto N^\alpha$ then the  exponent
$\alpha$ is determined by the low-frequency behavior of the noise spectrum.
This implies that a suitable choice of the  latter can even lead to a finite
conductivity! Such a scenario is less  unphysical than it may appear at a first
glance. Integrability of the motion implies that the only scattering mechanism
that determines the  heat resistance is the interaction with the baths. It is
therefore  reasonable that the actual way in which the latter transfer energy
among the modes plays a crucial role. 

{\it Modal fluxes - } We conclude this section with a discussion of the modal
heat fluxes $J_k$ as defined from Matsuda-Ishii formula (\ref{eq:fluxo}). Besides
providing a finer verification of the latter, the analysis is useful in
understanding the individual contributions of each ``channel'' to the heat
transport.  The spectra $J_k$ obtained for different chain lengths are reported
in  Fig.~\ref{f:flux-mais}. They have been scaled each to the maximum $J_M$,
while $k$  has been scaled to the wavenumber $k_M$ of the maximum itself. In
practice, this is asymptotically equivalent to scaling the vertical axis by a
factor $N^2$ and the horizontal axis by $1/\sqrt{N}$. We have preferred to
adopt this strategy in order to possibly get rid of the strong finite-size
corrections revealed by the previous analysis. Indeed, the good data collapse
(except for the right tail) noticeable in Fig.~\ref{f:flux-mais} is very
suggestive of the existence of an asymptotic spectrum. Additionally,  the
quadratic growth predicted for fixed b.c. is much more clear (see the inset)
than one could have expected from the scaling behavior of the total heat flux 
for the same chain lengths.

\begin{figure} 
\begin{center} 
\includegraphics*[width=7cm]{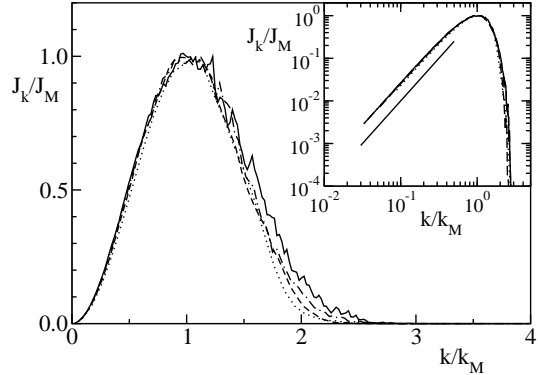}
\caption{The heat flux (scaled to the maximum value) versus the wavenumber $k$
(scaled to the position of the maximum) for different chain-lengths (dotted,
dashed, dot-dashed and solid curves refer to $N=32$, 64, 128 and 256,
respectively. The same quantities are reported in doubly logarithmic scales 
in the inset, to show the quadratic growth for small $k$.}
\label{f:flux-mais} 
\end{center} 
\end{figure} 

Besides looking at the average heat fluxes $J_k$, we have studied their
sample-to-sample fluctuations, by computing the variance $\sigma_J^2$. The
relative variance plotted versus the scaled (as in the previous figure)
wavenumber indicate that fluctuations are independent of the chain length
(see the almost overlapping curves in Fig.~\ref{f:varflu}): this means that
$J_k$ is not a self-averaging quantity in the thermodynamic limit and this
holds true even for low $k$-wave-numbers, whose behavior results, in principle,
from a spatial averaging over increasingly long spatial scales. More precisely,
we observe that the variance increases exponentially with $k$, starting
approximately from 0.26 for the longest wavelength. Moreover, we have compared
the theoretical results with non-equilibrium simulations with stochastic heat
baths and collision times distributed uniformly in the interval 
$[30 \div 60]$~. The
data plotted in Fig.~\ref{f:conf}  reveal a very good agreement over various
orders of magnitude if the coupling constant is set equal to $1/54$.

\begin{figure} 
\begin{center} 
\includegraphics*[width=7cm]{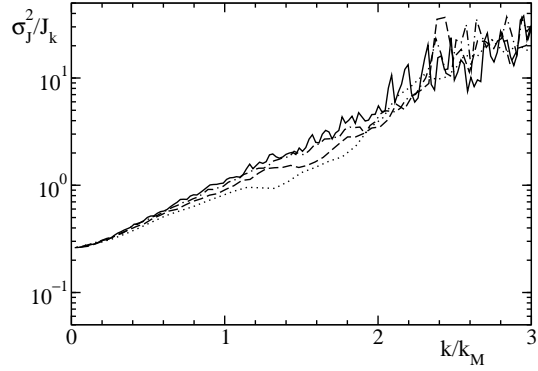}
\caption{The relative standard deviation versus the rescaled wavenumber
$k$ reveals a clear exponential growth.}
\label{f:varflu} 
\end{center} 
\end{figure} 

\begin{figure} 
\begin{center} 
\includegraphics*[width=7cm]{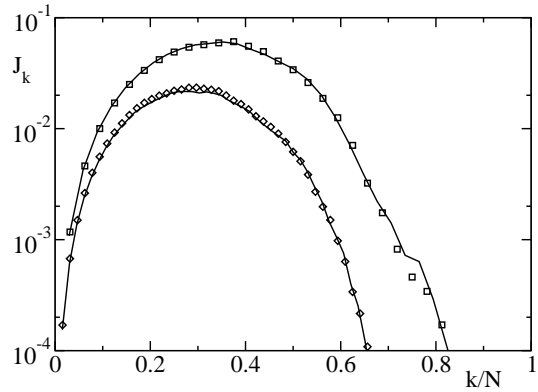}
\caption{Comparison between numerical results and the theoretical formula for
the heat flux (solid curve) (see eq.(\ref{eq:fluxo})~). The agreement is 
fairly good for both $N=32$ (squares) and $N=64$ (diamonds).}
\label{f:conf} 
\end{center} 
\end{figure}

\section{Linear response theory}
\subsection{The Boltzmann-Peierls equation}
In this section, we briefly review the ``traditional" approaches to the 
determination of thermal conductivity. Although they have a major importance 
in solid state applications, here, we limit ourselves to a very sketchy 
discussion, since it is sufficient to point out only those general issues that 
are of interest for our purposes. In this respect, our presentation is inspired 
by the review article of Jackson \cite{J78}.   

The most elementary picture of heat conductivity is based on the analogy with 
kinetic theory of gases where $\kappa=Cv_s \ell/3$, $C$ being the heat 
capacity, $v_s$ the sound velocity and $\ell$ the mean free path. In a lattice, 
one can imagine to replace particles with normal modes, but it is, of course, 
necessary to take into account that the latter have different group 
velocities, $v_{\bf k}={\partial \omega /\partial {\bf k}}$, depending on
their wavenumber. Accordingly, the above expression for $\kappa$ generalizes
to
\begin{equation}
\kappa \; = \; {1\over 3} \int d{\bf k} \, C_{\bf k}v^2_{\bf k} \tau_{\bf k} 
\quad ,
\end{equation} 
where we have introduced the relaxation time 
$\tau_{\bf k} = \ell_{\bf k}/v_{\bf k}$ that can be determined by
phenomenologically including all possible scattering mechanisms (anharmonicity, 
impurities, boundary effects, electrons etc.) that must be computed in some 
independent way. 

A less heuristic derivation of the above formula is obtained by solving the
Boltzmann equation in the relaxation time approximation \cite{calla}. In 1929 
R. Peierls proposed his celebrated theoretical approach based on the 
Boltzmann equation. The main idea is again taken from kinetic theory: 
lattice vibrations responsible for heat transport can be described as an 
interacting gas of phonons \cite{peierls}. Accordingly, one can introduce 
the time-dependent distribution function $N_k(x,t)$ of phonons with 
wavenumber $k$ in a macroscopically small volume around 
$x$.\footnote{Here, for simplicity, we are referring to a one-dimensional 
ordered crystal.} If we further limit ourselves to considering only the
cubic term in the interaction potential (three-phonon processes), the kinetic 
equations are of the form
\begin{eqnarray}
\label{cineq}
{\partial N_k \over \partial t} +  v_k {\partial N_k \over \partial x} \; &=& \;
\int \int dk' dk'' \{
\left[ N_k N_{k'} N_{k''} - (N_k+1) (N_{k'}+1) N_{k''}\right] W_{k k' k''} 
\nonumber \\
&+&{1\over 2 }\left[N_k (N_{k'}+1)(N_{k''}+1) - (N_k+1) N_{k'} N_{k''}\right] 
W_{k k' k''} 
\} \quad ,
\end{eqnarray} 
where the transition probability $W_{k k' k''}$ is basically obtained from the 
Fermi's golden rule. The r.h.s. is the collision integral, i.e. the difference 
between 
the number of processes (per unit time) that either increase or decrease the 
number of phonons in the state $k$. These nonlinear integro-differential 
equations are clearly impossible to solve in general. An approximate solution 
is obtained in the limit of small applied gradients, i.e. by looking for small 
perturbations of the equilibrium distribution $N_k = N_k^{\rm eq} + \delta N_k$,
where $N_k^{\rm eq}=(\exp(\hbar\omega_k/k_BT) - 1)^{-1}$.
This allows to write a linearized kinetic equation \cite{peierls,landau}
that, in the stationary case, is of the form
\begin{equation}
v_k {\partial N_k^{\rm eq} \over \partial T} {\partial T \over \partial x} \; = \; 
{\mathcal I}(\delta N) \quad,
\label{linboltz}
\end{equation}
where ${\mathcal I}$ is the linearized collision integral, which is a linear
functional of the $\delta N_k$s. The $\tau_k$ are thus determined as the 
eigenvalues of the problem.

Anyway, some useful information about thermal conductivity can be obtained by 
looking directly at the dynamics in Fourier space. Whenever third and fourth 
order terms are present in the equations of motion  (as in in the FPU model 
(\ref{fpu})), one can write
\begin{equation} 
\label{fourfpu}
\ddot Q_k \;=\; -\omega_k^2 Q_k - \sum_{k_1,k_2}\; 
V^{(3)}_{k k_1 k_2} Q_{k_1}Q_{k_2}  
-\sum_{k_1,k_2,k_3}\; V^{(4)}_{k k_1 k_2 k_3} Q_{k_1}Q_{k_2}Q_{k_3} \quad .
\end{equation} 
Accordingly, the harmonic part of the flux given by Eq.~(\ref{eq:flmod})
satisfies the dynamical equation
\begin{eqnarray}
\dot J_H &\;=\;&{im\over 3} \sum_{k k_1 k_2} 
\left[-v_k \omega_k + v_{k_1} \omega_{k_1}
+  v_{k_2} \omega_{k_2} \right] V^{(3)}_{-k k_1 k_2} Q_{k}Q_{k_1}Q_{k_2}
\nonumber \\
&+& {im\over 4} \sum_{-k k_1 k_2 k_3} \left[-v_k \omega_k + v_{k_1} \omega_{k_1}
+  v_{k_2} \omega_{k_2} +  v_{k_3} \omega_{k_3}\right] 
V^{(4)}_{-k k_1 k_2 k_3} Q_{k}Q_{k_1}Q_{k_2}Q_{k_3} .
\label{jhdot}
\end{eqnarray}
As expected, this implies that $J_H$ is a constant of motion in the 
harmonic case. In a perfect lattice, one has the selection rules
(remember that the mode indices range between $-N/2+1$ and $N/2$)
\begin{eqnarray}
& V^{(3)}_{-k k_1 k_2}  \ne 0 \qquad  &{\rm for }\qquad -k+k_1+k_2=0, \pm N 
\nonumber \\
& V^{(4)}_{-k k_1 k_2 k_3} \ne 0\qquad &{\rm for } \qquad-k+k_1+k_2+k_3=0, 
\pm N .
\label{selrule}
\end{eqnarray}
Peierls observed that if there is no dispersion, i.e.  $\omega_{k}=v_s|k|$, 
both three- and four-phonon contributions to (\ref{jhdot}) vanish when the sums 
in (\ref{selrule}) are equal to zero. Therefore, the only finite contributions 
to $\dot J_H$ are those arising from the so-called {\it Umklapp} processes 
corresponding to the above sums being equal to $\pm N$.

Besides the above general considerations, there are some specific comments 
regarding the role of dimensionality that can be drawn in the framework of 
perturbative theories. Indeed, by evaluating the r.h.s of (\ref{jhdot}) to 
lowest order (i.e. by replacing $Q_{k}$ with the harmonic solution 
(\ref{ak})) and averaging out the fast oscillations, one is left 
with the leading resonant terms that satisfy additional conditions
like 
\begin{equation}
-\omega_k+\omega_{k_1}+\omega_{k_2}=0  
\qquad -\omega_{k}+\omega_{k_1}+\omega_{k_2}+\omega_{k_3}=0
\label{selrule2}
\end{equation}
etc. Thus there is a big difference between three and four phonon processes
in 1-dimension, as in the former case the first of conditions (\ref{selrule})
and (\ref{selrule2}) cannot be simultaneously satisfied (see Ref.~\cite{L00} 
for some numerical results). The net result of this argument due to Peierls 
is that the lowest-order contributions to thermal resistance in 1-dimension 
arises from four phonon {\it Umklapp} processes. Of course, in higher 
dimensions, the situation is different, because the energy and momentum 
constraints can be satisfied also by three-phonon processes due to 
the existence of different (longitudinal and transverse) branches of the 
frequency spectrum. Let us finally notice that, although -- in the spirit of 
a perturbative calculation -- only the harmonic component of the flux $J_H$ 
has been considered, no basic difference arises upon including also 
the nonlinear component \cite{J78}. 

Similar conclusions can be drawn from the analysis of the high-temperature 
limit of the Boltzmann equation, following a standard argument originally due to
Pomeranchuk (see for example Chapter  VII in Ref. \cite{landau}). In this
limit, with reference to processes involving three long-wavelength phonons 
of wavenumber $k,k',k''$, the transition probability  $W_{k k' k''}\sim
kk'k''\sim k^3$, whereby $ N_k^{\rm eq}\approx k_BT/\hbar \omega_k \sim
1/k$.  From this consideration and from Eq.~(\ref{cineq}), one can estimate 
the linearized collision integral appearing in Eq.~(\ref{linboltz}) as 
${\mathcal I} \, \sim \,\delta N_k/\tau_k \,\sim \, k^{d+1} \delta N_k$,
in dimension $d$. Accordingly, the solution of Eq.~(\ref{linboltz})  
diverges as $\delta N_k\sim k^{-(2+d)}$. Finally, using the expression for 
the heat flux (\ref{eq:flmod}) with $\langle E_k\rangle = \hbar \omega_k N_k$ 
yields 
\begin{equation}
\langle J_H \rangle\;=\; 
\sum _{\bf k} \, \hbar\omega_{\bf k} v_{\bf k} \delta N_k 
\;\propto\; 
\int {dk \over k^2} .
\label{deltan}
\end{equation}
This means that the contribution of such processes would lead to a thermal
conductivity diverging like $1/k$ in any dimension. In order to avoid this 
divergence it is therefore necessary that long wavelength phonons are scattered
by short-wavelength ones. Following Ref.~\cite{landau}, let us consider the
process in which a short-wavelength phonon of index $k$ annihilates into two
phonons of index $k'$ and $k-k'-N$ with $k\ll N$ and $k\ll k'$. The condition
(\ref{selrule2}) requires $\omega_k=\omega_{k-k'}+\omega_{k'}$ that, in turn
can be satisfied only for $v_k\simeq v_s$. More generally, one can show  that,
in the absence of degeneration  points in the spectrum, the condition to be
fulfilled is that the group velocity of short wavelength phonons is {\it
larger} than the sound velocity i.e. $|v_{\bf k}|>v_s$. Once again this
constraint cannot be satisfied in one-dimensional homogeneous chains and one
would conclude that a finite conductivity can be possibly established only by
means of higher-order  processes.

The Boltzmann-Peierls approach is certainly one of the milestones of the 
theory of thermal transport in solids. Nonetheless, it is important to
recall that its derivation  is essentially based on second-order perturbation
theory (through the collision kernel $W_{k k' k''}$, which is evaluated by means  of
Fermi's golden rule) and involves the use of random phase approximation among the
phonon modes, which is certainly less appealing  than the {\it Stosszahlansatz}
originally introduced by Boltzmann for molecular collisions. It is however 
remarkable to notice how classical perturbative approaches are able to 
predict some peculiarities in low-dimensional anharmonic lattices.

\subsection{The Green-Kubo formula}
The other major tool, commonly used when dealing with transport
processes, is linear response theory. At variance with the response to
mechanical perturbations (e.g. an external electric field), heat conduction is
a process driven by boundary forces. Therefore, a conceptual difficulty arises, 
since there is no explicit small term in the Hamiltonian to be used as
an expansion parameter. This difficulty can be overcome at the price of a
stronger assumption, namely that {\it local equilibrium holds}. The hypothesis
looks physically reasonable, but it is far from being rigorously based even 
in simple mathematical models and it has been often devised as one of the weak 
points in the foundation of the whole theory. If local equilibrium holds, a 
temperature field $T({\bf x})$ can be defined accordingly, thus allowing to 
introduce a non-equilibrium distribution function
\begin{equation}
\rho \;=\; Z^{-1} \exp \left( -\int d{\bf x}\beta ({\bf x}) h({\bf x})\right)
,
\end{equation}
where $h({\bf x})$ is the Hamiltonian density, while $Z$ is the
partition function. By now assuming that the deviations from global 
equilibrium are small, we can write 
$\beta({\bf x})= \beta(1-\Delta T({\bf x})/T)$ and thus 
\begin{equation}
\rho \;=\; Z^{-1} \exp \left[ -\beta ({\mathcal H}+{\mathcal H}')\right],
\end{equation}
where ${\mathcal H}'$ is the perturbative Hamiltonian
\begin{equation}
{\mathcal H}' = -\frac{1}{T} \int d{\bf x} \Delta T({\bf x}) h({\bf x}) .
\end{equation}
It is therefore possible to proceed with a perturbative expansion, 
obtaining the well known Green-Kubo formula that in the classical case 
reads \cite{KT}
\begin{equation}
\label{realgk}
\kappa_{GK}\,=\,\frac{1}{k_BT^2}\lim_{t \to \infty} 
\int_0^t d\tau \, \lim_{V\to \infty} V^{-1}\langle \vJ (\tau) \vJ(0) \rangle 
\end{equation}
where $\vJ$ is the total heat flux defined by Eq.~(\ref{fluxgeneral}) 
and  $\kappa_{GK}$ should, more properly, 
be a tensor. However, in the simple case of isotropic homogeneous solids
made of atoms placed on a regular hyper-cubic lattice, the thermal 
conductivity tensor has a diagonal representation, with equal non-zero 
components: upon these assumptions, in dimension $d$, $\kappa_{GK}$ reduces 
to the  scalar quantity 
\begin{equation}
\label{realgk2}
\kappa_{GK}\,=\,\frac{1}{k_BT^2 d}\lim_{t \to \infty} 
\int_0^t d\tau \, \lim_{V\to \infty} V^{-1}\langle \vJ (\tau)\cdot \vJ(0) 
\rangle  .
\end{equation}

As often stated, Eq.~(\ref{realgk}) relates the non-equilibrium transport 
coefficient to the fluctuations of a system at equilibrium. 
It has to be reminded that its rigorous mathematical foundation 
is still lacking \cite{BLR00}. Besides this, 
there are several subtleties connected with a correct implementation of this
formula. First of all, one should notice that the infinite-volume limit
should be taken before the long-time limit, in order to avoid the 
problem of Poincar\'e recurrences. This is a particularly delicate matter 
whenever a slow decay of correlations is present. 

The next issue concerns the meaning of the ensemble average 
$\langle \cdot \rangle$. In the derivation {\it \'a la} Kubo, it denotes a 
canonical average, while the formally identical expression obtained by 
Green refers to the micro-canonical ensemble. In this latter case, if the total 
momentum $P$ is conserved, it has to be set equal to zero, otherwise 
$\langle \vJ \rangle \ne0$ and the integral in Eq.~(\ref{realgk}) would 
trivially diverge. Alternatively, as observed in \cite{BLR00}, one may compute 
the truncated correlation functions $\langle \vJ(t) \vJ(0)\rangle_T = 
\langle \vJ(t) \vJ(0) \rangle -\langle  \vJ \rangle^2$ for any
$P\ne0$.\footnote{Overlooking this point may lead to some confusion as in
Ref.~\cite{PC00}.}

Another way to see the problem was pointed out by Green himself \cite{G60}. He 
noticed that the microscopic 
expression of the heat flux to be employed in Eq.~(\ref{realgk}), {\it depends 
on the chosen ensemble}. Indeed, he showed that while Eq.~(\ref{fluxgeneral}) 
is the correct expression in the micro-canonical case, a ``counter-term" must 
be subtracted in the grand-canonical case. This is readily seen by letting 
$\dot\vx_i \to \dot\vx_i - {\bf v}$ where $\bf v$ is the velocity of the 
center of mass. Up to terms of order less than $\sqrt{N}$, one has
\begin{equation} 
\vJ_{gc}\;=\;\vJ_{mic} -  (E + pV){\bf v} \; =\;
        \vJ_{mic} - H {\bf v} \qquad ,
\end{equation}
where $E$ is the average energy, $p$ the pressure and $H$ the total enthalpy 
of the system. 
The micro-canonical and grand-canonical ensembles give the same results provided 
that the micro-canonical energy density is chosen to correspond to the canonical
temperature $T$. The reason for the different expressions to be 
used is that the same observable has different time-correlations in the 
different ensembles.

\subsection{Mode-coupling theory}
Despite the conceptual difficulties lying behind the derivation of 
Eq.~(\ref{realgk}), this formula provides a well defined prescription for 
determining the thermal conductivity  $\kappa_{GK}$
from the current-current correlation function at equilibrium. We 
claim that an effective method for estimating this correlation function is 
provided by the well known mode-coupling theory (MCT), introduced some 
decades ago to approach the problem of long-time tails in fluids \cite{PR75}. 
Since a rigorous proof of this statement is still lacking, we prefer to 
illustrate first some simple arguments to support the claim.
Afterwards, we introduce MCT in the simple context of the FPU model with 
cubic nonlinearity. Finally we briefly recall the major quantitative
results in various dimensions.

According to the classical perturbative approach outlined in the first section
of this chapter, the time scale for the relaxation process towards the
stationary state can be determined by linearizing the collision operator in 
Eq.~(\ref{cineq}). However, this may be insufficient, because of the possible 
existence of subtle dynamical correlations that escape the predicting ability 
of a perturbative approach: see, for instance, the necessity to invoke higher 
order processes at high temperatures, or the well known existence of slow
relaxation processes at low temperatures when the dynamics is almost
integrable.

A more powerful approach can be built starting from the observation that in 
solids, like in fluids, the slowest processes arise from the ``diffusion" of 
conserved quantities (such as energy and momentum). Actually, macroscopic 
conservation laws necessarily imply the existence of a hydrodynamic behavior, 
dominated by the time scales $1/\gamma_k$ associated with the dynamics of 
long-wavelength, i.e. low-{\bf k}, modes. It is then crucial to observe that
the damping factor $\gamma_k$ is expected to vanish in the limit $k \to 0$ 
both in crystals characterized by the existence of an 
acoustic band, as well as in fluids (e.g., hard spheres interacting via 
short-range potentials). In fact, as ${\bf k} = 0$ modes correspond to exactly 
conserved quantities, long-wavelength ones must, by continuity reasons, be
characterized by a slow dynamics. It is then of primary interest to notice 
that this is true independently of the strength of the perturbative terms. 
Accordingly, $\gamma_k$ may be very small also for strong nonlinearities, 
when no standard perturbative approach can be meaningfully implemented. 
As a matter of fact, numerical studies of several models of anharmonic 
crystals with confining nearest-neighbor interactions (like the FPU model) 
nicely confirm this scenario \cite{ACM95} also at high temperatures,
when high-$k$ modes behave like ``thermal" variables, rapidly
relaxing to equilibrium.\footnote{In the high--energy regime, the same 
models are known to exhibit a strongly chaotic and convincingly ergodic 
behavior. The time-scale separation between low-$k$ and high-$k$ modes seems 
to contradict this statement. This is not the case, since it has to be noticed 
that low-$k$ modes, although playing a major role in transport phenomena, are a 
negligible fraction of the spectrum in the thermodynamic limit. Accordingly,
equilibrium properties are dominated by thermal modes and ``hydrodynamic" 
deviations from ergodicity can be detected only as higher order corrections.}

According to the previous discussion, the long-time behavior of any 
current-current correlation function depends on how $\gamma_k \to 0$ in the 
limit $k \to 0$. In particular, if the damping factor $\gamma_k$
vanishes too rapidly with $k \to 0$, the temporal decay of the heat-flux
correlation function may be so slow that the integral in Eq.~(\ref{realgk}) 
diverges. 
In general this effect ought to depend also on the space dimension. Indeed,
almost conserved modes propagating with the sound velocity through the
lattice are expected to propagate more efficiently in low than in high 
dimensions, where the presence of transverse modes favors collision 
mechanisms. Actually, a well defined, i.e. finite, transport coefficient in 
anharmonic solids should emerge from an efficient dissipation of the energy 
of sound waves. In this sense, it is worth recalling that Fourier law follows
from the assumption that the temperature field obeys a diffusive equation.

As already anticipated, we illustrate an application of MCT to the FPU model 
with cubic nonlinearities. The procedure is an extension of linear response 
theory and represents a first step towards the construction of a formal 
approach for the description of transport properties in models of 1d solids. 
Moreover, it provides the theoretical 
background for arguing that the same features should be observed in all models 
where nonlinear effects can be ascribed to the two leading algebraic terms. 

In the framework of linear response theory, the dynamics of slow modes 
is described by generalized Langevin equations. These are linear stochastic 
equations  with memory terms and are usually derived with the projection method
introduced independently by Mori and Zwanzig \cite{KT}. 
To illustrate this in the present context, let us consider a one-dimensional 
chain like (\ref{acoustic}) with periodic boundary conditions.
The equations of motion for the normal coordinates (\ref{modi}) can be 
written as (see also  Eq.~(\ref{fourfpu}))
\begin{equation}
\ddot Q_k \;=\; -\omega_k^2 Q_k + {\mathcal F}_k
\label{newton}
\end{equation}
where ${\mathcal F}_k$ accounts for mutual interactions among the modes while 
$\omega_k$ denotes the normal-mode frequency. One can then define a projection 
operator ${\mathcal P}$, acting on the generic scalar observable $O$ as
\begin{equation}
 {\mathcal P} O  = \sum_k \left[\frac{\langle O Q_k^*\rangle}
 {\langle|Q_k|^2\rangle} Q_k + 
 \frac{\langle O  \dot Q_k^*\rangle}{\langle |\dot Q_k|^2 \rangle } \dot Q_k 
 \right] .
\end{equation}
Due to the conservation law of total momentum, we expect that 
the slow dynamics should be associated with the long-wavelength Fourier
modes $Q_k$ with $|k|\ll N/2$. Moreover, translational invariance implies that 
each mode is uncorrelated from the others, so that we can consider each 
mode separately.  The corresponding projected equations of motion (that are  
still exact) read as~\cite{L98}
\begin{equation}
{\ddot Q}_k  + \int_0^t \Gamma_k(t-s) \dot Q_k(s) \, ds
+ {\tilde \omega}_k^2 Q_k  
\;=\; R_k \quad,
\label{eqmost}
\end{equation}
where the random force $R_k(t) = (1 - {\mathcal P}) {\ddot Q}_k$ is related to
the memory kernel $\Gamma_k$ by the fluctuation-dissipation theorem
\begin{equation}
\Gamma_k(t) = \beta \langle  R_k(t) R_k^* (0) \rangle .
\label{gktr}
\end{equation}
The first effect of nonlinearities is to induce a temperature-dependent 
renormalization of the dispersion relation
\begin{equation}
{\tilde\omega}_k^2 = (\beta \langle |Q_k|^2 \rangle)^{-1} =
(1 + \alpha) \omega_k^2 \quad \quad , \quad \quad  
\alpha (\beta) = \frac{1}{\beta} \frac{\int \exp{-\beta V(x)} dx}
{\int x^2 \exp{-\beta V(x)} dx} -1 \quad.
\label{sound}
\end{equation}
This amounts to renormalizing the sound velocity from the ``bare'' value 
$v_s$ to  ${\tilde v}_s = v_s \sqrt{1 + \alpha}$.\footnote{Notice
that for $T\to 0$ ($\beta \to \infty$), $\alpha(\beta) \to 0$ as the integrals
in Eq.~(\ref{sound}) reduce to Gaussian integrals.}
A straightforward consequence of Eq.~(\ref{eqmost}) is that the normalized
correlation function ${\mathcal G}_k(t) = \beta {\tilde\omega}_k^2
\langle Q_k(t) Q_k^*(0) \rangle$ (${\mathcal G}_k(0)=1$) obeys the equation 
of motion
\begin{equation}
{\ddot{\mathcal G}}_k(t) + 
\int_0^t \Gamma_k(t-s) {\dot{\mathcal G}}_k(s) \, ds 
+ {\tilde\omega}_k^2 {\mathcal G}_k(t) 
\;=\; 0  \quad \quad .
\label{gkt}
\end{equation}

Up to here we performed an exact but formal manipulation of the equations of
motion. The crucial point is the explicit computation of the memory kernel
$\Gamma_k(t)$. MCT is an approximate, self-consistent method for obtaining such an
expression in terms of ${\mathcal G}_k(t)$. A first conceptual difficulty of the
projection approach is that $R_k$ does not evolve with the full
Liouvillean operator. One can bypass the problem with the replacement
\cite{PR75}
\begin{equation}
\langle R_k(t) R_k^*(0) \rangle \to 
\langle 
{\mathcal F}_k(t) 
{\mathcal F}_k^*(0) \rangle ,
\label{corfor}
\end{equation}
whose validity is based on the implicit hypothesis that the slow terms 
possibly contained in ${\mathcal F}_k(t)$ can be neglected in the 
thermodynamic limit. 
A second simplification amounts to factorizing multiple correlations.
For example, in the case of a quadratic force, one obtains~\cite{L98}
\begin{equation}
\Gamma_k(t) \;\propto\;  \, {\tilde\omega_k^2\over N} \sum_{k'} \,
{\mathcal G}_{k'}(t) \, {\mathcal G}_{k-k'}(t) \quad.
\label{memcub}
\end{equation}
This approximate expression of the memory kernel constitutes, together
with Eq. (\ref{gkt}), a closed system of equations for ${\mathcal G}_k$ that 
has to be solved self-consistently by introducing the 
Laplace transform of $\Gamma_k$,
\be
\Gamma_k(z)\;=\;\int_0^\infty e^{-izt} \Gamma_k(t) dt ,
\ee
and, analogously, ${\mathcal G}_k(z)$. One finds that 
(with $\dot{\mathcal G}_k(0)=0$)
\begin{equation}
{\mathcal G}_k(z) = \frac{
iz + \Gamma_k(z)}
{z^2 - {\tilde\omega_k}^2 -iz \Gamma_k(z)} \quad.
\end{equation}
As long as dissipation is small enough, ${\mathcal G}_k$ has the form 
\begin{equation}
{\mathcal G}_{k}(t)\sim\exp(i\lambda_k t) ,
\label{express}
\end{equation}
where the pole $\lambda_k$ of the above transform is approximately given by
\begin{equation}
\lambda_k \;=\; \pm {\tilde\omega}_k + i{\Gamma_k(|\tilde \omega_k|)\over 2} ,
\label{poles}
\end{equation}
which can be regarded as a generalized dispersion relation. The imaginary 
part $\gamma_k$ of $\lambda_k$  represents the effective relaxation rate 
of each Fourier mode as a consequence of its interaction with all the other 
modes.
 
As discussed in Ref.~\cite{PR75}, in one dimension, the self-consistent 
calculation predicts a singularity of the memory function at $z=0$: 
$\Gamma(z,q) \sim z^{-1/3} q^2$. Substituting this result into
the approximate dispersion relation (\ref{poles})  yields the non-analytic
dependence of the relaxation rate in the limit of small wave-numbers \cite{E91}
\begin{equation}
\gamma(q) \; \propto \; q^{5/3} .
\label{q53}
\end{equation}
The scaling behavior (\ref{q53}) has been confirmed in molecular-dynamics 
simulations performed at equilibrium for the FPU model \cite{L98} in a significative
range of chain sizes.
(see Fig.~\ref{f:gamma}).  
\begin{figure}
\begin{center}
\includegraphics*[width=8cm]{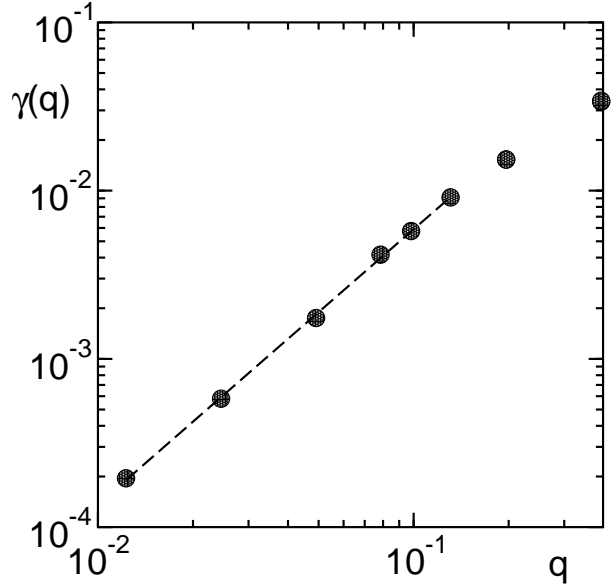}
\caption{The wavenumber dependence of the relaxation rates $\gamma(q)$ 
for the quartic FPU potential (\ref{fpu}) $g_4\varepsilon=8.8$ . All the
points were obtained from the initial decay of the envelope of 
${\mathcal G}_{1}$ for increasing values of $N$ up to $N=2048$. The dashed 
line is a power-law fit $q^{1.64}$.}
\label{f:gamma}
\end{center}
\end{figure} 
The above discussion can be generalized to the case in which a cubic force 
is present \cite{L98}. Although the dependence of the relaxation rate on 
the temperature depends on the specific form of the anharmonic
potential, the rate $q$ is expected to be the same for all
one-dimensional models, where the theory applies.
Application to 2d lattices has been also discussed \cite{LL00}.

More in general (see again \cite{E91}), the following dependences are found
for $d= 1 \div 3$,
\begin{eqnarray}
\lambda(q) &\simeq &c q - ic' q^{5/3} + \ldots
\quad \quad  (d = 1)  \nonumber \\
\lambda(q) &\simeq & c q - i c' q^2 \ln{q} + \ldots 
\quad \quad (d = 2) 
\label{omek} \\
\lambda(q) &\simeq & c q - ic' q^2 + \ldots 
\quad \quad  (d = 3) \nonumber
\end{eqnarray}
where $c'$ is a suitable multiplicative factor. Direct numerical evidence 
of the peculiar behavior of one-dimensional systems has been found also in
fluids \cite{DLQ89,NEHF93}.

In order to understand the consequences of the above reasoning for transport 
phenomena, it is convenient to look at the dynamics of Fourier modes. If we 
assume that memory effects are all contained in a mode-dependent
relaxation-time, Eq.~(\ref{eqmost}) effectively reduces to its Markovian 
limit, 
\begin{equation}
\ddot Q_k +\gamma_k \dot Q_k + \tilde\omega_k^2 Q_k  \;=\; R_k  
\label{markov}
\end{equation}
(referred to the finite-length case), where the random force is well 
approximated by a Gaussian white process
\begin{equation}
\langle R_k(t)R^*_{k}(t') \rangle
= {\gamma_k\over\beta}\delta(t-t') \quad ,
\end{equation}
and, for the sake of simplicity, we have neglected the small frequency
shift possibly arising from the solution of (\ref{poles}). At this level of
approximation, the physical implications of the mode-mode interactions are 
contained in the dispersion relation $\gamma_k$. Eq.~(\ref{markov}) provides
a convincing description of the numerical results reported in 
Refs.\cite{ACM95,L98}, where the Fourier-mode dynamics was studied for the 
FPU-$\beta$ model.

Let us now split $V$ into its harmonic and anharmonic parts and consequently 
write the flux (\ref{jtotal}) as $J=J_H+J_A$. For a strongly anharmonic system, 
like the one we consider here, we do not expect $J_A$ to be negligible. 
Nevertheless, in the spirit of Section 5.1, one can argue that the two terms 
exhibit the same leading asymptotic behavior, so that we can restrict 
ourselves to considering the autocorrelation of $J_H$ alone. This hypothesis 
has been also successfully tested in simulations \cite{LLP98b}.  

It is convenient to rewrite expression (\ref{vpere}) as 
\begin{equation}
J_H \;=\;    \, \sum_{k} \, v_k \left(E_k - \langle E_k \rangle\right) 
\;=\;\sum_{k} \, v_k  \delta E_k \quad
\end{equation}
(compare also with Eq. (\ref{deltan}) where $\delta E_k$ are the energy 
fluctuations of each mode). Notice also that, in view of the renormalization 
of the frequencies, $E_k$ should now be defined on the basis of the 
transformations (\ref{ak}) with $\omega_k$ replaced by $\tilde\omega_k$.
From Eq.~(\ref{markov}), one expects that, for small $\gamma_k$, energy 
fluctuations satisfy the Langevin equation
\begin{equation}
\dot{\delta E_k}\;=\; -\gamma_k \,{\delta E_k} + R'_k \quad ,
\end{equation}
with no oscillation for $\delta E_k$. In such a limit, $R'_k$ is well 
approximated by a 
Gaussian and delta-correlated random process and 
$\langle (\delta E_k)^2 \rangle =k_B^2T^2$. For large $N$, we obtain 
\cite{LLP98b}
\begin{equation}
\langle  J_H(t)J_H(0) \rangle \;\propto\; \sum_k {v_k}^2 
\langle (\delta E_k)^2 \rangle e^{-\gamma_k t}
\;=\; 
{Na\over 2\pi}k_B^2T^2 \int_{-\pi/a}^{\pi/a} dq \, v^2(q) \, e^{-\gamma(q) t}
\quad .
\label{jj}
\end{equation}
Since the integral is dominated by the low-$q$ contribution at large times, we 
can estimate the long-time behavior of Eq.(\ref{jj}) by letting 
$c(q)\simeq v_s$ 
and extending the integration to infinity. Furthermore, in accordance with 
Eq.~(\ref{q53}), we let $\gamma(q)=c' q^\delta$, obtaining
\begin{equation}
\langle  J_H(t)J_H(0) \rangle \propto
{ v_s^2 k_B^2T^2 N a \over (c' t)^{1\over \delta}} \,\left[ 1
+ {\mathcal O} \left(t^{-{2\over \delta}} \right)\right].
\label{corre}
\end{equation}
This result can be generalized to derive the following long-time
behavior for the heat--flux correlation function,
\begin{eqnarray}
\langle {\bf J}(t) {\bf J}(0)\rangle &\sim t^{-3/5} 
\quad \quad &(d = 1) \nonumber \\
\langle {\bf J}(t) {\bf J}(0)\rangle &\sim t^{-1} 
\quad \quad &(d = 2) \\
\langle {\bf J}(t) {\bf J}(0)\rangle 
&\sim t^{-3/2} 
\quad \quad &(d = 3) . \nonumber
\end{eqnarray}
The knowledge of the asymptotic behavior of 
$\langle {\bf J}(t){\bf J}(0)\rangle$ now allows determining the dependence
of $\kappa$ on $N$. In fact, upon restricting the integral in 
Eq.~(\ref{realgk}) to times smaller than the typical transit time $Na/v_s$, 
one obtains,
\begin{eqnarray}
\kappa &\sim N^{2/5} 
\quad \quad &(d = 1) \nonumber \\
\kappa &\sim \ln{N}
\quad \quad &(d = 2) \\
\kappa &\sim finite
\quad \quad &(d = 3) \nonumber
\end{eqnarray}
It is worth stressing again that these results can be derived without 
making any explicit reference to the details of the interactions among atoms 
in the lattice. In practice, this is tantamount to stating that all models
characterized by short range interactions and momentum conservation should
exhibit the same kind of anomalous behavior (for $d <3$). In the next
chapter we shall see that this is not completely correct, as for non 
confining potentials, a normal conductivity is found already in one dimension. 
In all other cases, in spite of the intrinsic approximations contained in 
the MCT, the predicted scaling behavior of $\kappa$ agrees with the numerical 
estimates obtained from both equilibrium and non-equilibrium simulations.
This is not surprising if one considers that, at relatively high energies,
the time scale associated with high-$k$ modes is well separated from the 
hydrodynamic ones, so that the leading term predicted by MCT is likely to
contain all the relevant information already in relatively small systems.

\section{Anharmonic chains with momentum-conserving potentials}
\subsection{Early results} 

This section is devoted to a historical review of molecular-dynamics
studies of thermal conduction in the class of models (\ref{acoustic}). The 
first simulations date back to the pioneering work of Payton, Rich and Visscher 
\cite{PRV67} and to the contribution of Jackson, Pasta and Waters
\cite{JPW68}. In both cases, the Authors performed non-equilibrium studies
of the FPU model (\ref{fpu}) with coupling constants $g_2$, $g_3$, and $g_4$
chosen in such a way to represent the leading terms of the expansion of 
the Lennard-Jones potential (\ref{lenjo}).
In order to study the effect of impurities in the crystal, either a disordered
binary mixture of masses \cite{PRV67} or random nonlinear coupling constants
\cite{JPW68} were considered. Ironically enough, those very first computer
studies attacked the problem from the most difficult side. In fact, even
before the effect of disorder was fully understood in harmonic chains,
they studied systems where anharmonicity and disorder are simultaneously 
present. Nevertheless, those early works have at least the merit to 
have showed how the interplay of the two ingredients can lead to unexpected 
results that, in our opinion, are still far from being fully understood. Indeed,
Ref.\cite{PRV67} revealed that the simple perturbative picture in
which anharmonicity and impurities provide two independent (and thus additive)
scattering mechanisms does not hold. More precisely, the Authors found even 
cases in which anharmonicity {\it enhances} thermal conductivity. A qualitative
explanation was put forward by claiming that anharmonic coupling induces an
energy exchange between the localized modes, thus leading to an increase of
the heat flux.

The limited computer resources available at that time prevented, however, 
addressing the issue whether the combined effect of anharmonicity and disorder 
can lead to a finite conductivity. On the other hand, it was payed attention at 
the temperature profile $T(x)$, noticing irregularities that depended on the 
realization of the disorder. While we know that $T(x)$ is not a self-averaging 
observable of disordered harmonic chains, it has not yet been clarified 
whether the dependence on the realization of the disorder persists over 
long enough time scales in anharmonic chains as well. 

Additional questions that have been investigated concern the concentration of 
impurities. Besides the obvious finding that disorder reduces the value 
of heat conductivity (for fixed finite-chain length), it was noticed an 
asymmetric behavior between the case of a few heavy atoms randomly added 
to an otherwise homogeneous light-atom chain and its converse. The smaller
values of the conductivity observed in the former cases were traced back to
the larger number of localized modes \cite{PRV67}.

Having recognized the difficulty of simultaneously coping with the effects of
nonlinearity and disorder, we now turn to the simpler case of anharmonic
homogeneous chains. Some early work in this direction was performed by 
Nakazawa \cite{N70} who considered equal-masses FPU and Lennard-Jones chains 
composed of 30 particles and coupled with Langevin baths at their boundaries. 
This setup required the integration of a set of stochastic differential 
equations, a task that was admittedly unfeasible with the computer resources 
available at that time. As a consequence, several attempts of designing 
artificial but easy-to-simulate models followed these first studies. Some 
examples are reviewed in Ref. \cite{V76}\footnote{Those attempts eventually 
led to the invention of the so-called ding-a-ling model 
described in the following Section as it belongs to the different class of
chains with external substrates.}
Let us mention among them the case of the harmonic hard-rod potential, i.e. 
a harmonic well delimited by an infinite barrier located at a given distance 
from the equilibrium position. A diverging conductivity was observed
with a method akin to Green-Kubo one \cite{V76}. 

The long period of time (almost a decade) during which the problem was 
practically forgotten signals perhaps the frustration encountered in the 
search for the
minimal and general requirements for building simple 1d models with good
transport properties. In the mid eighties several authors got again interested
in the problem, being able to perform non-equilibrium simulations of chains with 
smooth inter-particle potentials and a few hundreds of particles. In particular, 
a good deal of work was devoted to reconsidering the FPU model\cite{KM93,ONS94} 
and to studying the diatomic Toda chain \cite{MB83,JM89,NS90,GM92} whose 
Hamiltonian is of the type (\ref{acoustic}), with  
\begin{equation}  V(x)\;=\; {a\over b}\left[\exp(-bx) + ax\right] \quad ,  
\label{toda0} 
\end{equation}  
and $m_l$ being a sequence of alternating light and heavy masses with given
ratio $r$. At variance with the homogeneous case $r=1$, this model is no
longer integrable at equilibrium and can be thus considered as a meaningful 
candidate for testing the validity of the Fourier law. Notice that at variance 
with homogeneous models, (\ref{toda0}) admits also an optical branch 
in the harmonic limit. 

Out of the many conflicting results, Mareschal and Amellal \cite{MA88}
recognized that fluctuations of the heat current display peculiar features
that are normally absent in higher dimensions. More precisely, they computed 
the equilibrium  autocorrelation function of the heat flux (the Green-Kubo
integrand) for a Lennard-Jones chain of 200 particles. On a qualitative level, 
they noticed that the initial fast decay was followed by a very slow convergence
to zero. This is consistent with the possibility that such a long-time tail 
be responsible for a diverging transport coefficient, in close analogy with 
what happens in low-dimensional fluids \cite{PR75}. Additionally, in
Ref.~\cite{MA88} it was checked the robustness of this feature against
the introduction of further, extrinsic, scattering mechanisms. Indeed, the 
time tails survive the addition of a moderate fraction of impurities, 
introduced as either mass defects or variable interaction potentials. 
For instance, it was considered the case where every fourth particle has
a purely  repulsive interaction with its neighbors, of the type
\begin{equation} 
V_{\rm im}(x) \;=\; \epsilon \left({\sigma\over x}\right)^{12}
\quad .
\end{equation} 
A slow decay was finally observed also in the presence of an external 
sinusoidal field (akin to the Frenkel-Kontorova substrate potential considered
in the following). It has, however, to be recognized that this last observation 
contrasts the recent results obtained for several models with on-site forces
(see Sec. 7).  

\subsection{Divergence of heat conductivity} 

It took almost another decade for the first systematic studies on the size
dependence of the conductivity to appear. An extensive series of
non-equilibrium simulations were performed for the FPU chain with quadratic and
quartic \cite{LLP97,LLP98a,LLP98b} or cubic \cite{L00} interaction potential as well
as for the diatomic Toda one \cite{H99,V99}. 

\begin{figure}
\begin{center}
\includegraphics*[width=7cm]{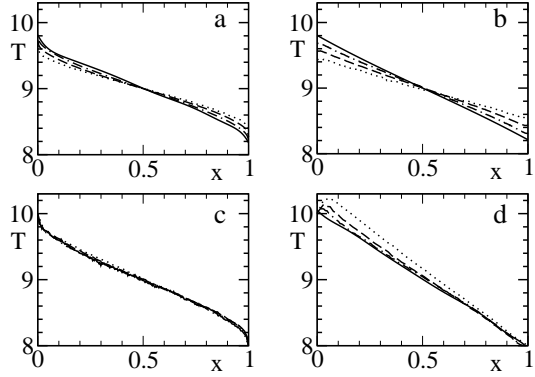}
\caption{Temperature profiles for the FPU$\beta$ model for $T_+=10$ and 
$T_- = 8$.  Panels (a) 
and (b) refer to stochastic reservoirs (acting through the randomization of
the velocity at random times uniformly distributed in the interval $[1,2]$).
Dotted, dashed, dot-dashed, and solid lines correspond to $N=128$, 256, 512,
and 1024, respectively. Panels (c) and (d) refer to Nos\'e-Hoover thermostats 
with $\Theta = 1$. In this case, dotted, dashed, dot-dashed, and solid lines 
correspond to $N=32$, 64, 128, and 256. In (a) and (c), fixed b.c. are 
imposed, while free b.c. are imposed in (b) and (d).}
\label{f:profs}
\end{center}
\end{figure}

Some evidence of anomalous transport properties is already given by the 
temperature profiles. While a fairly linear shape is obtained for free 
boundary conditions (see Fig.~\ref{f:profs}b,d), as predicted 
from the Fourier law, strong
deviations are observed for fixed boundary conditions (see
Fig.~\ref{f:profs}a,c). More important, such deviations persist
upon increasing the chain length: the nice overlap observed in panel (c)
indicates that the asymptotic temperature gradient is definitely non uniform. 
Altogether, this scenario is suggestive of the existence of long-range
effects.

Before discussing the divergence of $\kappa$, let us remark that, as
anticipated in Chap.~3, a much smaller boundary-resistance and thus smaller
finite-size corrections are found in Nos\'e-Hoover thermostats than in
stochastic ones (see Fig.~\ref{f:profs}).

As a result of the recent numerical studies, one can now safely claim that 
the conductivity of long but finite chains diverges as 
\begin{equation}
\kappa(N) \;\propto\; N^\alpha
\end{equation}
In Table I we compare the available estimates of the exponent $\alpha$
determined by different authors in various models. The numerical values 
range between 0.35 and 0.44, suggesting a nontrivial universal behavior. 
It is also remarkable to notice the overall consistency among the results
obtained with different thermostat schemes (ranging from deterministic to 
stochastic ones). 
\begin{table}
\begin{center}
\begin{tabular}{lccc}  
Model             &Reference     & $\alpha$ (NEMD) & $\alpha$ (GK)  \\
 			&		   &                 &      \\
			\hline
 			&		   &                 &      \\			
FPU-$\beta$       &\cite{LLP98a,LLP98b} & 0.37     & 0.37 \\
FPU-$\alpha$      &\cite{L00}    & $\lesssim$ 0.44 &   -  \\
Diatomic FPU  r=2 &\cite{V99}    & 0.43            & compatible  \\
Diatomic Toda r=2 &\cite{H99}    & 0.35-0.37       & 0.35 \\
	          &\cite{V99}	   & 0.39	         & compatible \\
Diatomic Toda r=8 &\cite{V99}    & 0.44            & compatible \\
Diatomic hard points&\cite{H99}    & 0.35	         &   -  \\
&		   &                 &      \\
			\hline
\end{tabular}
\vspace{0.5cm}
\caption{The estimated exponent $\alpha$ of divergence of the conductivity 
with size $N$, as obtained from both non-equilibrium molecular dynamics
(NEMD) simulations and through Green-Kubo (GK) equilibrium studies. 
Only the significative digits are reported as given in the quoted References.}
\end{center}
\end{table} 

In order to better appreciate the quality of the divergence rate that can be
numerically obtained, in Fig.~\ref{f:comparebc} we have plotted the
finite-length conductivity $\kappa(N) = JN/(T_+-T_-)$ versus the number of
particles in the FPU-$\beta$ model for fixed and free boundary conditions.
In the inset, one can see that the effective growth rate $\alpha_{eff}$ 
defined in (\ref{logder})
is basically the same in both cases, despite the clear differences in the 
actual values of the flux itself. 
Additionally, $\alpha_{eff}$ does not deviate significantly from the 
theoretical prediction ($\alpha=0.4$). 

\begin{figure}
\begin{center}
\includegraphics*[width=7cm]{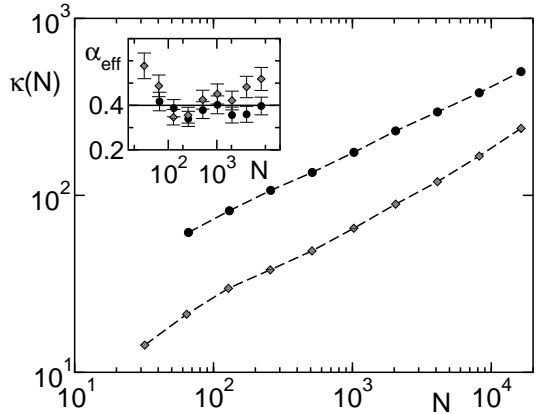}
\caption{Thermal conductivity of the FPU-$\beta$ model versus lattice length
$N$ for $T_+=0.11$, $T_-=0.09$, and $\Theta=1$. The inset shows the effective 
growth rate $\alpha_{eff}$ versus $N$. Circles and diamonds correspond
to free and fixed b.c., respectively.}
\label{f:comparebc}
\end{center}
\end{figure}

It is instructive to notice also that the now widely confirmed divergence
of the thermal conductivity with the chain length was already observed in
previous simulations in spite of opposite claims made by the authors
themselves.  We refer to a paper by Kaburaki and Machida, where it was 
conjectured a slow convergence \cite{KM93} of $\kappa(N)$ towards a finite
value in the FPU-$\beta$ model. By re-plotting their data in doubly logarithmic 
scales, a convincing power-law behavior is clearly seen instead, with 
even a quantitative agreement for the divergence exponent 
(see Fig.~\ref{f:kama}).

\begin{figure}
\begin{center}
\includegraphics*[width=7cm]{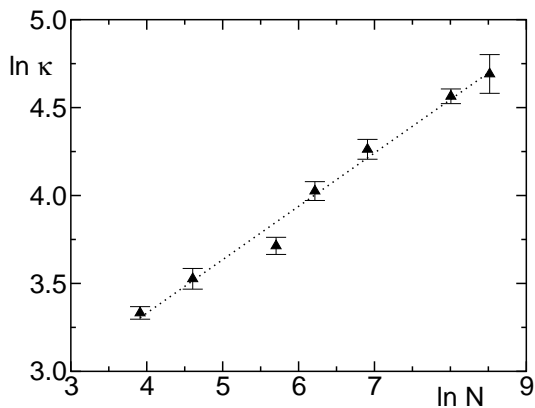}
\caption{Thermal conductivity of the FPU-$\beta$ model, $T_+=150$, $T_-=15$, 
fixed boundary condition. The data are taken from Ref.~\cite{KM93} }
\label{f:kama}
\end{center}
\end{figure}
Once the divergence is clearly established, the next question concerns the
universality of the divergence rate. The discussion of this point involves
considering a possible dependence on the temperature as well as on the leading
nonlinearities \cite{L00}. Both questions are addressed in 
Fig.~\ref{f:condalfa}, where $\kappa(N)$ is computed in the FPU-$\alpha$ model 
at a relatively low temperature. The convergence of the effective exponent 
towards 0.4 (see the inset in Fig.~\ref{f:condalfa}) suggests that the presence
of a quadratic nonlinearity in the force field does not modify the overall 
scenario observed in the FPU-$\beta$ model. Additionally, notice that 
changes in the temperature gradient, without modifying  the
average $T = (T_+ + T_-)/2$, modify the effective conductivity
only at relatively small sizes. In fact, we see in Fig.~\ref{f:condalfa} that 
the two sets of measures corresponding to $\Delta T = 0.1$ and 0.02 
(triangles and circles, respectively) approach each other for $N$ larger than
$10^3$.  In both cases $\kappa(N)$ increases linearly with $N$ for $N< 10^3$ 
and no sizeable temperature gradient forms along the chain. Both
facts hint at a weakness of anharmonic effects up to this time/length scales.
This is confirmed by the comparison with the results for a pure harmonic
chain (with the same setup and same parameters) that exhibit a clean linear
growth of $\kappa$ with $N$ (see the solid line in Fig.~\ref{f:condalfa}) 
and a few-percent differences in the initial size range. The fact that 
$\kappa$ is smaller for larger $\Delta T$ can be thus attributed to a
stronger boundary scattering that reduces the conductivity. 

\begin{figure}
\begin{center}
\includegraphics*[width=7cm]{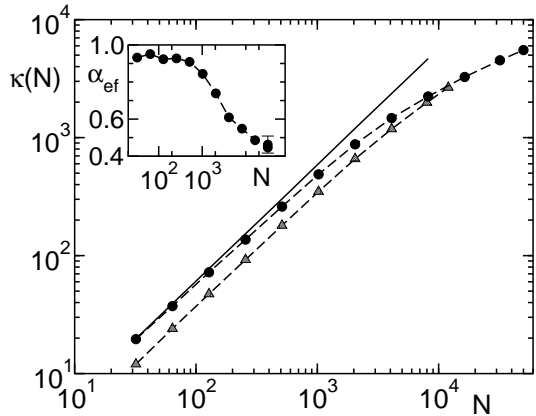}
\caption{Thermal conductivity of the FPU-$\alpha$ model versus lattice length
$N$ for $g=0.25$, $T=0.1$, and $\Theta=1$. Triangles and circles refer to
$\Delta T=0.1$ and $\Delta T=0.02$, respectively. The solid line corresponds
to the linear divergence observed in a harmonic chain with the same
temperatures. The inset shows the effective divergence rate $\alpha_{ef}$
versus $N$ for the data corresponding to full circles.} 
\label{f:condalfa}
\end{center}
\end{figure}

The overall scenario is confirmed by the computation of $\kappa$ through the
Green-Kubo formula. The very existence of an anomalous transport coefficient 
can be inferred from the slow decay of the corresponding autocorrelation 
function. In particular, if $\langle J(t)J(0)\rangle$ decays as 
$t^{-\beta}$ with $\beta < 1$, the integral in Eq.~(\ref{realgk})
diverges, thus signalling an infinite conductivity. Obviously, in finite
chains one expects that an exponential decay eventually sets in, so that
simulations should be performed for different chain lengths in order to
be sure to pick up the truly asymptotic scaling behavior. From a numerical
point of view, a simpler way to proceed consists in looking at the
low-frequency divergence of the power spectrum $S(\omega)$ of the total 
heat flux: by the Wiener-Khinchin theorem, a power-law decay of the
autocorrelations with exponent $\beta$ translates into an $\omega^{\beta-1}$
behavior of $S(\omega)$ at small $\omega$.

\begin{figure}
\begin{center}
\includegraphics*[width=7cm]{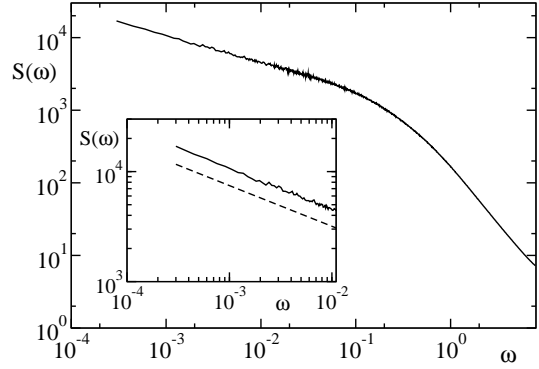}
\caption{Power spectrum $S(\omega)$ (in arbitrary units) of the global flux
for an FPU-$\beta$ chain of length $N=1024$ at a temperature $T=11.07$.
The curve results from an average over 1400 independent initial conditions.
A blow-up of the low-frequency region is reported in the inset: the dashed 
line is a shifted fit with slope -0.37.}
\label{f:spetflux}
\end{center}
\end{figure}

The spectrum plotted in Fig.~\ref{f:spetflux} is asymptotic in $N$ in the 
selected frequency range. The low-frequency divergence of $S(\omega)$
implies that the autocorrelation of $J$ decays with a rate $\beta \simeq
0.63$, i.e. that its time integral diverges. 

\begin{figure}
\begin{center}
\includegraphics*[width=7cm,angle=-90]{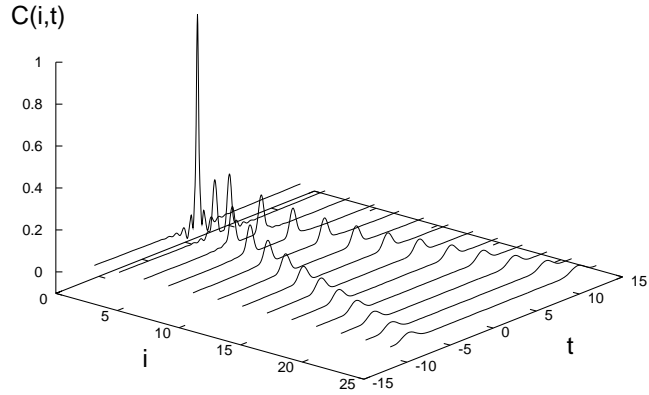}
\caption{The spatio-temporal correlation function  
$C(i,t) = \langle j_i(t)j_0(0)\rangle$ 
of the local flux for the FPU$\beta$ model. Micro-canonical simulations, 
energy density 8.8}
\label{crossco}
\end{center}
\end{figure}

A quantitative comparison with the previous results can be performed by
noticing that energy propagates with the constant sound velocity $v_s$.
This can be understood by, e.g., looking at the spatio-temporal correlation 
function  $C(i,t) = \langle j_i(t)j_0(0)\rangle$ of the local heat flux 
\cite{LLP98b} plotted in Fig.~\ref{crossco}. Accordingly, one can turn the
time divergence of $\kappa$ as determined from the Green-Kubo formula into
a divergence with $N$ by restricting the integral in  formula (\ref{realgk})
to times smaller than the ``transit time'' $Na/v_s$. This amounts to
ignoring all the contributions from sites at a distance larger than $N$. With
the above estimate of $C_J$, one obtains that  $\kappa \propto N^{1-\beta}$.
The latter exponent is the one reported in the last column of Table I.

It is instructive to repeat the modal analysis for the contribution to the
heat flux also in the nonlinear case. In spite of the fact that there are
no longer eigenmodes, one can determine the contribution of each mode to the 
heat flux (the modes to be considered being defined according to the imposed 
boundary conditions) \cite{FVMMC97}. The relevant difference with the harmonic case is that, 
because of the lack of integrability, the $k$-th mode does not only exchange 
energy with the left and right reservoirs, but also
with all other modes (this is precisely the mechanism that is eventually 
responsible - in 3d - for a normal conductivity). Accordingly, one can
write an energy balance equation for the $k$-th mode as
\begin{equation}
 J_k^+ + J_k^- + J_k^{nl} = 0
\end{equation}
where $J_k^{\pm}$ is a self-explanatory notation for the fluxes towards the
two heat baths, while $J_k^{nl}$ is the energy exchanged with the other modes
as a consequence of the non-integrable dynamics. Obvious global constraints
imply that $\sum_k J_k^{nl} =0$. Direct simulations performed with both
free and fixed boundary conditions suggest a much stronger property, namely
that each ``nonlinear'' flux vanishes, $J_k^{nl}=0$. The same simulations
indicate that the effect of the boundary conditions on the modal fluxes is 
qualitatively similar to that in harmonic systems. In fact, from
Fig.~\ref{f:fpuflumo}, we can see that for fixed boundary conditions, the 
contribution of low-$k$ modes is depleted and goes to 0 for $k \to 0$, while 
a growth, if not a divergence, is observed for free boundary conditions. 

From the hydrodynamic description put forward in the previous chapter,
one understands that the contribution to the anomalous behavior of
heat conductivity arises from the behavior of long-wavelength modes
in a similar way to the anomaly observed in disordered harmonic chains.
It is therefore natural to ask why boundary conditions are so important in disordered
harmonic chains that they may turn a diverging into a vanishing conductivity,
while the same does not occur in nonlinear systems. The question becomes
even more intriguing after having noticed that nonlinear systems are
characterized by a similar dependence of the modal fluxes to that found
in harmonic chains. A precise answer to this question would require combining 
in a single model internal relaxations (self-consistently described by 
mode-coupling theory) and dissipations due to the coupling with the external 
heat baths. Such a type of description is still lacking. 

\begin{figure}
\begin{center}
\includegraphics*[width=8cm]{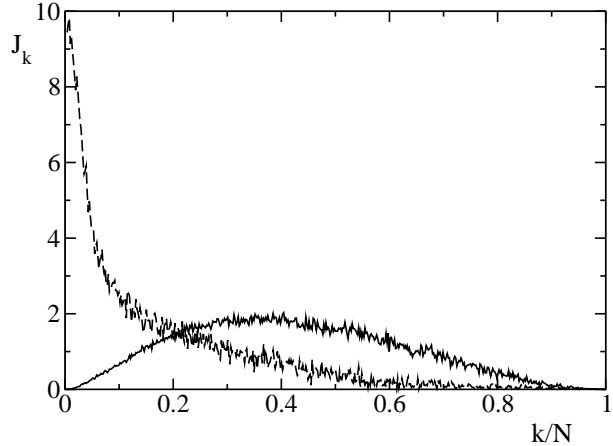}
\caption{Modal fluxes in an FPU chain of length 512 for fixed (solid line)
and free (dashed line) boundary conditions. Vertical units are fixed
in such a way that the total flux is normalized to unity in both cases.}
\label{f:fpuflumo}
\end{center}
\end{figure}
  
Finally, we discuss the role of the boundary resistance in connection with
the temperature dependence of conductivity. In fact, an interesting 
application of Eq.~(\ref{keff}) has been proposed in 
Ref.~\cite{AK01} with reference to the FPU-$\beta$ model. There, it has been
empirically found that the bulk conductivity scales with $L$ and $T$ as
\begin{eqnarray}
\label{eq-kap}
\kappa \simeq\cases{  1.2 \, L^\alpha T^{-1}&   $(T\lesssim 0.1)$\cr
    2 \, L^\alpha T^{1/4}&$ (T> 50)$\cr}\qquad   .
\end{eqnarray}
According to kinetic theory, the conductivity can also be expressed as 
$\kappa=\ell v_s C_v$. Since $C_v$ and $v_s$ are almost constant and of 
order 1 in a wide temperature range, $\ell\sim \kappa$. Hence, at low 
temperatures the boundary jumps dominate the thermal profile up to the size 
$L_*$ that can be estimated according to Eq. (\ref{keff}).
At low temperatures this effect is very strong since 
$L_*\sim (2\varepsilon/T)^{1\over 1-\alpha}$, while smaller boundary 
resistances are found at large temperatures, where 
$L_*\sim (2\varepsilon T^{1/4})^{1\over 1-\alpha}$.

\subsection{The hard-point gas} 

Although this review is basically devoted to analysing the behaviour of 
low-dimensional lattices, it is worth considering also fluid systems in so far
as no qualitative differences are expected for the scaling behaviour of their
transport properties. More specifically, in this section we discuss a set 
of point particles labelled by the index $i=1,...N$
moving along a one-dimensional box extending from $x=0$ to $x=L$. The mass,
position and velocity of the $i$th particle are denoted by $m_i$, $x_i$, and
$u_i$, respectively. Interaction occurs only through elastic collisions. After 
a collision between the $i$th and the $i+1$st particle, the respective 
velocities acquire the values
\begin{equation}
u_i'=\frac{m_i-m_{i+1}}{m_i+m_{i+1}} u_i+ \frac{2 m_{i+1}}{m_i+m_{i+1}}
u_{i+1} \quad , \quad
u_{i+1}'= \frac{2 m_{i}}{m_i+m_{i+1}}
u_i-\frac{m_i-m_{i+1}}{m_i+m_{i+1}} u_{i+1}, 
\label{coll}
\end{equation} 
as implied by momentum and energy conservation.
Between collisions, the particles travel freely with constant velocity. Notice 
also that they mantain their initial ordering (no crossing is allowed). 

The model is particularly suitable for numerical computation as it does not
require integration of nonlinear differential equations. Indeed, the dynamics
amounts simply to evaluating successive collision times and updating the
velocities according to Eqs.~(\ref{coll}). The only numerical errors are
those due to round-off. The coupling with heat baths at the boundaries can be
implemented in the usual way, e.g. by using Maxwell thermostats. Thus
whenever a particle of mass $m$ collides with a wall at temperature $T$, it is
reflected back with a velocity chosen from the distribution 
$P(u)= (m|u|/T) \exp[-m u^2/(2T)]$.  

In the limit where all the masses are equal, the system becomes integrable and
one expects the same behavior observed in harmonic chains (see Sec. 4.1). In
particular, the temperature profile is flat (with $T(x)=\sqrt{T_+ T_-}$), the 
heat current is independent of the system size and no local equilibrium is 
attained.  \footnote{In this respect, it is worth mentioning that the 
equal-mass case with dissipation has been studied by Du et al \cite{DLK95} as 
a toy model for a granular gas. They obtained a rather surprising stationary 
state which implied a breakdown of usual hydrodynamics.} However, as soon as 
the masses are different, the system is nonintegrable and (hopefully) ergodic, 
thus becoming a possible candidate for checking the validity of Fourier's law. 
Casati \cite{C86} considered the case of alternate masses ($m_{2i}= 1$,
$m_{2i+1}= (1+\delta)$ in suitable units\footnote{From the invariance of the
dynamics described by Eq.~(\ref{coll}) under mass rescaling ($m_i \to \nu 
m_i$), it follows indeed that the only independent parameters are the mass 
ratio $(1+\delta)$ and the ratio of the boundary temperatures $T_+/T_-$. In 
fact the temperature profile does not change under  $m_i \to \nu m_i$. Also 
from the boundary conditions, it is easily shown that 
$T( \nu T_+, \nu T_-, x )= \nu T(T_+,T_-,x)$.}). While from his numerical
results it was not possible to draw any definite conclusion about the scaling 
behaviour of the conductivity, more recent simulations by Hatano \cite{H99} 
suggest a divergence rate consistent with what found for FPU (see Table I). 
Although we do not see any reason why coupled rotors and FPU should belong to  
different universality classes, the behaviour of conductivity in the 
hard-point gas appears to be still quite a controversial issue: simulations 
performed by Dhar \cite{D01b} point to a slow divergence 
($\kappa\sim L^{\alpha}$ with 
$\alpha < 0.2$); equilibrium simulations discussed in Ref.~\cite{GHN01} have
led the authors even to conjecture a normal behavior; finally, further direct 
numerical studies confirm instead the 
existence of a divergence \cite{CP01}. Accordingly, it seems reasonable to
hypothesize that the uncertainty is due to strong finite-size effects that 
slow down the convergence to the expected asymptotic behaviour. 

Besides the dependence of $\kappa$ on $L$, in Ref.~\cite{D01b} it has been 
studied the shape of the temperature profile. The simulations have been 
performed 
for different values of $\delta$ and for $N$ up to $1281$ (adjusting the
system size $L$ so as to keep the average density of particles equal to $2$). 
The number of particles has been chosen to be odd, so that the two particles in 
contact with the heat baths have always the same mass. Moreover, the averaging
of the various observables has been performed over a time span corresponding
to $10^9- 10^{10}$ collisions. In Fig.~(\ref{del1}), it is plotted the 
steady-state temperature profile for different values of $N$ and the same 
$\delta=0.22$. The profile is smooth except at the boundaries, where two
temperature drops are observed whose amplitude decreases with the system size.

Upon increasing $N$, the temperature profile approaches a limiting form. 
Quite amazingly, this shape is quite close to the one that would be predicted 
by kinetic theory. In fact, let us recall that kinetic theory applied to a 
one-dimensional gas predicts the Fourier law with a conductivity
$\kappa \sim \sqrt{T}$. By then integrating the equation 
$\sqrt{T}dT/dx = c.nt$ with suitable boundary conditions, one obtains 
\begin{equation}
T_k(x)\;=\;\left[T_+^{3/2} (1-{x\over L}) + T_-^{3/2}{x\over L} \right]^{2/3}.
\label{kineq}
\end{equation} 
This corresponds to the solid curve in Fig.~(\ref{del1}). The agreement with
the numerical simulations is surprising, since kinetic theory predicts a
finite conductivity. We are inclined to interpret this result as a
confirmation of the existence of strong finite-size corrections.

The hard-point gas is interesting also for the possibility to investigate the
nearly integrable regime when $\delta \ll 1$. The smaller is $\delta$, the 
larger has to be the system size in order to generate the same temperature 
profile. In Ref.~\cite{D01b} it has been conjectured that $T(x,N,\delta)$ 
depends on $\delta$ and $N$ only through the scaling combination 
$\delta ^2 N$ and it has been proposed the following scaling form
\begin{equation}
T(x,N,\delta)=T_k(x)+\frac{1}{(\delta^2 N)^{\gamma}} g(x).
\label{scaleq}
\end{equation}
The above relation can be tested with reference to the data
reported in Fig.~(\ref{del1}). The good data collapse that can be appreciated
in the inset of Fig.~(\ref{del1}) supports the validity of the scaling
relation (\ref{scaleq}) with $\gamma=0.67$.  

\begin{figure}
\begin{center}
\includegraphics*[width=7cm]{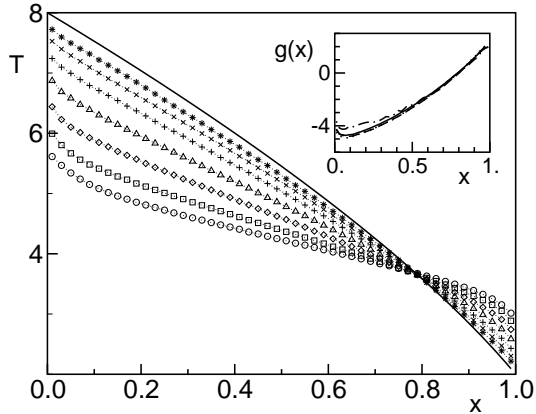}
\caption{Temperature profiles of the hard point gas for $\delta=0.22$, 
$T_+=8, T_-=2$ and sizes $N=21,41,81,161,321,641$ and $1281$ (from top to
bottom). The solid line corresponds to Eq.~(\ref{kineq}). In the inset,
it is plotted $g(x)$ (see Eq.~(\ref{scaleq})) with the data for
$N=161,321,641$ and $1281$}
\label{del1}
\end{center}
\end{figure}

\subsection{The coupled-rotor model}

The simplest example of a classical-spin 1d model with nearest neighbor 
interactions lies in the class (\ref{acoustic}) with \cite{GLPV00,GS00}
\begin{equation}
V(x) = 1 - \cos x .
\end{equation}
This model can be read also as a chain of $N$ coupled pendula, 
where the $p_i$'s and the $q_i$'s represent 
action-angle variables, respectively. It has been extensively studied 
\cite{BGG85,LPRV87,EKLR94} as an example of a chaotic dynamical system  
that becomes integrable both in the small and high energy limits, when it
reduces to a harmonic chain and free rotors, respectively. In the two 
integrable limits, the relaxation to equilibrium slows down very rapidly for 
most of the observables of thermodynamic interest (e.g., the specific heat) 
\cite{LPRV87,EKLR94}. As a consequence, the equivalence between ensemble and time 
averages is established over accessible time scales only inside a limited 
interval of the energy density $e$. Here, we focus our attention mainly on
heat conduction in the strongly chaotic regime. 

In Refs.~\cite{GLPV00,GS00}, it has been shown that, contrary to the
expectations, this model exhibits a finite conductivity in spite of the
existence of an acoustic branch in its spectrum in the harmonic limit. 
In Ref.~\cite{GLPV00}, simulations have been performed for $T_+ = 0.55$, 
$T_-= 0.35$, and chain 
lengths ranging from $N=32$ to 1024 with fixed boundary conditions and
Nos\'e-Hoover thermostats. The equations of motion have been integrated with 
a 4-th order Runge-Kutta algorithm and a time step $\Delta t = 0.01$. The 
results, reported in Fig.~\ref{f:cond-rot} 
clearly reveal a convergence to a value of $\kappa$ approximately equal to 7
(see the circles). The dotted line in good agreement with the numerical data
is the best fit with the function $a + b/N$\footnote{Notice that this 
function is asymptotically equivalent to expression (\ref{keff}), derived 
under the assumption that finite-size corrections are due only to the boundary 
resistance.}. However, more important than assessing the convergence 
properties of $\kappa(N)$ is to notice its finiteness for $N \to \infty$.

\begin{figure}
\begin{center}
\includegraphics*[width=7cm]{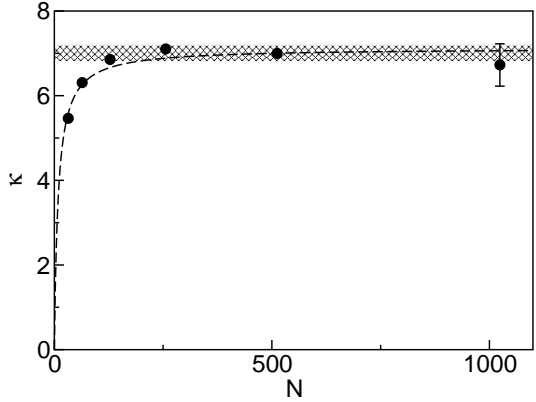}
\caption{Conductivity $\kappa$ versus chain length $N$ as obtained from 
non-equilibrium molecular dynamics. Circles correspond to the rotator model 
with temperatures $T_+ = 0.55$ and $T_- = 0.35$. The dashed line represents 
the best fit with the function $a + b/N$. The shaded region represents the 
uncertainty about the conductivity on the basis of the Green-Kubo formula.}
\label{f:cond-rot}
\end{center}
\end{figure}

In fact, this is the first system where normal heat conduction has been 
convincingly ascertained in the absence of an external field. Precisely 
because of this 
atypical behavior, it is important to confirm this result with a computation 
of thermal conductivity through the Green-Kubo formula. In Ref.~\cite{GLPV00}, 
micro-canonical simulations have been performed
in a chain with periodic boundary conditions. In the absence of thermal baths, 
the equations of motion are symplectic; accordingly, the Authors  
have made use of a 6-th order McLachlan-Atela integration 
scheme \cite{MA92}, fixing the energy density equal to $e=0.5$, a value that
corresponds to $T \approx 0.46$, close enough to the average temperature 
in the nonequilibrium simulations. 

The correlation function has been computed by exploiting the Wiener-Khinchin 
theorem, i.e. by anti-transforming the Fourier power spectrum of the total
flux. The curve plotted in Fig.~\ref{f:corr-rot} indicate both a clean 
exponential decay and an independence of the behavior of $N$ for $N\ge256$
(at least in the reported time range). This allows for an accurate 
determination of the integral of $C_j(t)$.
\begin{figure}
\begin{center}
\includegraphics*[width=8cm]{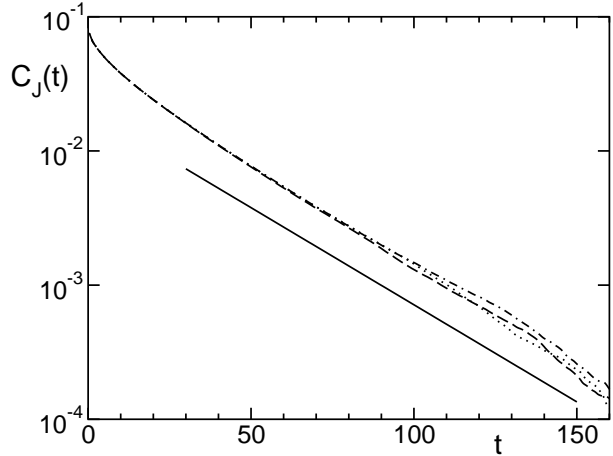}
\caption{The autocorrelation function of the total heat flux in a chain of
coupled rotors with periodic boundary conditions and energy density $e=0.5$.
Dashed, dot-dashed and dotted lines correspond to $N=256$, 512, and 1024,
respectively. The solid line, corresponding to $C_j = \exp(-t/30)$ has been
drawn for reference.}
\label{f:corr-rot}
\end{center}
\end{figure}
The gray region in Fig.~\ref{f:cond-rot} corresponds to the resulting value of 
$\kappa$ taking into account statistical fluctuations. The quantitative
agreement between the two estimates of the heat conductivity is important in
that it confirms also the finiteness of $\kappa$ in a context where this was
not a priori obvious.

In order to emphasize the difference between the dynamics of the present
model and that of the previous systems, it is instructive to look at the 
power spectrum of the low-$k$ Fourier modes. In Fig.~\ref{f:alte-mas} 
it is possible to compare the spectra of some low-$k$ modes in coupled rotors
with those in a diatomic FPU-$\beta$ chain. In the latter case, sharp peaks
are clearly visible (notice also that the peaks become increasingly narrow
upon decreasing $k$): this is a signal of an effective propagation of  
correlations \cite{LLP98b}. Conversely, in the rotors, the low-frequency 
part of the spectrum is described very well by a Lorentzian with half-width 
$\gamma = D k^2$ ($D \approx 4.3$). This represents an independent proof that 
energy diffuses, as one expects whenever the Fourier's law is established.

\begin{figure}
\begin{center}
\includegraphics*[width=7cm]{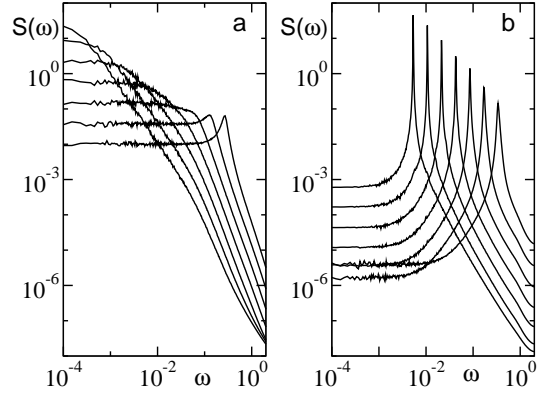}
\caption{Power spectra of the 1st, 2nd, 4th, 8th, 16th, 32nd, and 64th Fourier
mode in arbitrary units for a chain of $N=1024$ particles. Panel (a) refers
to a chain of rotors with energy density $e=0.5$ (the wavenumber increases
from left to right); panel (b) refers to a diatomic FPU-$\beta$ chain with  
masses 1, 2 and energy density $e=8.8$ (the wavenumber increases from top
to bottom in the low-frequency region. In both cases the curves result from
an average over 1000 independent simulations.}
\label{f:alte-mas}
\end{center}
\end{figure}

In the attempt to explain the striking difference in the transport behavior
exhibited by this model with respect to that of the previous models in the 
same class, one cannot avoid noticing that the pair potential 
$V(q_{i+1}-q_{i})$ possesses infinitely many equivalent valleys.
As long as $(q_{i+1}-q_{i})$ remains confined to the same valley, there is
no reason to expect any qualitative difference with, e.g., the FPU-$\beta$ 
model. Phase slips (jumps of the energy barrier), however, may very well
act as localized random kicks, that contribute to scattering of the 
low-frequency modes, thus leading to a finite conductivity. In order to test
the validity of this conjecture, one can study the temperature dependence of 
$\kappa$ for low temperatures when jumps across barriers become increasingly
rare. The data plotted in Fig.~\ref{f:cond-div} indicate that the thermal 
conductivity behaves as $\kappa \approx \exp (\eta/T)$ with 
$\eta \approx 1.2$. The same scaling behavior is exhibited by the average
escape time $\tau$ (see triangles in Fig.~\ref{f:cond-div}) though with a
different $\eta \approx 2$. The latter behavior can be explained by 
assuming that the phase slips are the results of activation processes. 
Accordingly, the probability of their occurrence is proportional to 
$\exp(-\Delta V/T)$, where
$\Delta V$ is the barrier height to be overcome. The behavior of $\tau$ is 
thus understood, once we notice that $\Delta V = 2$. In the absence of phase 
slips, the dependence of the conductivity on the length should be the same 
as in FPU-systems, i.e. $\kappa \approx N^{2/5}$. In the presence of phase 
slips, it is natural to expect that the conductivity is limited by the
average distance $\overline N$ between consecutive phase slips. Under the 
further assumption of a uniform distribution of the slips, their spatial and 
temporal separation has to be of the same order, thus implying 
that $\kappa(T)$ exhibits the same divergence as $\tau$ for $T \to 0$, though
with a different rate $\kappa \approx \exp[2\Delta V/(5T)]$. Therefore,
at least on a qualitative level, one can indirectly confirm that phase 
slips are responsible for the normal heat transport. On a quantitative level,
however, there is a discrepancy  between the observed and the expected value
of the exponent $\eta$ (1.2 vs. 0.8). Among the possible explanations for the
difference, we mention the presence of space-time correlations in 
the pattern of phase-slips and the existence of ever increasing deviations
from the asymptotic law $\kappa \approx N^{2/5}$ for $T \to 0$, due to the
vanishing of nonlinear terms. 
\begin{figure}
\begin{center}
\includegraphics*[width=7cm]{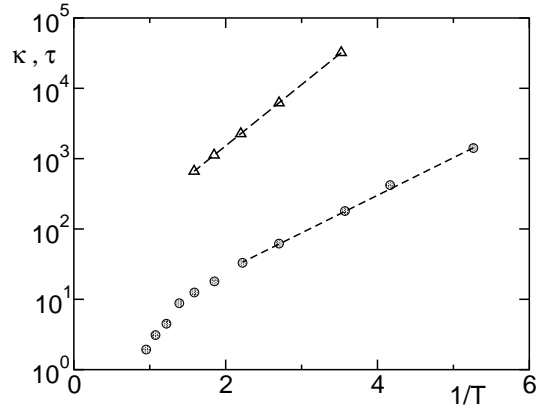}
\caption{Thermal conductivity $\kappa$ versus the inverse temperature $1/T$ 
in the rotor model (open circles). Triangles correspond to the average time 
separation between consecutive phase slips in the same system.}
\label{f:cond-div}
\end{center}
\end{figure}

In order to further test the conjecture that jumps between adjacent valleys of
the potential are truly responsible for a normal heat transport, some other 
models have been studied as well.

Let us start by considering an asymmetric version of the rotor model, namely
\begin{equation}
V(x) = A - \cos x + 0.4 \sin 2x
\end{equation}
where $A$ is fixed in such a way that the minimum of the potential energy is
zero. Simulations performed at a temperature corresponding to one quarter 
of the barrier-height again indicate that the conductivity is finite, 
confirming the empirical idea that jumps are responsible for breaking 
the coherence of the energy flux \cite{GLPV00}. 

It is instructive to look more closely at the behavior of this model.  
In view of the asymmetric potential, one might expect that the average force 
$\phi = \langle \sum f_i \rangle/N$ is non-zero (like, e.g., in the 
FPU-$\alpha$ case). Nonetheless, micro-canonical simulations show that
although the distribution of forces is definitely asymmetric, their average
value is numerically zero. This can be understood by noticing that in view 
of the boundedness of the potential, the system cannot withstand any 
compression.

Accordingly, we are led to introduce yet another model where the existence
of more than one valley is accompanied by an unbounded potential. The
simplest way to achieve this is by considering the double-well potential
\begin{equation}
V(x) = -x^2/2 + x^4/4 
\label{e:quasifi4}
\end{equation}
(it the same as in FPU-$\beta$ with the opposite sign for the harmonic term). 
The results of direct simulations \cite{V01} performed with Nos\'e-Hoover
thermostats are reported in Fig.~\ref{f:ieri} for three different values of 
the temperature below the barrier height. One can see that the growth of the 
conductivity, after an initial slowing down, increases towards, presumably, 
the same asymptotic behavior observed in the FPU model. This is at variance
with the preliminary simulations reported in Ref.~\cite{GLPV00}, where it
was conjectured a normal heat transport. These results thus indicate that
jumps alone from one to another valley are not sufficient to destroy 
the coherence of low-$k$ modes dynamics. It is necessary that subsequent
jumps be independent of one another. In the double-well potential 
(\ref{e:quasifi4}), $q_{i+1}-q_i$ cannot exhibit two consecutive jumps to 
the right. 
\begin{figure}
\begin{center}
\includegraphics*[width=7cm]{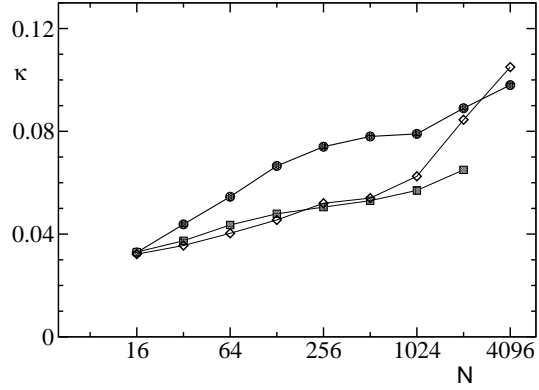}
\caption{Thermal conductivity $\kappa$ versus chain length for the potential
(\ref{e:quasifi4}) and three different average temperatures: 1/16 (circles), 
3/32 (squares), and 1/8 (diamonds). In all cases the temperature difference
is chosen equal to 1 tenth of the average temperature.}
\label{f:ieri}
\end{center}
\end{figure}

\section{Anharmonic chains with external substrate potentials}
In this section we consider the class of models described by the Hamiltonian
(\ref{optical}), with a non-vanishing substrate potential $U$. This means that 
translational invariance breaks down and momentum is no longer a constant of 
the motion.  Accordingly, the dispersion relation is such that $\omega(q)\ne0$
for $q=0$. 

\subsection{Ding-a-ling and related models}

The so-called ding-a-ling model was first introduced by Dawson \cite{D62}
as a toy model for a 1d plasma. It can refer to different contexts: (i) a set
of identical charge-sheets embedded in a fixed neutralizing background; (ii)
a system of harmonic oscillators with the same frequency and equilibrium 
positions sitting on a periodic lattice and undergoing elastic collisions 
that exchange their velocities. Notice that in the low-energy limit, it 
reduces to the 1d Einstein crystal, i.e. set of independent harmonic 
oscillators all having the same frequency (no dispersion). 
 
Independently of \cite{D62}, Casati et. al \cite{CFVV84} introduced a 
modified version, where the harmonic oscillators (say the even-numbered 
particles) alternate with {\it free} particles of the same (unit)
mass. The latter are only constrained to lie between the two adjacent 
oscillators. The Hamiltonian can be symbolically written as
\begin{equation}
{\mathcal H}\;=\; \frac12 \sum_l^N \left[p_l^2 + \omega_l^2 q_l^2\right]
\,+\, {\rm ``hard \; point \; core "} \quad ,
\label{dingling}
\end{equation}
where $\omega_l=\omega$ for even $l$ and zero otherwise. 
A common feature of this class of models is that within collisions the 
motion of the particles can be determined analytically so that the basic
requirement is the computation of the occurrence times of the collision events. 
Therefore, the dynamics naturally reduces to a discrete mapping.

For the isolated system (e.g. a chain with periodic boundary conditions) the
dynamics depends only on the dimensionless parameter
$\varepsilon=e/(\omega\, a)^2$ where $e$ is the energy per particle and $a$ the
lattice spacing. The Authors of Ref.~\cite{CFVV84} studied the dynamical 
behavior of the model by fixing $e=1$ and changing $\omega$. They concluded 
that, for $\omega$ and $N$ large enough, the dynamics is strongly chaotic and 
soliton-like pulses are sufficiently attenuated \cite{C86}. This renders the 
model a good candidate to check the validity of Fourier's law.

{\it Finite thermal conductivity -}  
The validity of the Fourier's law was first established by performing a series
of non-equilibrium simulations, where the freely moving end-particles were put 
in contact with two Maxwellian reservoirs. The average flux $J$ was then
computed by summing the amounts of energy $\delta E$ exchanged with one of the 
reservoirs in all collisions during the simulation time. The average 
temperature gradient was estimated with a linear fit (to get rid of 
boundary effects). By evaluating the thermal conductivity as a function of the 
lattice length up to $N=18$ for $T_+=2.5$, $T_-=1.5$ and $\omega=1$, it was 
concluded that $\kappa(N)$ attains a constant limiting value already for 
$N>10$~. 

After having established the existence of a finite value of the transport 
coefficient, the Authors have compared the value of $\kappa$ with the result 
of linear response theory. To this aim, because of the discontinuities due 
to the collision processes, it was preferred to express the Green-Kubo formula 
in terms of the integral quantity
$$
\Delta Q(t,\tau) = \int_{t}^{t+\tau} J(t_0)dt_0 \qquad .
$$
From Eq.~(\ref{realgk}), recalling that the ensemble average is equivalent to
a time average (and understanding the limit $N \to \infty$), it is 
straightforward to show that (for more details see Ref.~\cite{PR92})
\begin{equation} 
\kappa_{GK}(T) \;=\;  \lim_{\tau\to\infty}{1\over 2NT^2 \tau }
\left\langle (\Delta Q(t,\tau))^2  \right\rangle_t \quad, 
\label{gkdingling}
\end{equation}  
where the subscript $t$ indicates that the average is performed over the time
variable $t$.

Additionally, it was decided to compute the total heat flux not by summing
up the $p_l^3$ local contributions as, for instance, done in Ref.~\cite{GHN01} 
for the hard point gas,\footnote{Recall, indeed, that the substrate potential does
not contribute to the flux.} but, more directly, determining 
$\Delta Q(t,\tau)$ as the amount of energy exchanged in all collisions 
occurred in the interval $[t,t+\tau]$.

Casati {\it et al.} provided a convincing numerical evidence that 
the limit (\ref{gkdingling}) exists for a closed chain of 48 particles 
and $\omega=10$. In this case, the energy transport is diffusive and they 
showed that the $\kappa_{GK}$-value obtained in this way is in good agreement 
with the one obtained from direct simulations. 
\footnote{As for the similar comparisons discussed in the previous chapter
with reference to the coupled rotors, micro-canonical simulations have to
be performed for an energy density that corresponds to the average temperature 
in the non-equilibrium simulations.}

The results of Casati {\it et al.} have been lately reconsidered by Mimnagh 
and Ballentine \cite{MB97} who performed a detailed series of simulations 
with longer chains and in a wider parameter range. Curiously enough, they found 
that the value of $\kappa$ reported in Ref.~\cite{CFVV84} is not the true
asymptotic value (achieved only for $N > 200$) but a minimum of $\kappa(N)$.
\begin{figure}
\begin{center}
\includegraphics*[width=5cm]{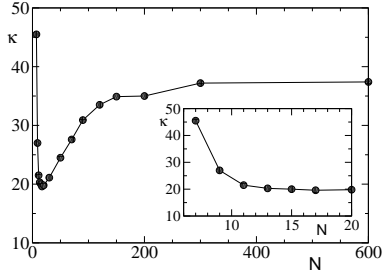}
\caption{Thermal conductivity of the ding-a-ling model. Size dependence of 
$\kappa$ for $\omega=1$ and $\varepsilon=1.5$ (from \cite{MB97}). In the 
inset, it is presented an expanded view in the range of sizes considered
in Ref.~\cite{CFVV84}}
\label{f:ding}
\end{center}
\end{figure} 
This can be seen in Fig.~\ref{f:ding}, where one can also appreciate how
the unfortunate choice of working with $N<20$ may give the false impression of 
a convergence towards a smaller value. This fact does not affect the 
correctness of the conclusion reported in Ref.~\cite{CFVV84} and the importance 
of the results, but signals again how cautious one should be in drawing 
conclusions from the study of relatively short systems!    
Motivated by this observation, Mimnagh and Ballentine carefully studied 
finite-size corrections in a wide range of $\varepsilon$-values, by plotting 
the resistivity $\rho(N) = 1/\kappa(N)$ versus $N$. In all cases, the data 
are well fitted by 
\begin{equation}
\rho(N)\;=\; \rho_\infty \left(1 + {\mu \over \sqrt{N}} \right)
\end{equation}
for $N$ large enough. Accordingly, the minimal chain length required 
to obtain an estimate of $\rho_\infty$ with a fixed relative accuracy 
is proportional to $\mu^2$, a quantity which is found to 
increase dramatically for $\varepsilon>\varepsilon_c=0.04$ (see  
Fig.~\ref{f:ding2}). This is readily understood as $\varepsilon\to\infty$ is
an integrable limit that corresponds to a gas of bouncing free particles.
\begin{figure}
\begin{center}
\includegraphics*[width=5cm]{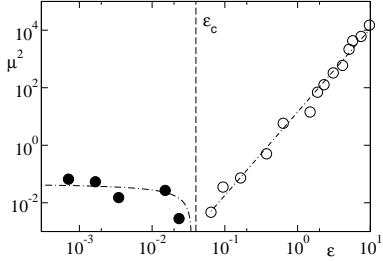}
\caption{The coefficient $\mu^2$ as a function
of $\varepsilon$ for the ding-a-ling model (from \cite{MB97})}
\label{f:ding2}
\end{center}
\end{figure} 
The asymptotic resistivity $\rho_\infty$ is a monotonously decreasing function 
of $\varepsilon$, displaying a crossover from a slower to a faster decay
at $\varepsilon_c$ (see Fig.~\ref{f:ding3}). Since a similar crossover is 
found when looking at both the collision rate and the maximum Lyapunov 
exponent, it is natural to interpret the phenomenon as a sort of transition 
from strong to weak chaos, in close analogy with what found, e.g., in
the FPU model \cite{PCS91}. 
\begin{figure}
\begin{center}
\includegraphics*[width=5cm]{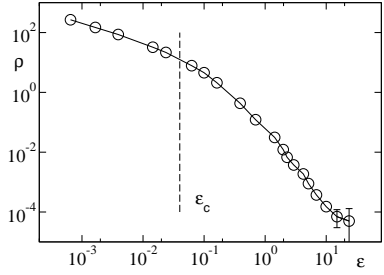}
\caption{The asymptotic resistivity $\rho_\infty$ versus $\varepsilon$ 
(from \cite{MB97})}
\label{f:ding3}
\end{center}
\end{figure} 

Finally, as for the temperature profile, the nonlinear shape reported in 
Fig.~\ref{f:prding} reveals a sizeable temperature dependence of the 
conductivity.
\begin{figure}
\begin{center}
\includegraphics*[width=5cm]{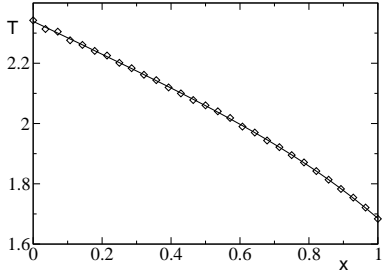}
\caption{Temperature profile for the ding-a-ling model, 
$\omega=2$ (from \cite{MB97}) }
\label{f:prding}
\end{center}
\end{figure} 
 
{\it Modified models - } 
Prosen and Robnik \cite{PR92} considered the
original Dawson system where, as already mentioned, the free particles are
removed (i.e. they set $\omega_l=1$ for all particles in (\ref{dingling})).
They named it ding-dong model to distinguish it from the one presented above. 
Their careful numerical study confirmed the validity of Fourier's law in a wide
temperature range. Besides direct non-equilibrium simulations with
Maxwellian thermostats and the Green-Kubo formula, they also implemented an
efficient transient method that allowed them to explore the high temperature
regime ($T>3$), where, because of the nearly integrable dynamics, a slow 
convergence of the averages with time and/or size is observed. In the 
opposite, low-temperature, limit ($T<0.1$), they are able to prove that the 
conductivity vanishes as $\exp(-1/4T)$ finding a reasonable agreement with  
numerical results.   

Posch and Hoover \cite{PH98} investigated a modified ding-a-ling model where 
the harmonic potential is replaced by a gravitational one  
\begin{equation} 
{\mathcal H}\;=\;
\sum_l^N \left[{p_l^2 \over 2m}+ mg_l |q_l|\right] \,+\, {\rm ``hard \; point
\; core "} \quad . 
\label{dinggrav} 
\end{equation} 
where $g_l$ is a constant acceleration for even $l$ and zero otherwise.
Their simulations further confirm that thermal conductivity is finite for 
this class of models. Moreover, they computed the spectrum of Lyapunov 
exponents in the non-equilibrium steady state, showing that the microscopic 
dynamics takes place on a strange attractor. Finally they observed that the
heat flux is proportional to the difference between the phase-space dimension
and that of the strange attractor.

Finally, it is worth mentioning the exactly solvable model studied by Kipnis
et al. \cite{KMP82} that could be regarded as the "stochastic version" of  the
ding-ling model. As in the latter, it consists of a linear array of
harmonic oscillators but the interaction occurs via a random redistribution
of the energy between  nearest neighbors rather than through deterministic
collisions. More precisely, upon denoting with $\xi_l = p_l^2+q_l^2$ the
energy of the $l$-th oscillator (in suitable units), the dynamics is given by
the updating rules
\begin{equation}
\xi_l' \;=\; P(\xi_l + \xi_{l+1}) , \qquad
\xi_{l+1}' \;=\; (1-P)(\xi_l + \xi_{l+1})
\quad ,
\end{equation}
where $P$ is a random variable uniformly distributed in the interval $[0,1]$. 
The total
energy is thus kept constant except for the two oscillators at the chain
extrema, that are in contact with reservoirs at different temperatures $T_\pm$
according to a Glauber dynamics. The Authors were able to rigorously show that
a unique stationary non-equilibrium measure exists and to compute both
the temperature profile and the heat flux in the steady state:
\begin{eqnarray}
&&T(x) \;=\; T_-\left({1-x \over 2}\right) + T_+\left({1+x \over 2}\right), 
\qquad -1\le x \le 1 \\
&&j \;=\; -\frac{k_B}{4} \left( T_+ - T_- \right) \quad .
\end{eqnarray}
The last results imply that Fourier's law holds and that the thermal
conductivity is equal to $k_B/2$ in the chosen units.

\subsection{Klein-Gordon chains}

An important subclass of models (\ref{optical}) is the one in which 
the inter-particle potential is harmonic                  
\begin{equation} 
{\mathcal H} = \sum_{l=1}^N \left[{p_l\over 2m}^2+ U(q_l) + 
\frac12 C (q_{l+1}-q_{l})^2\right] \quad ; 
\label{kleing}
\end{equation}  
it is often referred to as the Klein-Gordon lattice. The latter has
recently received a great attention as a prototypical system where
strong discreteness effects may come into play in the limit of small $C$.

The first and most complete study of the transport problem in this class of 
models has been carried on by Gillan and Holloway \cite{GH85} for the 
Frenkel-Kontorova potential
\begin{equation}
U(x) \;=\; -U_0\, \cos\left({2\pi x\over a}\right) .
\label{fk}
\end{equation}
The model can be interpreted as a chain of either coupled particles in an
external periodic field or torsion pendula subject to gravity. In the 
latter case $a=2\pi$ and $q_l$ represents the angle with respect to the 
vertical direction: it can be read as the discretized (and 
non-integrable) version of the well-known sine-Gordon field equation.

Besides energy, the dynamics admits a further conserved quantity, the  winding
number ${\mathcal P}$, which is an integer defined by the boundary condition 
$q_{l+N}=q_l+a{\mathcal P}$. In the particle interpretation ${\mathcal P}$ represents the number of
potential wells, while for the pendula it can be viewed as the degree of
built-in twist in the system.\footnote{In physical terms, one has to apply 
equal and opposite forces (in the particle version) or torques (pendulum 
version) to the two end particles, in order to maintain the required value of 
${\mathcal P}$ which can be also interpreted as the net number of kink excitations (i.e.
the number of kinks minus the number of anti-kinks).} 
According to the general theory of irreversible processes \cite{DGM}, 
transport involves thus the flow of both particles and energy caused by 
gradients of the number density and temperature. Accordingly, a $2\times 2$ 
matrix of transport coefficients is required, but because of the Onsager 
relations, there exist only 3 independent transport coefficients, that
are chosen to be the thermal conductivity, the diffusion coefficient and the
heat of transport. Gillan and Holloway computed the thermal conductivity
numerically in the general case of non-vanishing winding number with three
different methods: (i) attaching two heat baths; (ii) through the Green-Kubo
formula; (iii) by adding an external field (see Chap. 3). All the methods 
give consistent results and clearly indicate that the thermal conductivity 
is finite. 

Their results were later confirmed by a numerical study by Hu {\it et al.} 
\cite{HLZ98} who investigated the dependence of the transport coefficient
on the lattice length (for $P=0$). In Ref.~\cite{HLZ98} it was also shown 
that the same holds for a more general version of the Frenkel-Kontorova 
model with an anharmonic inter-site potential.  

In order to illustrate the type of behavior that is observed in this class
of systems we show in Fig.~\ref{f:condphi4} some data for the $\phi^4$ chain 
\begin{equation}
U(x) \;=\;  \frac a2 x^2 + \frac b4 x^4 \quad.
\label{phi4}
\end{equation}
In panel (a), we present a case of fast convergence to a small $\kappa$ 
value for a single-well potential; panel (b) refers instead to a
low-temperature regime characterized by large thermal conductivity.

\begin{figure}
\begin{center}
\includegraphics*[width=7cm]{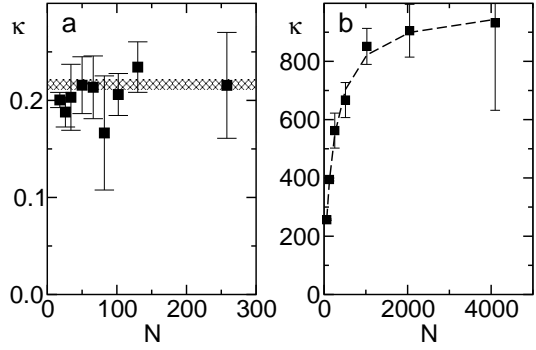}
\caption{Thermal conductivity versus chain length in $\phi^4$ chains with 
Nos\'e-Hoover thermostats ($\Theta = 10$). Panel (a) refers to the single
well case ($a=b=1$ in Eq.~\ref{phi4}): the results have been obtained for 
$C=1.$, $T_+ = 8$, and $T_- = 6$. The shaded region represents the value 
obtained from the Green-Kubo formula with its statistical uncertainty. 
Panel (b) refers to the double-well case ($a=-1$, $b=1$) for an average 
temperature $T= 0.37$ and a temperature difference 0.002. The dashed line
is just a guide for the eyes.}  
\label{f:condphi4}
\end{center}
\end{figure}
Evidence of a finite conductivity for the case $a=0$ has been reported in 
Refs.~\cite{HLZ99,AK00}. The conductivity is found to decrease with temperature
according to the law \cite{AK00} 
\begin{equation}
\kappa(T)\propto  T^{-1.35(2)}
\label{eq:AK}
\end{equation}
that is reminiscent of what often experimentally observed in insulating 
crystals. 
Upon increasing $\Delta T$, there exists a ``transition'' to a nonlinear regime 
characterized by a non uniform local temperature gradient. By including the 
empirical law (\ref{eq:AK}) into Fourier's law, it is possible to check the 
consistency with the measured $T(x)$: a good agreement with the simulations
is found, provided boundary jumps are taken into account. 

Finally, Tsironis {\it et al.} \cite{TBSZ99} obtained further numerical 
evidence of the existence of a finite thermal conductivity for systems like 
(\ref{kleing}). Beside reconsidering the Frenkel-Kontorova potential 
(\ref{fk}), two further examples were analyzed, the sinh-Gordon and 
bounded single-well potentials,
\begin{equation}
U(x) \;=\;  \cosh x \, -1 \qquad , \qquad 
U(x) \;=\;\frac 12 \left(1-{\rm sech}^2x\right)
\label{cosh}
\end{equation}
as representatives of the classes of hard and soft anharmonicity,
respectively.

\section{Integrability and ballistic transport} 
 
When the equilibrium dynamics of a lattice can be decomposed into that of 
independent ``modes'', the system is expected to behave as an ideal conductor.
The simplest such example is obviously the harmonic crystal, that has infinite 
conductivity and cannot, therefore, support any temperature gradient. However, 
this applies also to the broader context of integrable nonlinear systems. They 
are mostly one-dimensional models characterized by the presence of 
``mathematical solitons", whose stability is determined by the interplay of 
dispersion and nonlinearity. This interplay is expressed by the existence of a 
macroscopic number of {\it conservation laws} constraining the dynamical 
evolution. Thereby, the existence of stable nonlinear excitations in
integrable systems is expected to lead to ballistic rather than to diffusive 
transport.
As pointed  out by Toda \cite{T79}, solitons travel freely, no temperature 
gradient can be maintained and the conductivity is thus infinite. From the point
of view of the  Green-Kubo formula, this ideal conducting behavior is
reflected by the existence of a nonzero flux autocorrelation at arbitrarily
large times. According to the discussion reported in Section 6.2, this, in turn,
implies that the finite-size conductivity diverges linearly with the size.

Although integrable models are, in principle, considered to be exactly solvable,
the actual computation of dynamic correlations is technically involved. A 
more straightforward approach is nevertheless available to evaluate the 
asymptotic value of the current autocorrelation. This is accomplished by means 
of an inequality due to Mazur \cite{M69} that, for a generic observable $A$, 
reads as 
\begin{equation}
\lim_{\tau \rightarrow \infty} \frac{1}{\tau} \int_0^ \tau 
\langle A(t)A(0)\rangle  \, dt \; \geq \; \sum_n
\frac{\langle A Q_n\rangle^2}{\langle Q_n^2\rangle} \quad,
\label{mazur}
\end{equation}
where $\langle \ldots \rangle$ denotes the (equilibrium) thermodynamic average, 
the sum is performed over a set of conserved and mutually orthogonal 
quantities ${Q_n}$ ($\langle Q_n Q_m\rangle=\langle Q_n^2\rangle\delta_{n,m}$).
Furthermore, it is assumed that $\langle A\rangle=0$.

Zotos \cite{Z01} applied the above result to the equal-masses Toda chain 
with periodic boundary conditions, defined, in reduced units, by the
Hamiltonian
\begin{equation}
{\mathcal H} \;=\; \sum_{l=1}^N \left[ \frac{p_l^2}{2}+\exp(-r_l) \right]
\label{toda}
\end{equation}
where $r_l=x_{l+1}-x_l$ is the relative position of neighboring particles.
As is known \cite{Toda}, the model is completely integrable as admits $N$ 
independent constants of th motion, the first among which are 
\begin{eqnarray}
\label{law1}
Q_1&=&\sum_{l=1}^N p_l\\
Q_2&=&\sum_{l=1}^N \frac{p_l^2}{2}+e^{-r_l}\\
Q_3&=&\sum_{l=1}^N \frac{p_l^3}{3}+(p_l+p_{l+1})e^{-r_l}\\
Q_4&=&\sum_{l=1}^N \frac{p_l^4}{4}+(p_l^2+p_lp_{l+1}+p_{l+1}^2)e^{-r_l}
+\frac{1}{2}e^{-2r_l}+e^{-r_l}e^{-r_{l+1}}\\
Q_5&=&\sum_{l=1}^N \frac{p_l^5}{5}+(p_l^3+p_l^2p_{l+1}+p_lp_{l+1}^2+p_{l+1}^3)
e^{-r_l}\\
&+&(p_l+p_{l+1})e^{-2r_l}+(p_l+2p_{l+1}+p_{l+2})e^{-r_l}e^{-r_{l+1}}...
\label{laws}
\end{eqnarray}
The ``trivial" conserved quantities $Q_1$ (the total momentum) and 
$Q_2$ (the total energy) are of course present in all translationally invariant
systems of the form (\ref{acoustic}), irrespective of their integrability.

Let us consider the fixed temperature-pressure thermodynamic ensemble,
\begin{equation}
\langle A(t)A(0)\rangle=Z^{-1}\int \prod_{l=1}^L dp_l dr_l A(t)A(0) 
e^{-\beta({\mathcal H}+PL)}
\label{ave}
\end{equation}
where $Z=\int \prod_{l=1}^N dp_l dr_l e^{-\beta({\mathcal H}+PL)}$, 
$L=\sum_{l=1}^N r_l$ is the  ``volume" of the chain and $P$ is the pressure. 
In this ensemble, equal-time correlation functions can be calculated 
analytically. For instance, the average inter-particle distance is 
\begin{equation}
\langle r\rangle \; =\; \ln \beta -\Psi(\beta P)
\label{q}
\end{equation}
where $\Psi(z)$ is the digamma function.

The total heat flux is given by (see Eqs. (\ref{hf2}) and (\ref{jtotal})~)
\begin{equation}
J \; = \; \sum_{l} \left[p_l h_l+\frac{(p_{l+1}+p_l)}{2} \, r_l e^{-r_l} 
\right]
\label{je}
\end{equation}
where $h_l=\frac{p_l^2}{2}+\frac{1}{2}(e^{-r_l}+e^{-r_{l-1}})$.
In order to apply the Green-Kubo formula in the chosen ensemble, one has to 
consider a ``shifted" flux (see the discussion at the end of Sec. 5.2)
\begin{equation}
\tilde J \; = \; J-\frac{\langle Q_1 J\rangle}{\langle Q_1^2\rangle}Q_1 .
\label{jet}
\end{equation}
This is equivalent to removing the contribution of $Q_1$ in the right hand 
side of the Mazur inequality (\ref{mazur}) for $A=J$.

Lower bounds on the long time value of $\langle J(t)J(0)\rangle$ can thus be
calculated by the inequality (\ref{mazur}), using the first $m$ conservation 
laws \cite{Z01}. As $Q_3$ has a structure very similar to the energy current, 
a large contribution from this term has to be expected. Moreover, $Q_n$ with 
even $n$ are uncoupled with $\tilde J$, so that it suffices to consider odd 
values of $n$. 

In order to utilize the inequality (\ref{mazur}), it is not necessary to
orthogonalize the conserved quantities (\ref{law1}-\ref{laws}). One can,
indeed, replace the sum of the first $m$ terms in the r.h.s. of
Eq.~(\ref{mazur}) with 
\begin{equation}
C^m \;=\; \langle\tilde J|Q\rangle\langle Q|Q\rangle^{-1}
\langle Q|\tilde J\rangle
\label{ortho}
\end{equation}
where $\langle Q|Q\rangle$ is the $m\times m$ overlap matrix of $Q_{n'n}$ and 
$\langle Q|\tilde J\rangle$ is the overlap vector of $\tilde J$ with the 
$Q_n's$. The ratio $C^m /\langle\tilde{J}^2\rangle$, representing a lower
bound to the conductivity $\kappa$, is found to increase monotonously with the 
temperature. At low $T$, the growth is linear with a slope comparable to the 
density of solitons $N_s/N=\ln 2/\pi^2 T$. This trend is interpreted as an 
evidence for the increasing contribution of thermally excited nonlinear 
modes to ballistic transport.

\section{Two-dimensional lattices}
It is well known that many properties of statistical systems depend
on the dimension $d$ of the space, where they are embedded. In Chapter 5, we
have seen that transport properties are not expected to violate this 
rule. In this Chapter we discuss some results of molecular-dynamics
simulations in two dimensions. In fact, as soon as the dimension of the 
physical-space is set to a value larger than 1, the direct investigation 
of sufficiently large systems becomes problematic. 

First, we briefly discuss the numerical studies appeared so far in the 
literature, presenting them in a historical perspective. Then, we more
extensively discuss some recent numerical experiments that have allowed
verifying the predictions of mode--coupling theory also in two dimensions.

\subsection{Early results}
To our knowledge, the first attempts of investigating the heat conduction 
problem in $2d$ lattices with (at that time) heavy numerical simulations 
are two papers by Payton, Rich and Visscher \cite{PRV67,RVP71}, that appeared 
more than three decades ago.  These Authors investigated the combined effect
of nonlinearity and disorder on heat conduction in 2d harmonic and
anharmonic lattices, with Lennard-Jones pair potentials.
Their studies aimed also at analyzing the dependence of heat conductivity
$\kappa$
on the temperature and on the concentration of impurities, as a measure of 
disorder. They found evidence of an increase of $\kappa$ in disordered 
nonlinear systems compared to the harmonic case (see also Chap. 6).

On the other hand, the dependence of $\kappa$ on the system size 
was ignored, probably because the Authors did not consider this a problem of 
major concern. In fact, according to the classical view of 
Peierls \cite{peierls}, phonon-phonon scattering processes were assumed to be
sufficiently efficient to ensure normal transport properties in the presence 
of strong nonlinearity and disorder.

Later, Mountain and MacDonald \cite{MM83} performed a more careful study on 
the dependence of $\kappa$ on the temperature $T$. 
They considered a $2d$ triangular lattice of unit-mass atoms, interacting via 
a Lennard-Jones 6/12 potential. At variance with the previous investigations,
no disorder was included, and their numerical results were consistent with 
the expected classical law $\kappa \sim T^{-1}$. Again, the dependence 
of $\kappa$ on the system size was not investigated.

The first contribution in this direction is the paper by Jackson and 
Mistriotis \cite{JM89}. These Authors compared measurements
of $\kappa$ in the $1d$ and $2d$ FPU lattices: they concluded that in both 
cases there was no evidence that the transport coefficient is finite in the 
thermodynamic limit. It is worth mentioning an interesting remark by these 
Authors: ``the dependence of $\kappa$ on the system size cannot be adequately 
described in the high temperature, i.e., classical, limit by Peierls' model
of the diluted phonon gas, because the perturbative Umklapp processes 
cannot account for the genuine nonlinear effects that characterize 
such a dependence''. 

Conversely, more recent molecular--dynamics simulations of the 2d 
Toda-lattice \cite{NKS92} have been interpreted in favor of the finiteness 
of $\kappa$ in the thermodynamic limit. Further confirmation in this 
direction can be found in appendix B of the interesting paper by 
Michalski \cite{M92}, where heat conductivity in models of amorphous 
solids was thoroughly investigated.

The interplay of disorder and anharmonicity, that inspired the first 
contributions by Payton, Rich and Visscher has been reconsidered by 
Poetzsch and B\"ottger \cite{PB94,PB98} who investigated percolating and
compositionally disordered $2d$ systems. In particular, they tried to pinpoint 
the role of third- and fourth-order anharmonicity, concluding that, at equal 
temperature, the latter yields a larger value of $\kappa$ than the former one.
Moreover, the Authors reported also about the dependence of $\kappa$ on the 
system size. They assumed the same point of view of Michalski \cite{M92}, but 
a careful inspection of Fig.~5 in \cite{PB94} shows that their data are also 
compatible with a systematic increase of $\kappa$ with the system size.

In a more recent and accurate investigation, Dellago and Posch \cite{DP97} 
studied, by molecular--dynamics techniques, heat conduction as well
as Lyapunov instability in a generalized version of the $XY$--model. Besides 
various interesting
results, the manuscript contains a very clear indication of the finiteness 
of $\kappa$ in the thermodynamic limit. In the light of what discussed in 
Section 6.4, this result is not surprising, since $\kappa$ is finite 
already in $1d$ for models of this type.
A further interesting remark contained in Ref.~\cite{DP97} concerns the 
behavior of $\kappa$ below the transition temperature of the $XY$-model when 
the diffusive behavior of energy transport is lost and anomalous behavior 
seems to set in.

\subsection{Divergence of heat conductivity}
Heat conduction in $2d$ models of oscillators coupled through anharmonic, 
momentum--conserving interactions is expected to exhibit different properties
from those of $1d$ systems. In fact, MCT predicts a logarithmic divergence of
$\kappa$ with the system size $N$ at variance with the power--law predicted
for the $1d$ case (see Section 5.3).

Following \cite{LL00}, we discuss the results of molecular--dynamics
simulations of the FPU--$\beta$ potential (see Eq.~(\ref{fpu}), $g_3=0$) and 
the LJ-(6/12) potential (see Eq.~(\ref{lenjo})). For the
sake of simplicity we introduce the shorthand notations $V_1(z)$ and
$V_2(z)$ to denote, respectively, the two models. The goal of this twofold 
choice is to verify that the prediction of MCT is truly independent of the 
potential, provided it belongs to the class of anharmonic momentum--conserving 
interactions. 

While $V_1$ does not contain natural scales for both distances and energies,
the natural length scale of $V_2(z)$ is the equilibrium distance $a$, while
its energy scale is the well depth $\epsilon$. Therefore, after having
arbitrarily fixed $g_2 = 1$ and $g_4 = 0.1$ in $V_1(z)$, $a$ and $\epsilon$
have been determined ($a= 25$, $\epsilon 8.6$) by imposing that the 
coefficients of the second and fourth order terms of the Taylor expansion of 
$V_2$ around its minimum coincide with $g_2$ and $g_4$, respectively. Notice, 
however, that, at variance with $V_1(z)$, the Taylor expansion of $V_2(z)$ 
incorporates also a non-vanishing cubic term. 
Despite both models are characterized by the same shortest harmonic time scale, 
$\tau_{min} = \pi/\sqrt{2}$, the L-J model has been integrated with 
a slightly shorter time step ($\Delta t = 5 \cdot 10^{-3}$, compared to 
$10^{-2}$ for the FPU model) in order to ensure a sufficient accuracy when 
the dynamics experiences the strong nonlinearities of $V_2(z)$ (nearby the 
divergence of the Lennard-Jones potential).

In both cases, simulations have been performed with reference to a square 
lattice containing $N_x \times N_y$ atoms of equal masses $m$. The equilibrium 
position of each atom has been chosen to coincide with a lattice site, 
labelled by a pair of integer indices $(i,j)$. 
The origin of the Cartesian reference frame is fixed in such a way that 
$1< i < N_x$ and $1< j < N_y$. Accordingly, the components of the 2d-vector 
of the equilibrium position, ${\vr}^0_{ij}$, are given by the interger pairs 
$(i,j)$. Moreover, in analogy to the $1d$ case, each atom has been assumed to 
interact with its nearest-neighbors (herein identified with the von Neumann 
neighborhood).

The model is thus represented by the Hamiltonian
\begin{equation}
{\mathcal H} = \sum_{i=1}^{N_x} \sum_{j=1}^{N_y} 
\left[ \frac{|{\vp}_{ij}|^2}
{2m} + V(|\vq_{i+1j} - \vq_{ij}|) + V(|\vq_{ij+1} - \vq_{ij}|)
\right]
\label{Ham2}
\end{equation}
where $\vq_{ij}(t) = \vr_{ij}(t) - \vr^0_{ij}$,
$\vr_{ij}(t)$ is the instantaneous position vector of the $(i,j)$-atom,
and $\vp_{ij}(t)$ is the corresponding momentum vector. 

For what concerns the definition of local temperature, the following
definitions are all equivalent (see also Sec. 2.2), 
\begin{equation}
k_B T_{ij} = {\langle (p^{(x)}_{ij})^2 \rangle \over m}= 
{\langle (p^{(y)}_{ij})^2 \rangle \over m} = 
{ \langle {(p^{(x)}_{ij})^2 + (p^{(y)}_{i,j})^2}\rangle \over 2m} ,
\label{temp}
\end{equation}
where $p^{(x)}_{ij}$ and $p^{(y)}_{ij}$ are the $x$ and $y$ components
of the momentum vector $\vp_{ij}$, respectively.

Moreover, tedious but straightforward calculations, akin to those 
presented in Section 2.3 , allow to express 
the $x$ and $y$ components of the local heat flux $\vj_{ij}$ as
\begin{eqnarray}
j^{(x)}_{ij} &=  \frac {a}{4m} \left[f^{xx}_{ij}\left(p^{(x)}_{ij} + 
p^{(x)}_{i+1j}\right) + f^{yx}_{ij}\left(p^{(y)}_{ij} + p^{(y)}_{i+1j}
\right)\right] \cr
j^{(y)}_{ij} &=  \frac {a}{4m} \left[f^{xy}_{ij}\left(p^{(x)}_{ij} + 
p^{(x)}_{ij+1}\right) + f^{yy}_{ij}\left(p^{(y)}_{ij} + p^{(y)}_{ij+1}
\right)\right]
\end{eqnarray}
where the components of the local forces are defined as 
$$ 
f^{xx}_{ij} = - \frac{\partial V \left(\left| \vq_{i+1j} - 
\vq_{ij}\right|\right)}
{\partial q^{(x)}_{ij}} \quad f^{yx}_{ij} = - 
\frac{\partial V \left(\left| \vq_{i+1j} 
- \vq_{ij}\right|\right)}{\partial q^{(y)}_{ij}}
$$
$$ 
f^{xy}_{ij} = - \frac{\partial V \left(\left| \vq_{ij+1} - 
\vq_{ij}\right|\right)} {\partial q^{(x)}_{ij}} \quad f^{yy}_{ij} = - 
\frac{\partial V \left(\left| \vq_{ij+1} 
- \vq_{ij}\right|\right)}{\partial q^{(y)}_{ij}} .
$$
Finally, in analogy with Eq. (\ref{jtotal}), the total heat flux vector is 
defined as
\begin{equation}
{\vJ}  \;=\; \sum_{i,j} {\vj}_{ij} .
\label{avflux}
\end{equation}

The non-equilibrium simulations have been performed by coupling all atoms on
the left (right) edge of the $2d$ lattice with the same thermal bath operating
at temperature $T_+$ ($T_-$). In the numerical studies reported hereafter, 
thermal baths have been simulated by applying the Nos\'e-Hoover 
method. Nonetheless, the Authors of Ref.~\cite{LL00} verified that the same 
results are obtained upon using stochastic thermal baths, as well.
Periodic and fixed boundary conditions have been adopted in the direction
perpendicular ($y$) and parallel ($x$) to the thermal gradient, respectively.
Finally, let us notice that, for the investigation of the thermodynamic limit, 
the simulations for different lattice sizes should be performed by keeping the 
ratio $R = N_y/N_x$ constant. From the numerical point of view, it is 
convenient to choose small $R$ values, since for a given longitudinal length 
$aN_x$, the simulations are less time consuming. However, too small ratios 
would require considering larger system sizes to clearly observe $2d$ features.
In \cite{LL00} it was checked that $R=1/2$ is a good compromise for both 
$V_1$ and $V_2$ choices.

With the above physical setup, heat equation (\ref{fourier}) implies that
a constant thermal gradient should establish through the lattice in the
$x$-direction with $\langle J^{(x)} \rangle > 0 $ and $\langle
J^{(y)} \rangle = 0$. The time span needed for a good convergence of the 
time-average $\langle \cdot \rangle$ increases with $N_x$: for instance, 
${\mathcal O}(10^5)$ units proved sufficient for $N_x = 16$, while 
${\mathcal O}(10^7)$ units are needed when $N_x = 128$ (for not too small 
energy densities). 

The detailed analysis of temperature profiles performed in Ref.~\cite{LL00} 
has revealed deviations from the linear shape predicted by Fourier law
(this is particularly true in the case of the Lennard-Jones potential $V_2$),
but one cannot exclude that this is to be attributed to the relatively
large temperature differences adopted in order to have nonnegligible 
heat fluxes. Anyway, despite such deviations, simulations provide
convincing evidence that the temperature gradient scales like $N_x^{-1}$.
Accordingly, the dependence of $\kappa$ with the system size $aN_x$ can be
determined by plotting $\kappa \propto \langle j^{(x)} \rangle N_x$ versus $N_x$. 
The data reported in Fig.~\ref{kappafpu} support the MCT prediction of a 
logarithmic growth both for the FPU and Lennard-Jones potentials for two 
rather different choices of heat bath temperatures.

\begin{figure}[tcb]
\begin{center}
\includegraphics*[width=8cm]{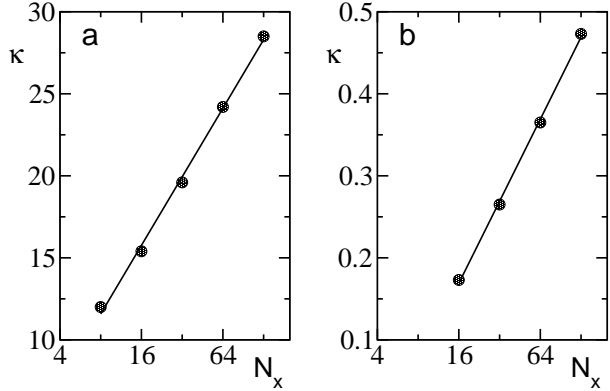}
\caption{Heat conductivity $\kappa$ versus the system size $N_x$ for 
the 2d FPU $\beta$ (a) and Lennard-Jones (b) models. In panel
(a) $T_+ = 20$ and $T_- = 10$; in panel (b) $T_+ = 1$ and $T_- = 0.5$.
In both cases, statistical errors have the size of the symbols.}
\label{kappafpu}
\end{center}
\end{figure}
Since also in the $2d$ case the temperature gradient vanishes in the 
thermodynamic limit, one is allowed to conjecture that linear response theory
should reproduce the behavior of sufficiently large systems.
 
According to the general discussion carried on in Sec. 5.2,
\begin{equation}
\kappa  \;=\; \kappa_{xx}\; = \; \frac{1}{k_B T^2}\, 
\lim_{t\to\infty}\int_0^t \frac{\langle J^{(x)}(\tau) J^{(x)}(0) \rangle}
        {V} d\tau ,
\label{GKx}
\end{equation}
where $V=Ra^2 N_x^2$ is the system volume and $J^{(x)}$ is the $x$-component of 
the total heat flux vector (\ref{avflux}). Notice that simulations have to be
performed for a sufficiently large size $N_x$, since the thermodynamic limit 
has to be taken before the infinite time limit in the above formula.

Numerical simulations at constant energy, with $R = 1/2$ and periodic boundary 
conditions in both directions \cite{LL00} convincingly suggest a logarithmic 
divergence (in time) of the correlation integral appearing in (\ref{GKx}).
Once more, this scaling behavior is consistent with the outcome of
direct non-equilibrium simulations. Indeed, by assuming that Eq.~(\ref{GKx})
reproduces the correct size dependence if the integral is cut-off at a time
$t =a N_x/v_s$ \cite{LLP98b}, the temporal logarithmic divergence translates 
into an analogous divergence with the system size (see also Section 6.2).

Let us finally mention that recent results by Shimada et al. \cite{SMYSI00}
confirm the overall scenario and, furthermore, provide the direct confirmation
that $\kappa$ is finite in $3d$ for this class of models.

\section{Conclusions}
While this review, hopefully, provides a rather complete account of the 
existing dynamical approches to heat conduction in low dimensional lattices, 
it certainly does not solve all open questions. Some of the most interesting 
issues requiring further investigations are briefly summarized in this 
concluding chapter.

The first problem concerns heat transport at low
temperatures. It is well known that many of the Hamiltonian models used for
describing anharmonic crystals with nearest--neighbor interactions exhibit very
slow relaxation to equilibrium below a typical energy density $e_c$ (or 
temperature) that depends on the model and on the space dimension.  For
instance, the 1d FPU $\beta$--model shows a crossover between fast and slow
relaxation at a value of the energy density $ e \approx 1$, with all the 
parameters of the model set to unity \cite{PCS91}. The same holds for the 1d
Lennard--Jones potential at a close value of the energy density for $\epsilon$
and $a$ (see Eq.~(\ref{lenjo})~) chosen in such a way that the coefficients of
the second and fourth order terms of the  Taylor series expansion around the
equilibrium position coincide  with those of the FPU $\beta$--model. When
passing to 2d, for both models, the value of $e_c$ decreases: for instance, in
the Lennard--Jones 6/12 potential  $e_c \approx 0.3$ \cite{BT83}.  This
peculiar behavior in the low--temperature regime can be attributed to long
living meta-stable states that slow--down dramatically the relaxation
process\footnote{It would be interesting to investigate the possibility of
experimental tests of such a phenomenon in real solids, where  transient
effects can be usually resolved by fast spectroscopic techniques. Some authors
have also suggested strong analogies with glassy dynamics \cite{CG99}, a
subject that has recently become of primary interest for theoretical end
experimental investigations in out--of--equilibrium physics.}. A similar
scenario can be observed in the 1d rotor model described in Section 6.4. When
approaching the two integrable limits of this model  (the harmonic and the
``free rotors'' for small and large temperatures  respectively)  again slow
relaxation mechanisms set in.  An even more interesting situation concerns the
2d version of the rotator model, akin to the $XY$--model (see e.g. \cite{LR01}
and references therein). In fact, it is
characterized by the presence of the  so called Kosterlitz--Thouless phase
transition at finite temperature between a disordered high temperature phase
and a  low temperature one, where vortices condensate. From a dynamical point
of view, there are analogies with the above mentioned examples, namely also 
this low temperature phase exhibits slow relaxation dynamics  to equilibrium
\cite{DP97}. The possible relation with  topological changes of the phase space
due to the  presence of a phase transition should be investigated.

Upon all what we have discussed in this review, in particular in Section
5, one might expect that the slowing-down of relaxation processes
at low temperatures should be even more relevant for transport properties.
In fact, recent numerical investigations \cite{LL00} have shown 
that in 2d FPU--like models the heat--flux correlation function seemingly
exhibits the resurgence of a power--law divergence of
the heat conductivity in the thermodynamic limit\footnote{It is worth 
recalling that for this class of 2d models, the heat conductivity shows a 
logarithmic divergence with the system size, see chapter 9.}. Anyway,
it is quite difficult to conclude only on the basis of numerical
investigations if this has to be attributed to finite size and finite
time effects (see the discussion reported in Ref.~\cite{LL00}). 
The problem could be better tackled in the 1d FPU model, where 
finite--size effects are expected to be even more relevant
below the crossover temperature. This question is closely related to
other finite--size effects observed in these models \cite{L00}.

Another issue that remains partially unexplored concerns the
possible role of nonlinear excitations in transport.
The idea that solitons may play a role in heat conduction dates back to Toda 
\cite{T79}. For instance, it has been invoked to explain the anomalous
behavior of the FPU model as a consequence of ballistic transport due to
solitons of the modified Korteweg--deVries equation. Actually, such an
equation can be obtained as a continuum limit of the FPU lattice model.
Numerical experiments indicate that such solutions may persist as long
living states of the FPU dynamics. On the other hand, upon what reported
in the previous chapters, the leading contribution to the divergence of 
heat conductivity is given by the slow--relaxation properties
of long wavelength modes. We cannot however exclude that also nonlinear 
excitations like solitons or kinks, according to the model at hand,
may contribute to the divergence of $\kappa$.
Recently, it has been proposed that transport
properties should be affected also by the presence of periodic, spatially
localized lattice waves denoted as {\sl breathers} \cite{TBSZ99}. 
Anyway, the effect of any kind of nonlinear excitation is quite difficult
to be detected.  Even if we could assume that the the energy flux 
is the sum of a phononic and a solitonic contribution, $J=J_{ph}+J_{sol}$
how can we hope to distinguish the latter if the phononic 
part already yields anomalous behavior? In general, the chaotic features 
of the dynamics prevent the possibility of disentangling the contribution
of nonlinear waves from that of extended modes. 
In this respect some better insight on the role of nonlinear
excitations might be obtained from the analysis of integrable 
systems, as in the problem of ballistic spin transport 
(see \cite{Z01} and references therein).
Altogether, nonlinear excitations are one of the possible ingredients 
affecting the divergence of heat conductivity. Although it seems rather natural
to conejcture that all models characterized by momentum conservation fall in 
the same universality class (with the only exception of bounded potentials), 
the numerical discrepancies among the various systems are at least suggestive 
of relatively strong finite-size corrections. If, on the other hand, we 
remind that there is no way to control the approximations implicitely contained 
in the self-consistent mode-coupling theory, we realize that even the problem
of determining the asymptotic growth rate of heat conductivity in homogeneous 
systems may have not yet come to an end. 
 
A further remark concerns the combination of disorder and nonlinearity.  This
has been considered sporadically in the past \cite{PRV67,RVP71,PB94,PB98} in
relation with the problem of heat transport (see also section 9.1). We
should say that poor progress has been made in this direction during  the last
decades. In fact, few results are available and in general no  clear conclusion
can be drawn about the dependence of the thermal conductivity  on the system
size. Even the most recent results contained in a contribution by Hu et al. 
\cite{LZH01} are not convincing in this respect. In Fig.~\ref{f:condis} we
compare the finite--length heat conductivity $\kappa(N)$ for free and fixed
boundary conditions in the same model and for the same parameter values 
considered in \cite{LZH01}. Upon these results, one can only conclude that 
the dynamical regime explored in that paper is practically indistinguishable 
from the disordered harmonic case (see Section 4.2). Despite the already heavy 
numerical efforts needed to produce the data in Fig.~\ref{f:condis}, much 
longer time scales and system sizes have to be explored in order to fully
appreciate the role of nonlinear terms. 

\begin{figure}[tcb]
\begin{center}
\includegraphics*[width=7cm]{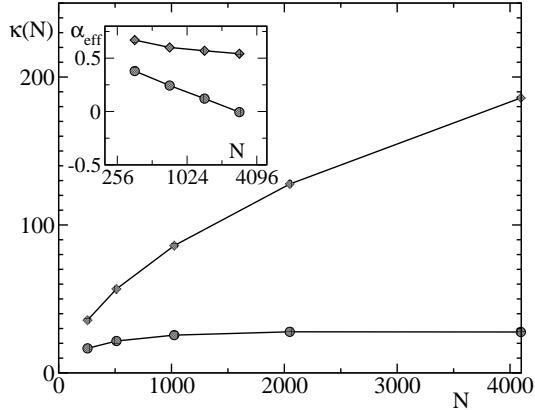}
\caption{Finite-length conductivity in disordered FPU chains as from
from Ref. \cite{LZH01}, $T_+= 1 \cdot 10^{-3}$,  $T_-= 5 \cdot 10^{-4}$. 
Diamonds refer to free boundary
conditions while full dots refer to fixed b.c.. In the inset we report
the effective exponent defined in (\ref{logder}).}
\label{f:condis}
\end{center}
\end{figure}                                                                    

We conclude this section by addressing the reader to a final interesting open
question about the study of heat conduction in structurally disordered
lattices, as well as in models of amorphous materials \cite{M92} or
quasicrystals \cite{M00}. When the crystal structure of a lattice is destroyed,
the phononic contribution to anomalous heat transport is expected to
play a much less relevant role. Nonetheless, peculiar transport properties are
known to arise in real amorphous materials \cite{AF93}. The most studied
examples are real glasses, where viscosity exhibits a dramatic increase below a
transition temperature specific of the material at hand. It is still unclear if
phenomena like this should be ascribed to mechanisms other than phononic
contributions. The effectiveness of mode--coupling theory in describing
thermodynamic-limit divergences in models of solids, as well as glassy 
dynamics in models of real glasses, indicate the extremely fascinating 
perspective of a possible unified theory of anomalous transport in condensed 
matter systems. In this respect, we should also remark that much remains to 
be done in both cases in order to clarify the reliability of the analytical 
estimates based on the mode--coupling approach. In particular, the many 
approximations adopted in the derivation of the scaling laws reported in 
section 5.3 are justified by quite rough arguments. A closer inspection 
of their validity by performing analytical as well as more accurate numerical 
calculations would be highly desirable. Furthermore, the problem of 
thermodynamic-limit 
divergences is not the only open question in this context. The other crucial 
facet of the problem is the temperature dependence of heat conductivity in 
such materials, where various dynamical solutions associated with the 
non--homogeneous structure of
the systems should be considered as responsible of deviations from the expected
classical laws.

\addcontentsline{toc}{section}{Acknowledgments} 
\section*{Acknowledgments} 
We thank L. Ballentine, A. Dhar, D. Mimnagh, M. Vassalli for providing us 
numerical data.
We acknowledge useful discussions with the members of the research group {\it
Dynamics of Complex Systems} in Florence as well as a partial support by
the INFM project {\it Equilibrium and nonequilibrium dynamics in condensed
matter}. This work is also part of the EC network LOCNET, Contract No.
HPRN-CT-1999-00163 and of COFIN00 project {\it Caos e localizzazione in
meccanica classica e quantistica}. 

\appendix
\section{A rigorous definition of temperature}
In this appendix we discuss a rigorous dynamical definition of temperature.
The starting point is the entropy $S$ since in the $\mu$-canonical ensemble, it
plays the role of a generalized thermodynamic potential which allows 
determining (through the computation of suitable derivatives) any other
thermodynamic observable.

In particular, the temperature can be defined from the well known thermodynamic
relation, 
\begin{equation}
{1\over T}=\left({\partial S\over \partial E}\right)_V 
\label{stherm}
\end{equation}
where the subscript $V$ indicates partial derivative at constant
volume.

Upon assuming that the phase space is equipped with a uniform undecomposable 
probability measure, $S$ is given by the logarithm of the volume covered by
all micro-states with energy ${\mathcal H}\leq E$,
\begin{eqnarray}
S(E,N,V) \equiv \ln \Omega (E,N,V) =  
 \ln \int_{{\mathcal H}\leq E}d\Gamma ,
\label{entrop1}
\end{eqnarray}  
where we have neglected an irrelevant multiplicative factor in
front of $\Omega$ necessary only to make the argument of the logarithm 
dimensionless.

From Eqs.~(\ref{stherm}), (\ref{entrop1}) one obtains  
\begin{equation}
{1\over T}= \frac{1}{\Omega} \int_{{\mathcal H}=E}{\dekinchin}
\label{tempa1}
\end{equation}
where the integral is the ``area'' of the constant energy hyper-surface
${\mathcal H} = E$. Upon now introducing a vector ${\bf u}$ such that 
$\nabla\cdot {\bf u} =1$ and using the divergence theorem, we obtain
\begin{equation}
\frac{1}{T} = \frac{\int_{{\mathcal H}=E}\dekinchin}
                  {\int_{{\mathcal H}<E}\nabla\cdot {\bf u} d\Gamma}  =
               \frac{\int_{{\mathcal H}=E}\dekinchin}
                  {\int_{{\mathcal H}=E} \nabla {\mathcal H} \cdot{\bf u}
                 \dekinchin}
\label{Tkin}
\end{equation}
It is easy to recognize that the above equation coincides with 
Eq.~(\ref{tempvir}) which can thus be obtained by following a purely
geometrical approach. More details about the derivation of this formula 
can be found in \cite{GL99}.

\section{Exact solution for the homogeneous harmonic chain}
In this appendix we closely follow the procedure adopted in Ref.~\cite{RLL67} 
to solve the Fokker-Planck (\ref{eq:liouv}) equation for a 
homogeneous harmonic chain. 
Starting from the equilibrium solution (\ref{eq:coreq}), let us define 
\begin{eqnarray}
\overline{\bf U}  & \equiv & {\bf U}_e + \frac{k_B(T_+-T_-)}{2\omega^2}{\bf U} 
 \nonumber \\ 
\overline{\bf V} & \equiv & {\bf V}_e + \frac{k_B(T_+-T_-)}{2}{\bf V} \\ 
\overline{\bf Z} & \equiv & \frac{k_B(T_+-T_-)}{2\lambda}{\bf Z} \quad, 
\nonumber
\label{eq:corst}
\end{eqnarray}
From Eq.~(\ref{eq:FPasy}), it follows that $\bf U$, $\bf V$ and $\bf Z$ satisfy 
the equations,
\begin{eqnarray}
{\bf Z} & = & -{\bf Z}^\dag  \label{eq:corst1} \\
{\bf V} & = & {\bf U} {\bf G} + {\bf ZR} \label{eq:corst2} \\
2{\bf S} -{\bf VR}-{\bf RV} &=& \nu[{\bf GZ}-{\bf ZG}] \label{eq:corst3} 
\end{eqnarray}
where $\nu = \omega^2/\lambda^2$ is the only, dimensionless, parameter
that matters. In addition, $\bf U$ and $\bf V$ are required to be symmetric.
From the peculiar structure of the matrices $\bf R$ and $\bf S$, it follows
that the l.h.s. of Eq.~(\ref{eq:corst3}) is a {\it bordered}
matrix (i.e., its only nonvanishing elements are located on the external
columns and rows). Accordingly, the r.h.s. must be bordered as well, i.e.
in the bulk, $\bf Z$ commutes with $G$. The most general structure of a
matrix commuting with $\bf G$ in the interior is the linear combination
of a matrix $M_{ij}^d$ with equal elements along the diagonals ($i+j$ constant)
and a matrix $M_{ij}^c$ with equal elements along the cross-diagonals
($i-j$ constant). The antisymmetry requirement for $\bf Z$ (see 
Eq.~(\ref{eq:corst1})), implies that no contribution of the second type
is present and, more precisely, that 
\begin{equation} 
Z_{ij} = \phi(j-i)
\label{eq:zsol}
\end{equation} 
with the further constraint $\phi(j) = -\phi(-j)$. The quantities $\phi(j)$ 
are fixed by equating the border elements of the commutator $[G,Z]$
(multiplied by $\nu$) with those of the l.h.s. of Eq.~(\ref{eq:corst3}),
\begin{equation} 
\nu \phi(j) = \delta_{j1} - V_{1j} = \delta_{j1} + V_{N,N-j+1} 
\label{eq:fireq}
\end{equation} 
where $\phi_N \equiv 0$ by definition.
From Eq.~(\ref{eq:corst2}) and its transposed expression it follows that 
$\bf U$ satisfies a similar relation to that for $Z$,
\begin{equation} 
 {\bf GU} - {\bf UG} = {\bf RZ} + {\bf ZR} .
\label{eq:addcor}
\end{equation} 
Accordingly, also $\bf U$ commutes with $G$ in the bulk. The different
symmetry property of $\bf U$ with respect to $\bf Z$ implies, however, that
$\bf U$ is constant along the cross-diagonals. It is easy to verify that
a solution of Eq.~(\ref{eq:addcor}) is given by
\begin{equation} 
U_{ij} = \cases{ \phi(i+j-1) &   \hbox{if}  $i+j \le N $ \cr
                 \phi(2N+1-i-j)  & \hbox{if}  $i+j \ge N $} .
\label{eq:usol}
\end{equation} 
In principle, this is not the only solution of Eq.~(\ref{eq:addcor}), as one 
can add any symmetric matrix commuting with $G$; however, one can check a 
posteriori that the addition of any such matrix would eventually violate
the symmetry properties of $\bf V$.

As a result of Eq.~(\ref{eq:usol}), also the matrix $\bf X$ can be expressed
in terms of the auxiliary variables $\phi(j)$. By replacing the $\bf Z$ and
$\bf X$ solutions in the r.h.s. of Eq.~(\ref{eq:corst2}), we both obtain
an equation for the vector $\phi(j)$,
\begin{equation} 
  \sum_{j=1}^{N-1} K_{ij}\phi(j) = \delta_{1i},
\label{eq:phieq}
\end{equation} 
where ${\bf K} = {\bf G} + \nu {\bf I}$, and the following expression for 
$\bf V$,
\begin{equation} 
 {\bf V} = {\bf S} - \nu {\bf U} .
\label{eq:Vexp}
\end{equation} 
The problem of finding a solution for the heat transport in a homogeneous
chain is accordingly reduced to solving Eq.~(\ref{eq:phieq}) that can be
written as the recursive relation
\begin{equation} 
 \phi(j+1) = (\nu+2)\phi(j) - \phi(j-1)
\label{eq:recphi}
\end{equation} 
which has to be complemented by suitable initial and final conditions.
From the above equation, it follows that $\phi(j)$ is the linear
combination of two exponentials $\exp(\pm\alpha j)$ with
\begin{equation} 
 \hbox{e}^{-\alpha} = 1 + \frac{\nu}{2} - \sqrt{\nu +\frac{\nu^2}{4}}
\label{eq:musol}
\end{equation} 
Upon imposing the appropriate initial conditions, we finally obtain
\begin{equation} 
 \phi(j) = \frac{\sinh(N-j)\alpha}{\sinh N\alpha}
\label{eq:newparal}
\end{equation} 
which completes the solution for the stationary probability distribution.

\end{document}